\documentclass[iop]{emulateapj}

\usepackage{natbib,aas_macros}
\usepackage{multirow,color}

\begin{document}
\shorttitle{
ALMA Census of FIR Size and Luminosity
}
\shortauthors{Fujimoto et al.}
\slugcomment{ApJ in press}

\title{%
Demonstrating A New Census of INfrared Galaxies with ALMA (DANCING-ALMA).\\
I. FIR Size and Luminosity Relation at $\lowercase{z}=0$$-$6 Revealed with 1034 ALMA Sources.
}

\author{%
Seiji Fujimoto\altaffilmark{1}, 
Masami Ouchi\altaffilmark{1,2}, 
Takatoshi Shibuya\altaffilmark{1}, 
and Hiroshi Nagai \altaffilmark{3}
}

\email{sfseiji@icrr.u-tokyo.ac.jp}

\altaffiltext{1}{%
Institute for Cosmic Ray Research, The University of Tokyo,
Kashiwa, Chiba 277-8582, Japan
}
\altaffiltext{2}{%
Kavli Institute for the Physics andMathematics of the Universe 
(Kavli IPMU), WPI, The University of Tokyo, 
Kashiwa, Chiba 277-8583, Japan
}
\altaffiltext{3}{%
National Astronomical Observatory of Japan, Mitaka, 
Tokyo 181-8588, Japan
}

\def\aj{AJ}%
\def\actaa{Acta Astron.}%
\def\araa{ARA\&A}%
\def\apj{ApJ}%
\def\apjl{ApJ}%
\def\apjs{ApJS}%
\def\ao{Appl.~Opt.}%
\def\apss{Ap\&SS}%
\def\aap{A\&A}%
\def\aapr{A\&A~Rev.}%
\def\aaps{A\&AS}%
\def\azh{AZh}%
\def\baas{BAAS}%
\def\bac{Bull. astr. Inst. Czechosl.}%
\def\caa{Chinese Astron. Astrophys.}%
\def\cjaa{Chinese J. Astron. Astrophys.}%
\def\icarus{Icarus}%
\def\jcap{J. Cosmology Astropart. Phys.}%
\def\jrasc{JRASC}%
\def\mnras{MNRAS}%
\def\memras{MmRAS}%
\def\na{New A}%
\def\nar{New A Rev.}%
\def\pasa{PASA}%
\def\pra{Phys.~Rev.~A}%
\def\prb{Phys.~Rev.~B}%
\def\prc{Phys.~Rev.~C}%
\def\prd{Phys.~Rev.~D}%
\def\pre{Phys.~Rev.~E}%
\def\prl{Phys.~Rev.~Lett.}%
\def\pasp{PASP}%
\def\pasj{PASJ}%
\def\qjras{QJRAS}%
\def\rmxaa{Rev. Mexicana Astron. Astrofis.}%
\def\skytel{S\&T}%
\def\solphys{Sol.~Phys.}%
\def\sovast{Soviet~Ast.}%
\def\ssr{Space~Sci.~Rev.}%
\def\zap{ZAp}%
\def\nat{Nature}%
\def\iaucirc{IAU~Circ.}%
\def\aplett{Astrophys.~Lett.}%
\def\apspr{Astrophys.~Space~Phys.~Res.}%
\def\bain{Bull.~Astron.~Inst.~Netherlands}%
\def\fcp{Fund.~Cosmic~Phys.}%
\def\gca{Geochim.~Cosmochim.~Acta}%
\def\grl{Geophys.~Res.~Lett.}%
\def\jcp{J.~Chem.~Phys.}%
\def\jgr{J.~Geophys.~Res.}%
\def\jqsrt{J.~Quant.~Spec.~Radiat.~Transf.}%
\def\memsai{Mem.~Soc.~Astron.~Italiana}%
\def\nphysa{Nucl.~Phys.~A}%
\def\physrep{Phys.~Rep.}%
\def\physscr{Phys.~Scr}%
\def\planss{Planet.~Space~Sci.}%
\def\procspie{Proc.~SPIE}%
         
\def\tcr{\textcolor{red}}
\def\tcb{\textcolor{black}}
\def\tcm{\textcolor{black}}
\def\tck{\textcolor{black}}

\def\rme{\rm e}
\def\rmFIR{\rm FIR}
\def\itHubble{\it Hubble}
\def\rmyr{\rm yr}

\begin{abstract}
We present the large statistics of the galaxy effective radius $R_{\rme}$ in the rest-frame far-infrared (FIR) wavelength $R_{\rme(FIR)}$ obtained from
1627 Atacama Large Millimeter/submillimeter Array (ALMA) 1-mm band maps that become public by 2017 July.
Our ALMA sample consists of 1034 sources with the star-formation rate $\sim100$$-$$1000\,M_{\odot}{\rmyr^{-1}}$ and the stellar mass $\sim$$10^{10}$$-$$10^{11.5}\,M_{\odot}$ at $z=0-6$. 
We homogeneously derive $R_{\rme(FIR)}$ and FIR luminosity $L_{\rmFIR}$ of our ALMA sources via the $uv$-visibility method with the exponential disk model, 
carefully evaluating selection and measurement incompletenesses by realistic Monte-Carlo simulations.
We find that there is a positive correlation between $R_{\rme(FIR)}$ and $L_{\rmFIR}$ at the $>99\%$ significance level. 
The best-fit power-law function, $R_{\rme(FIR)}\propto\,L_{\rmFIR}^{\alpha}$, provides $\alpha=0.28\pm0.07$, 
and shows that $R_{\rme(FIR)}$ at a fixed $L_{\rmFIR}$ decreases toward high redshifts. 
The best-fit $\alpha$ and the redshift evolution of $R_{\rme(FIR)}$ are similar to those of $R_{\rme}$ in the rest-frame UV (optical) wavelength $R_{\rme(UV)}$ ($R_{\rme(Opt.)}$) revealed by {\itHubble\,\,Space\,\,Telescope} (HST) studies. 
We identify that our ALMA sources have significant trends of $R_{\rme(FIR)}\lesssim$$\,R_{\rme(UV)}$ and $R_{\rme(Opt.)}$, 
which suggests that the dusty starbursts take place in compact regions. 
Moreover, $R_{\rme(FIR)}$ of our ALMA sources is comparable to $R_{\rme(Opt.)}$ of quiescent galaxies at $z\sim$1$-$3 as a function of stellar mass, 
supporting the evolutionary connection between these two galaxy populations. 
We also investigate rest-frame UV and optical morphologies of our ALMA sources with deep HST images, 
and find that $\sim$30$-$40\% of our ALMA sources are classified as major mergers.  
This indicates that dusty starbursts are triggered not only by the major mergers but also the other mechanism(s).
\end{abstract}


\section{Introduction}
\label{sec:intro}

The study of galaxy sizes provides key insights into the galaxy formation and the evolution. 
The galaxy sizes are defined by the effective radius of $R_{\rm e}$. 
The $R_{\rm e}$ value in the rest-frame ultra-violet (UV) bands, $R_{\rm e(UV)}$, traces the area of the young star formation with the small amount of dust. 
The $R_{\rm e}$ value in the rest-frame optical bands, $R_{\rm e(Opt.)}$, shows the old star regions that allow us to understand the positional star-formation histories and the stellar migrations. 
The $R_{\rm e}$ value in the rest-frame far-infrared (FIR) bands, $R_{\rm e(FIR)}$, indicates the active star-forming regions that are obscured by the large amount of dust.  
The size-luminosity relation and the size evolution in all UV, optical, 
and FIR wavelengths are important to understand the cosmic stellar mass assembly in galaxies complementary. 

The $R_{\rm e(UV)}$ and $R_{\rm e(Opt.)}$ have been measured for the galaxies selected at the rest-frame UV and optical bands with Advanced Camera for Surveys (ACS)  and Wide Field Camera 3 (WFC3), respectively, on-board {\it Hubble\,Space\,Telescope} (HST) at $0\lesssim z\lesssim8$ \citep[e.g.,][]{shen2003,ferguson2004,hathi2008,oesch2010,ono2013,vanderwel2014,shibuya2015}. 
One of the most extensive size studies is conducted by \cite{shibuya2015} with the sample of $\sim$190,000 galaxies at $z=0-10$ based on the deep HST images. 
The study shows that $R_{\rm e(UV)}$ and rest-frame UV luminosity ($L_{\rm UV}$) have a positive power-law correlation of  
$R_{\rm e(UV)}\propto L_{\rm UV}^{0.27\pm0.01}$, 
and that there is a $R_{\rm e(UV)}$ evolution of $R_{\rm e(UV)}\propto(1+z)^{-1.3\sim-1.0}$. 
The best-fit power-low slope and the redshift evolution are explained by the disk-formation model \citep[e.g.,][]{fall1983,fall2002,mo1998} 
and the dark matter halo assembly \citep[e.g.,][]{ferguson2004,hathi2008}, respectively.

The Atacama Large Millimeter/submillimeter Array (ALMA) enables us to conduct the $R_{\rm e(FIR)}$ measurements with the capabilities of the high angular resolution and sensitivity. 
Bright submillimeter (submm) galaxies (SMGs; $S_{\rm 1 mm}\,\gtrsim$ 1 mJy) have been observed with the high-resolution ($0\farcs16-0\farcs3$) ALMA 870 $\mu$m, and spatially resolved \citep{ikarashi2015,simpson2015a,hodge2016}. 
Recent studies obtain the $R_{\rm e(FIR)}$ of $\sim0.7-2.4$ kpc. 
Compared to the $R_{\rm e(Opt.)}$ values of $\sim$ 4.4 kpc for the SMGs \citep{chen2015}, 
the recent results indicate that the intense starbursts occur in the very compact regions \citep{simpson2015a}.  
\cite{tadaki2016} and \cite{barro2016} also report that the most of the rest-frame FIR sizes are smaller than the rest-frame optical sizes 
based on the similar high-resolution ALMA 870 $\mu$m observations for massive star-forming galaxies at $z\sim2.2-2.5$. 

In the $R_{\rm e(FIR)}$ studies for FIR galaxies fainter than the SMGs, 
\cite{rujopakarn2016} perform ALMA 1.3 mm blind deep imaging in Hubble Ultra Deep Field (HUDF; see also \citealt{dunlop2017}).
This HUDF study identifies 16 sources with the flux range of $S_{\rm 1 mm}\sim$ $0.16-1.0$ mJy, 
and estimates $R_{\rm e (FIR)}$ to be $1.4-4.3$ kpc. 
Another approach is carried out by \cite{gonzalez2017} 
who observe three massive galaxy clusters of Hubble Frontier Fields (HFF) with the ALMA 1.1 mm band for gravitationally lensed sources behind the clusters. 
Twelve sources are identified with the intrinsic flux of $S_{\rm 1 mm}\sim$ $0.14-1.6$ mJy, 
and the intrinsic $R_{\rm e(FIR)}$ values are estimated to be $\sim$ $0.3-2.5$ kpc under the assumption that these twelve sources reside at $z=2$. 
Although these results in HUDF and HFF extend the $R_{\rm e(FIR)}$ studies to the faint end, 
the $R_{\rm e(FIR)}-$FIR luminosity ($L_{\rm FIR}$) relation is still not well determined. 
This is because the source number is not enough to perform a reliable size statistic. 

The $R_{\rm e(FIR)}$ evolution is also the remaining issue. 
Although the previous ALMA studies newly evaluate the $R_{\rm e(FIR)}$ values in the wide $L_{\rm FIR}$ range, 
the sample with a wide redshift range is also necessary to investigate the $R_{\rm e(FIR)}$ evolution. 
Moreover, the technique of the size and flux measurements are different among the previous studies. 
A homogeneous measurement based on a large dataset is required to compare the different redshift samples and 
determine the $R_{\rm e(FIR)}$ evolution. 

In this paper, we make the dataset from the ALMA archive by 2017 July.  
The 1627 deep ALMA maps in Band 6/7 provide us the largest dataset of multi-field ALMA ever made, 
which is composed of 1614 maps from the independent fields and 13 maps from the 3 lensing clusters. 
We analyze this large dataset to investigate the $R_{\rm e(FIR)}-L_{\rm FIR}$ relation and the $R_{\rm e(FIR)}$ evolution 
by a systematic $uv$-visibility based technique. 
This is the first challenge for the statistical investigation of the $R_{\rm e(FIR)}$ properties in the systematic way. 
In Section \ref{sec:data_reduction}, we show the observations and the data reduction. 
Section \ref{sec:data_analysis} describes the method of the source extraction, the flux and size measurements, our simulations, and our sample properties for deriving the $R_{\rm e(FIR)}-L_{\rm FIR}$ relation and the $R_{\rm e(FIR)}$ evolution. 
We report the results of the redshift distribution, the $R_{\rm e(FIR)}-L_{\rm FIR}$ relation, and the $R_{\rm e(FIR)}$ evolution, comparing with the previous studies in Section \ref{sec:result}. 
In Section \ref{sec:discussion}, we discuss the physical origins of the dusty starbursts. 
A summary of this study is presented in Section \ref{sec:summary}. 

Throughout this paper, we assume a flat universe with 
$\Omega_{\rm m} = 0.3$, 
$\Omega_\Lambda = 0.7$, 
$\sigma_8 = 0.8$, 
and $H_0 = 70$ km s$^{-1}$ Mpc$^{-1}$. 
We use magnitudes in the AB system (Oke \& Gunn 1983).

\section{Data and Reduction} 
\label{sec:data_reduction}
\subsection{Data}
\label{sec:data}
To accomplish a complete $R_{\rm e(FIR)}$ study, 
we make full use of ALMA archival data in cycles $0-3$ that became public by 2017 July. 
We collect 1627 continuum maps in Band 6/7 from the ALMA archival data in the regions of HUDF, HFF, 
Cosmic Evolution Survey (COSMOS; \citealt{scoville2007}), 
Subaru/$XMM-Newton$ Deep Survey (SXDS; \citealt{furusawa2008}), 
and The Great Observatories Origins Deep Survey South (GOODS-S; \citealt{vanzella2005}), 
that have rich multi-wavelength data. 
Tables \ref{tab:sub_data} and \ref{tab:our_alma} 
present the summary and the details, respectively, for those of 1627 continuum maps.

We divide these 1627 continuum maps into 10 sub-datasets based on the angular resolutions (Table \ref{tab:sub_data}), 
because the angular resolutions would make different systematics in the $R_{\rm e(FIR)}$ estimates. 
Here, we regard the angular resolutions as the circularized beam size $\theta_{\rm circ}$ given by 
\begin{equation}
\theta_{\rm circ} = \sqrt{ab}, 
\end{equation}
where $a$ and $b$ represent the full width half maximums (FWHMs) of the major- and minor-axis of the synthesized beam, respectively. 
We list up the 1627 continuum maps with the observing details in Table \ref{tab:our_alma}. 

Note that the 1627 continuum maps are taken in two types of the observation modes that are single pointing observations and mosaic observations with several pointing. 
We refer the single pointing and mosaic observations as "single-field data" and "mosaic data", respectively, that are presented in Table \ref{tab:our_alma}.

\begin{table}
\begin{center}
\caption{Summary of Our ALMA Maps
\label{tab:sub_data}}
\begin{tabular}{ccc}
\hline
\hline
Sub-Dataset & $\theta_{\rm circ}$ &Number of Maps \\
(1) & (2) & (3)\\
\hline
 SB1 & 0$''$ $-$ 0$\farcs$2 & 48 \\
 SB2 & 0$\farcs$2 $-$ 0$\farcs$4 & 303 \\ 
 SB3 & 0$\farcs$4 $-$ 0$\farcs$6 & 394 \\
 SB4 & 0$\farcs$6 $-$ 0$\farcs$8 & 203 \\ 
 SB5 & 0$\farcs$8 $-$ 1$\farcs$0 & 254 \\
 SB6 & 1$\farcs$0 $-$ 1$\farcs$2 & 154 \\ 
 SB7 & 1$\farcs$2 $-$ 1$\farcs$4 & 228 \\
 SB8 & 1$\farcs$4 $-$ 1$\farcs$6 & 11 \\ 
 SB9 & 1$\farcs$6 $-$ 1$\farcs$8 & 18 \\
 SB10& 1$\farcs$8 $-$ 2$\farcs$0 & 14 \\ \hline
\hline
\end{tabular}
\end{center}
\footnotesize{Notes: 
(1): ALMA maps with the circularized beam sizes $\theta_{\rm circ}$ from 0$''$ to $2\farcs0$ with a step of $0\farcs2$ 
are referred to as SB1 to SB10, respectively.
(2): Criterion of the $\theta_{\rm circ}$ range for each sub-dataset.  
(3): Number of the ALMA maps in each sub-dataset.
}
\end{table}

\begin{table*}
\caption{Examples of Our ALMA Maps 
\label{tab:our_alma}}
\begin{tabular}{ccccccccc}
\hline
\hline
 Map ID & Project ID & Target & $\lambda_{\rm obs}$ & $\nu_{\rm obs}$ (Band) & Ant. Dist. & $\sigma_{\rm f}$ & Beam Size  & $\theta_{\rm circ}$\\
              &                 &            &        (mm)                   &           (GHz)                   &    (m)        &     (mJy beam$^{-1}$)& ($''\times''$) & $('')$ \\
              &     (1)   &             (2)                  &     (3)                               &     (4)      &     (5)        &    (6)      & (7)  & (8)   \\
\hline
	&			     &				      &          & Single-Field Data &          &        &                 &          \\ \hline
1 & 2011.1.00097.S & COSMOSLowz$\_$64$\_$0 & 0.88 & 342 (7) & 21$-$384 & 0.45 & 0.58 $\times$ 0.42 & 0.49 (SB3) \\
2 & 2011.1.00097.S & COSMOSLowz$\_$64$\_$1 & 0.88 & 342 (7) & 21$-$384 & 0.44 & 0.59 $\times$ 0.42 & 0.49 (SB3) \\
3 & 2011.1.00097.S & COSMOSLowz$\_$64$\_$10 & 0.88 & 342 (7) & 21$-$384 & 0.47 & 0.61 $\times$ 0.41 & 0.5 (SB3) \\
4 & 2011.1.00097.S & COSMOSLowz$\_$64$\_$11 & 0.88 & 342 (7) & 21$-$384 & 0.46 & 0.61 $\times$ 0.42 & 0.5 (SB3) \\
5 & 2011.1.00097.S & COSMOSLowz$\_$64$\_$12 & 0.88 & 342 (7) & 21$-$384 & 0.5 & 0.62 $\times$ 0.41 & 0.5 (SB3) \\
6 & 2011.1.00097.S & COSMOSLowz$\_$64$\_$13 & 0.88 & 342 (7) & 21$-$384 & 0.47 & 0.62 $\times$ 0.41 & 0.5 (SB3) \\
7 & 2011.1.00097.S & COSMOSLowz$\_$64$\_$14 & 0.88 & 342 (7) & 21$-$384 & 0.48 & 0.62 $\times$ 0.42 & 0.51 (SB3) \\
8 & 2011.1.00097.S & COSMOSLowz$\_$64$\_$15 & 0.88 & 342 (7) & 21$-$384 & 0.46 & 0.63 $\times$ 0.41 & 0.5 (SB3) \\
9 & 2011.1.00097.S & COSMOSLowz$\_$64$\_$16 & 0.88 & 342 (7) & 21$-$384 & 0.47 & 0.63 $\times$ 0.41 & 0.5 (SB3) \\
10 & 2011.1.00097.S & COSMOSLowz$\_$64$\_$17 & 0.88 & 342 (7) & 21$-$384 & 0.53 & 0.64 $\times$ 0.42 & 0.51 (SB3) \\
11 & 2011.1.00097.S & COSMOSLowz$\_$64$\_$18 & 0.88 & 342 (7) & 21$-$384 & 0.51 & 0.7 $\times$ 0.4 & 0.52 (SB3) \\
12 & 2011.1.00097.S & COSMOSLowz$\_$64$\_$19 & 0.88 & 342 (7) & 21$-$384 & 0.49 & 0.71 $\times$ 0.4 & 0.53 (SB3) \\
13 & 2011.1.00097.S & COSMOSLowz$\_$64$\_$2 & 0.88 & 342 (7) & 21$-$384 & 0.47 & 0.59 $\times$ 0.42 & 0.49 (SB3) \\
14 & 2011.1.00097.S & COSMOSLowz$\_$64$\_$20 & 0.88 & 342 (7) & 21$-$384 & 0.51 & 0.71 $\times$ 0.4 & 0.53 (SB3) \\
15 & 2011.1.00097.S & COSMOSLowz$\_$64$\_$21 & 0.88 & 342 (7) & 21$-$384 & 0.51 & 0.71 $\times$ 0.4 & 0.53 (SB3) \\
16 & 2011.1.00097.S & COSMOSLowz$\_$64$\_$22 & 0.88 & 342 (7) & 21$-$384 & 0.5 & 0.71 $\times$ 0.4 & 0.53 (SB3) \\
17 & 2011.1.00097.S & COSMOSLowz$\_$64$\_$23 & 0.88 & 342 (7) & 21$-$384 & 0.53 & 0.71 $\times$ 0.4 & 0.53 (SB3) \\
18 & 2011.1.00097.S & COSMOSLowz$\_$64$\_$3 & 0.88 & 342 (7) & 21$-$384 & 0.47 & 0.59 $\times$ 0.42 & 0.49 (SB3) \\
19 & 2011.1.00097.S & COSMOSLowz$\_$64$\_$4 & 0.88 & 342 (7) & 21$-$384 & 0.45 & 0.59 $\times$ 0.42 & 0.49 (SB3) \\
20 & 2011.1.00097.S & COSMOSLowz$\_$64$\_$5 & 0.88 & 342 (7) & 21$-$384 & 0.44 & 0.59 $\times$ 0.42 & 0.49 (SB3) \\
21 & 2011.1.00097.S & COSMOSLowz$\_$64$\_$6 & 0.88 & 342 (7) & 21$-$384 & 0.47 & 0.61 $\times$ 0.42 & 0.5 (SB3) \\
22 & 2011.1.00097.S & COSMOSLowz$\_$64$\_$7 & 0.88 & 342 (7) & 21$-$384 & 0.46 & 0.61 $\times$ 0.42 & 0.5 (SB3) \\
23 & 2011.1.00097.S & COSMOSLowz$\_$64$\_$8 & 0.88 & 342 (7) & 21$-$384 & 0.47 & 0.61 $\times$ 0.42 & 0.5 (SB3) \\
24 & 2011.1.00097.S & COSMOSLowz$\_$64$\_$9 & 0.88 & 342 (7) & 21$-$384 & 0.5 & 0.61 $\times$ 0.41 & 0.5 (SB3) \\
25 & 2011.1.00097.S & COSMOS$\_$lowz$\_$64$\_$0 & 0.88 & 342 (7) & 21$-$382 & 0.28 & 0.58 $\times$ 0.47 & 0.52 (SB3) \\
26 & 2011.1.00097.S & COSMOS$\_$lowz$\_$64$\_$1 & 0.88 & 342 (7) & 21$-$382 & 0.28 & 0.58 $\times$ 0.47 & 0.52 (SB3) \\
27 & 2011.1.00097.S & COSMOS$\_$lowz$\_$64$\_$10 & 0.88 & 342 (7) & 21$-$382 & 0.31 & 0.57 $\times$ 0.47 & 0.51 (SB3) \\
28 & 2011.1.00097.S & COSMOS$\_$lowz$\_$64$\_$11 & 0.88 & 342 (7) & 21$-$382 & 0.31 & 0.57 $\times$ 0.47 & 0.51 (SB3) \\
29 & 2011.1.00097.S & COSMOS$\_$lowz$\_$64$\_$12 & 0.88 & 342 (7) & 21$-$382 & 0.31 & 0.57 $\times$ 0.48 & 0.52 (SB3) \\
30 & 2011.1.00097.S & COSMOS$\_$lowz$\_$64$\_$13 & 0.88 & 342 (7) & 21$-$382 & 0.31 & 0.57 $\times$ 0.48 & 0.52 (SB3) \\
31 & 2011.1.00097.S & COSMOS$\_$lowz$\_$64$\_$2 & 0.88 & 342 (7) & 21$-$382 & 0.31 & 0.58 $\times$ 0.47 & 0.52 (SB3) \\
32 & 2011.1.00097.S & COSMOS$\_$lowz$\_$64$\_$3 & 0.88 & 342 (7) & 21$-$382 & 0.31 & 0.58 $\times$ 0.47 & 0.52 (SB3) \\
33 & 2011.1.00097.S & COSMOS$\_$lowz$\_$64$\_$4 & 0.88 & 342 (7) & 21$-$382 & 0.29 & 0.58 $\times$ 0.47 & 0.52 (SB3) \\
34 & 2011.1.00097.S & COSMOS$\_$lowz$\_$64$\_$5 & 0.88 & 342 (7) & 21$-$382 & 0.3 & 0.58 $\times$ 0.47 & 0.52 (SB3) \\
35 & 2011.1.00097.S & COSMOS$\_$lowz$\_$64$\_$6 & 0.88 & 342 (7) & 21$-$382 & 0.31 & 0.58 $\times$ 0.47 & 0.52 (SB3) \\
36 & 2011.1.00097.S & COSMOS$\_$lowz$\_$64$\_$7 & 0.88 & 342 (7) & 21$-$382 & 0.31 & 0.57 $\times$ 0.47 & 0.51 (SB3) \\
37 & 2011.1.00097.S & COSMOS$\_$lowz$\_$64$\_$8 & 0.88 & 342 (7) & 21$-$382 & 0.29 & 0.57 $\times$ 0.47 & 0.51 (SB3) \\
38 & 2011.1.00097.S & COSMOS$\_$lowz$\_$64$\_$9 & 0.88 & 342 (7) & 21$-$382 & 0.32 & 0.57 $\times$ 0.47 & 0.51 (SB3) \\
39 & 2013.1.00151.S & COSMOS$\_$medz$\_$107$\_$0 & 0.88 & 342 (7) & 21$-$382 & 0.17 & 0.57 $\times$ 0.48 & 0.52 (SB3) \\
40 & 2013.1.00151.S & COSMOS$\_$medz$\_$107$\_$1 & 0.88 & 342 (7) & 21$-$382 & 0.26 & 0.57 $\times$ 0.48 & 0.52 (SB3) \\
41 & 2013.1.00151.S & COSMOS$\_$medz$\_$107$\_$2 & 0.88 & 342 (7) & 21$-$382 & 0.25 & 0.56 $\times$ 0.48 & 0.51 (SB3) \\
42 & 2013.1.00151.S & COSMOS$\_$medz$\_$107$\_$3 & 0.88 & 342 (7) & 21$-$382 & 0.24 & 0.56 $\times$ 0.48 & 0.51 (SB3) \\
43 & 2013.1.00151.S & COSMOS$\_$medz$\_$107$\_$4 & 0.88 & 342 (7) & 21$-$382 & 0.25 & 0.56 $\times$ 0.48 & 0.51 (SB3) \\
44 & 2011.1.00097.S & COSMOShighz$\_$113$\_$0 & 0.88 & 342 (7) & 21$-$375 & 0.21 & 0.68 $\times$ 0.49 & 0.57 (SB3) \\
45 & 2011.1.00097.S & COSMOShighz$\_$113$\_$1 & 0.88 & 342 (7) & 21$-$375 & 0.21 & 0.68 $\times$ 0.49 & 0.57 (SB3) \\
46 & 2011.1.00097.S & COSMOShighz$\_$113$\_$2 & 0.88 & 342 (7) & 21$-$375 & 0.22 & 0.68 $\times$ 0.5 & 0.58 (SB3) \\
47 & 2011.1.00097.S & COSMOShighz$\_$113$\_$3 & 0.88 & 342 (7) & 21$-$375 & 0.22 & 0.68 $\times$ 0.5 & 0.58 (SB3) \\
48 & 2011.1.00097.S & COSMOShighz$\_$113$\_$4 & 0.88 & 342 (7) & 21$-$375 & 0.23 & 0.68 $\times$ 0.51 & 0.58 (SB3) \\
49 & 2011.1.00097.S & COSMOShighz$\_$113$\_$5 & 0.88 & 342 (7) & 21$-$375 & 0.25 & 0.68 $\times$ 0.51 & 0.58 (SB3) \\
50 & 2011.1.00097.S & COSMOShighz$\_$113$\_$6 & 0.88 & 342 (7) & 21$-$375 & 0.24 & 0.68 $\times$ 0.51 & 0.58 (SB3) \\
51 & 2011.1.00097.S & COSMOShighz$\_$138$\_$0 & 0.88 & 342 (7) & 21$-$384 & 0.2 & 0.64 $\times$ 0.51 & 0.57 (SB3) \\
52 & 2011.1.00097.S & COSMOShighz$\_$138$\_$1 & 0.88 & 342 (7) & 21$-$384 & 0.2 & 0.63 $\times$ 0.51 & 0.56 (SB3) \\
53 & 2011.1.00097.S & COSMOShighz$\_$138$\_$2 & 0.88 & 342 (7) & 21$-$384 & 0.19 & 0.62 $\times$ 0.51 & 0.56 (SB3) \\
54 & 2011.1.00097.S & COSMOShighz$\_$138$\_$3 & 0.88 & 342 (7) & 21$-$384 & 0.2 & 0.62 $\times$ 0.51 & 0.56 (SB3) \\
55 & 2011.1.00097.S & COSMOShighz$\_$138$\_$4 & 0.88 & 342 (7) & 21$-$384 & 0.19 & 0.61 $\times$ 0.51 & 0.55 (SB3) \\
56 & 2011.1.00097.S & COSMOShighz$\_$138$\_$5 & 0.88 & 342 (7) & 21$-$384 & 0.19 & 0.61 $\times$ 0.51 & 0.55 (SB3) \\
57 & 2011.1.00097.S & COSMOShighz$\_$138$\_$6 & 0.88 & 342 (7) & 21$-$384 & 0.18 & 0.61 $\times$ 0.51 & 0.55 (SB3) \\
58 & 2011.1.00097.S & COSMOShighz$\_$141$\_$0 & 0.88 & 342 (7) & 21$-$384 & 0.2 & 0.63 $\times$ 0.48 & 0.54 (SB3) \\
59 & 2011.1.00097.S & COSMOShighz$\_$141$\_$1 & 0.88 & 342 (7) & 21$-$384 & 0.2 & 0.63 $\times$ 0.49 & 0.55 (SB3) \\
60 & 2011.1.00097.S & COSMOShighz$\_$141$\_$2 & 0.88 & 342 (7) & 21$-$384 & 0.21 & 0.63 $\times$ 0.49 & 0.55 (SB3) \\
.  & .  & .  & .  & .  &.  & .  &. &.  \\
.  & .  & .  & .  & .  &.  & .  &. &.  \\
\end{tabular}
\end{table*}

\begin{table*}
\begin{tabular}{ccccccccc}
\hline
\hline
 Map ID & Project ID & Target & $\lambda_{\rm obs}$ & $\nu_{\rm obs}$ (Band) & Ant. Dist.  & $\sigma_{\rm f}$ & Beam Size & $\theta_{\rm circ}$  \\
              &                 &            &        (mm)                   &           (GHz)                   &     (m)       &     (mJy beam$^{-1}$)& ($''\times''$) & $('')$ \\
\hline
.  & .  & .  & .  & .  &.  & .  &. &.  \\
.  & .  & .  & .  & .  &.  & .  &. &.   \\
1569 & 2015.1.00039.S & ALESS103.3 & 0.99 & 304 (7) & 15$-$310 & 0.1 & 1.25 $\times$ 0.85 & 1.03 (SB6) \\
1570 & 2015.1.00039.S & ALESS116.2 & 0.99 & 304 (7) & 15$-$310 & 0.1 & 1.26 $\times$ 0.85 & 1.03 (SB6) \\
1571 & 2015.1.00039.S & ALESS124.1$\_$124.4 & 0.99 & 304 (7) & 15$-$310 & 0.1 & 1.26 $\times$ 0.84 & 1.02 (SB6) \\
1572 & 2015.1.00039.S & ALESS14.1 & 0.99 & 304 (7) & 15$-$310 & 0.11 & 1.24 $\times$ 0.84 & 1.02 (SB6) \\
1573 & 2015.1.00039.S & ALESS2.2 & 0.99 & 304 (7) & 15$-$310 & 0.11 & 1.24 $\times$ 0.84 & 1.02 (SB6) \\
1574 & 2015.1.00039.S & ALESS23.1 & 0.99 & 304 (7) & 15$-$310 & 0.11 & 1.24 $\times$ 0.84 & 1.02 (SB6) \\
1575 & 2015.1.00039.S & ALESS3.1 & 0.99 & 304 (7) & 15$-$310 & 0.11 & 1.24 $\times$ 0.84 & 1.02 (SB6) \\
1576 & 2015.1.00039.S & ALESS37.2 & 0.99 & 304 (7) & 15$-$310 & 0.1 & 1.24 $\times$ 0.84 & 1.02 (SB6) \\
1577 & 2015.1.00039.S & ALESS55.2 & 0.99 & 304 (7) & 15$-$310 & 0.1 & 1.24 $\times$ 0.84 & 1.02 (SB6) \\
1578 & 2015.1.00039.S & ALESS69.2$\_$69.3 & 0.99 & 304 (7) & 15$-$310 & 0.11 & 1.25 $\times$ 0.84 & 1.02 (SB6) \\
1579 & 2015.1.00039.S & ALESS72.1 & 0.99 & 304 (7) & 15$-$310 & 0.1 & 1.25 $\times$ 0.84 & 1.02 (SB6) \\
1580 & 2015.1.00039.S & ALESS76.1 & 0.99 & 304 (7) & 15$-$310 & 0.1 & 1.24 $\times$ 0.84 & 1.02 (SB6) \\
1581 & 2015.1.00039.S & ALESS79.4 & 0.99 & 304 (7) & 15$-$310 & 0.11 & 1.25 $\times$ 0.84 & 1.02 (SB6) \\
1582 & 2015.1.00039.S & ALESS87.3 & 0.99 & 304 (7) & 15$-$310 & 0.1 & 1.25 $\times$ 0.84 & 1.02 (SB6) \\
1583 & 2015.1.00039.S & ALESS88.2 & 0.99 & 304 (7) & 15$-$310 & 0.1 & 1.25 $\times$ 0.84 & 1.02 (SB6) \\
1584 & 2015.1.00039.S & ALESS9.1 & 0.99 & 304 (7) & 15$-$310 & 0.11 & 1.24 $\times$ 0.84 & 1.02 (SB6) \\
1585 & 2015.1.00039.S & ALESS99.1 & 0.99 & 304 (7) & 15$-$310 & 0.1 & 1.25 $\times$ 0.85 & 1.03 (SB6) \\
1586 & 2015.1.00540.S & UVISTA-169850 & 1.29 & 233 (6) & 15$-$331 & 0.03 & 1.43 $\times$ 1.08 & 1.24 (SB7) \\
1587 & 2015.1.00540.S & UVISTA-238225 & 1.29 & 233 (6) & 15$-$331 & 0.03 & 1.44 $\times$ 1.08 & 1.24 (SB7) \\
1588 & 2015.1.00540.S & UVISTA-279127 & 1.29 & 233 (6) & 15$-$331 & 0.03 & 1.43 $\times$ 1.09 & 1.24 (SB7) \\
1589 & 2015.1.00540.S & UVISTA-304384 & 1.29 & 233 (6) & 15$-$331 & 0.03 & 1.43 $\times$ 1.09 & 1.24 (SB7) \\
1590 & 2015.1.00540.S & UVISTA-304416 & 1.29 & 233 (6) & 15$-$331 & 0.03 & 1.44 $\times$ 1.08 & 1.24 (SB7) \\
1591 & 2015.1.00540.S & UVISTA-65666 & 1.29 & 233 (6) & 15$-$331 & 0.03 & 1.44 $\times$ 1.08 & 1.24 (SB7) \\
1592 & 2015.1.00664.S & KMOS3DCOS4-10347 & 1.13 & 265 (6) & 15$-$460 & 0.07 & 0.79 $\times$ 0.69 & 0.73 (SB4) \\
1593 & 2015.1.00664.S & KMOS3DCOS4-13174 & 1.13 & 265 (6) & 15$-$460 & 0.07 & 0.79 $\times$ 0.69 & 0.73 (SB4) \\
1594 & 2015.1.00664.S & KMOS3DCOS4-13701 & 1.13 & 265 (6) & 15$-$460 & 0.07 & 0.79 $\times$ 0.69 & 0.73 (SB4) \\
1595 & 2015.1.00664.S & KMOS3DCOS4-15813 & 1.13 & 265 (6) & 15$-$460 & 0.07 & 0.79 $\times$ 0.69 & 0.73 (SB4) \\
1596 & 2015.1.00664.S & KMOS3DCOS4-15820 & 1.13 & 265 (6) & 15$-$460 & 0.07 & 0.78 $\times$ 0.69 & 0.73 (SB4) \\
1597 & 2015.1.00664.S & KMOS3DCOS4-19680 & 1.13 & 265 (6) & 15$-$460 & 0.07 & 0.78 $\times$ 0.69 & 0.73 (SB4) \\
1598 & 2015.1.00664.S & KMOS3DCOS4-24763 & 1.13 & 265 (6) & 15$-$460 & 0.07 & 0.78 $\times$ 0.69 & 0.73 (SB4) \\ \hline
	&			     &				      &          & Mosaic Data\tablenotemark{$\dagger$}   &          &          &      &                         \\ \hline
1599 & 2012.1.00756.S & Field-1$\_$0 & 1.13 & 265 (6) & 20$-$650 & 0.05 & 0.51 $\times$ 0.41 & 0.45 (SB3) \\
1600 & 2012.1.00756.S & Field-1$\_$1 & 1.13 & 265 (6) & 20$-$650 & 0.05 & 0.5 $\times$ 0.41 & 0.45 (SB3) \\
1601 & 2012.1.00756.S & Field-1$\_$2 & 1.13 & 265 (6) & 20$-$650 & 0.05 & 0.5 $\times$ 0.41 & 0.45 (SB3) \\
1602 & 2012.1.00756.S & Field-1$\_$3 & 1.13 & 265 (6) & 20$-$650 & 0.05 & 0.5 $\times$ 0.42 & 0.45 (SB3) \\
1603 & 2012.1.00756.S & Field-1$\_$4 & 1.13 & 265 (6) & 20$-$650 & 0.05 & 0.5 $\times$ 0.42 & 0.45 (SB3) \\
1604 & 2012.1.00173.S & HUDF$\_$0 & 1.36 & 221 (6) & 15$-$1268 & 0.03 & 0.47 $\times$ 0.38 & 0.42 (SB3) \\
1605 & 2012.1.00173.S & HUDF$\_$1 & 1.36 & 221 (6) & 15$-$1268 & 0.03 & 0.47 $\times$ 0.38 & 0.42 (SB3) \\
1606 & 2012.1.00173.S & HUDF$\_$10 & 1.25 & 241 (6) & 15$-$1268 & 0.01 & 0.93 $\times$ 0.69 & 0.8 (SB5) \\
1607 & 2012.1.00173.S & HUDF$\_$2 & 1.36 & 221 (6) & 15$-$1268 & 0.03 & 0.48 $\times$ 0.38 & 0.42 (SB3) \\
1608 & 2012.1.00173.S & HUDF$\_$3 & 1.36 & 221 (6) & 15$-$1268 & 0.03 & 0.47 $\times$ 0.38 & 0.42 (SB3) \\
1609 & 2012.1.00173.S & HUDF$\_$4 & 1.25 & 241 (6) & 15$-$1268 & 0.01 & 0.65 $\times$ 0.51 & 0.57 (SB3) \\
1610 & 2012.1.00173.S & HUDF$\_$5 & 1.36 & 221 (6) & 15$-$1268 & 0.03 & 0.47 $\times$ 0.38 & 0.42 (SB3) \\
1611 & 2012.1.00173.S & HUDF$\_$6 & 1.36 & 221 (6) & 15$-$1268 & 0.03 & 0.47 $\times$ 0.38 & 0.42 (SB3) \\
1612 & 2012.1.00173.S & HUDF$\_$7 & 1.25 & 241 (6) & 15$-$1268 & 0.01 & 0.87 $\times$ 0.66 & 0.75 (SB4) \\
1613 & 2012.1.00173.S & HUDF$\_$8 & 1.25 & 241 (6) & 15$-$1268 & 0.01 & 0.71 $\times$ 0.56 & 0.63 (SB4) \\
1614 & 2012.1.00173.S & HUDF$\_$9 & 1.25 & 241 (6) & 15$-$1268 & 0.01 & 1.08 $\times$ 0.75 & 0.9 (SB5) \\
1615 & 2013.1.00999.S & Abell$\_$2744$\_$0 & 1.14 & 263 (6) & 15$-$820 & 0.06 & 0.65 $\times$ 0.5 & 0.57 (SB3) \\
1616 & 2013.1.00999.S & Abell$\_$2744$\_$1 & 1.14 & 263 (6) & 15$-$820 & 0.06 & 0.59 $\times$ 0.47 & 0.52 (SB3) \\
1617 & 2013.1.00999.S & Abell$\_$2744$\_$2 & 1.14 & 263 (6) & 15$-$820 & 0.06 & 0.66 $\times$ 0.5 & 0.57 (SB3) \\
1618 & 2013.1.00999.S & Abell$\_$2744$\_$3 & 1.14 & 263 (6) & 15$-$820 & 0.06 & 0.63 $\times$ 0.5 & 0.56 (SB3) \\
1619 & 2013.1.00999.S & Abell$\_$2744$\_$4 & 1.14 & 263 (6) & 15$-$820 & 0.06 & 0.64 $\times$ 0.5 & 0.56 (SB3) \\
1620 & 2013.1.00999.S & Abell$\_$2744$\_$5 & 1.14 & 263 (6) & 15$-$820 & 0.06 & 0.65 $\times$ 0.5 & 0.57 (SB3) \\
1621 & 2013.1.00999.S & Abell$\_$2744$\_$6 & 1.14 & 263 (6) & 15$-$820 & 0.06 & 0.63 $\times$ 0.5 & 0.56 (SB3) \\
1622 & 2013.1.00999.S & Abell$\_$2744$\_$7 & 1.14 & 263 (6) & 15$-$820 & 0.06 & 0.63 $\times$ 0.5 & 0.56 (SB3) \\
1623 & 2013.1.00999.S & MACSJ0416.1-2403$\_$0 & 1.14 & 263 (6) & 15$-$349 & 0.07 & 1.53 $\times$ 0.85 & 1.14 (SB6) \\
1624 & 2013.1.00999.S & MACSJ0416.1-2403$\_$1 & 1.14 & 263 (6) & 15$-$349 & 0.06 & 1.54 $\times$ 0.86 & 1.15 (SB6) \\
1625 & 2013.1.00999.S & MACSJ0416.1-2403$\_$2 & 1.14 & 263 (6) & 15$-$349 & 0.06 & 1.54 $\times$ 0.85 & 1.14 (SB6) \\
1626 & 2013.1.00999.S & MACSJ0416.1-2403$\_$3 & 1.14 & 263 (6) & 15$-$349 & 0.06 & 1.5 $\times$ 0.86 & 1.13 (SB6) \\
1627 & 2013.1.00999.S & MACSJ1149.5+2223$\_$0 & 1.14 & 263 (6) & 15$-$349 & 0.07 & 1.2 $\times$ 1.13 & 1.16 (SB6) \\
\hline
\end{tabular}
\footnotesize{Notes: 
(1) ALMA project code.
(2) Target name in the initial ALMA observation. 
(3) Wavelength in the observed frame. 
(4) Frequency in the observed frame. 
(5) Range of the antenna distances. 
(6) One sigma noise measured after the CLEAN process. 
(7) Synthesized beam size (weighting="natural"). 
(8) Circularized beam size. Classification of the sub-dataset is presented in the parentheses. 
}
$\dagger$ We cut out the mosaic maps based on the source positions identified in the dirty maps to perform the data analyses (Section \ref{sec:reduction}). 
\footnotesize{(The complete table is available in a machine-readable form in the online journal.)}
\end{table*}

\begin{figure*}
\begin{center}
\center{\includegraphics[trim=0.0cm 0.4cm 0cm 0.6cm, clip,angle=0,width=1.0 \textwidth]{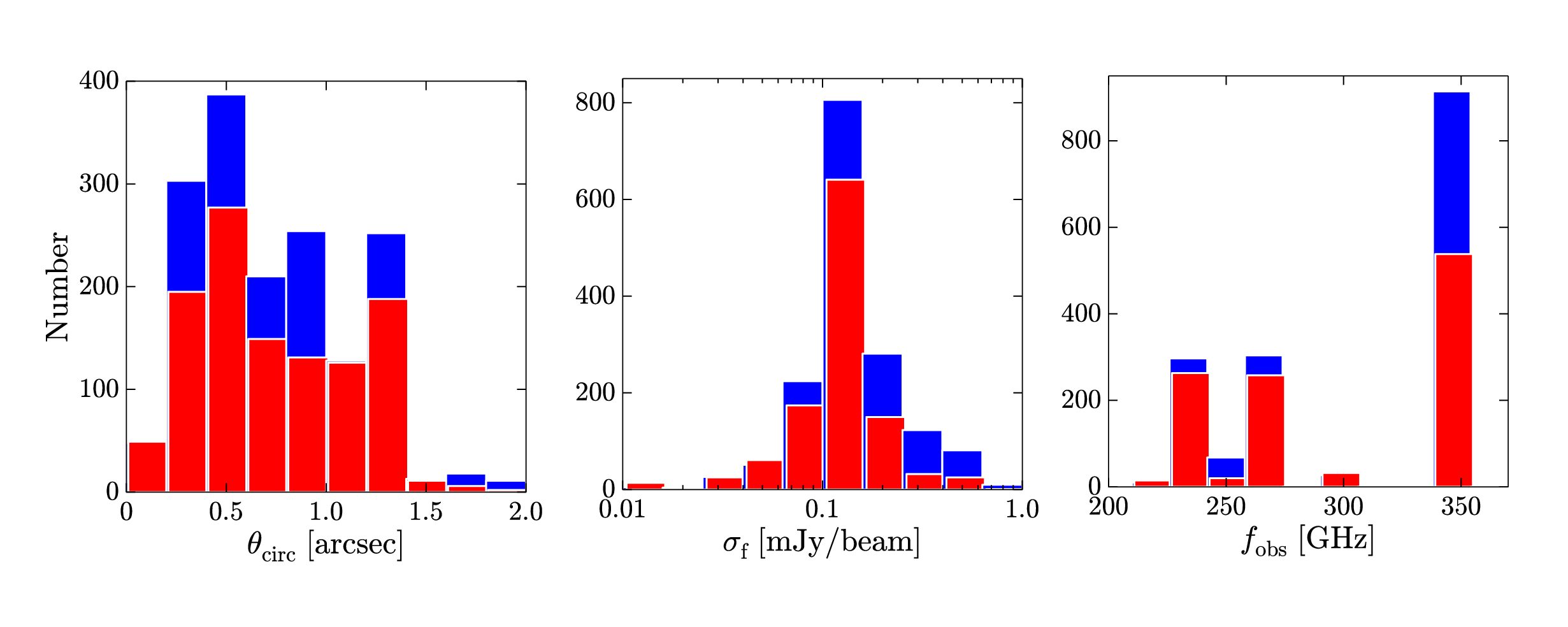}}
\caption[]{
Histograms of the circularized beam size (left), the sensitivity (middle), and the frequency in the observed frame (right) for our ALMA maps. 
The blue histograms denote our 1627 ALMA maps. 
The red histograms represent our ALMA maps with the ALL5S source(s) whose peak pixel value(s) exceeds the $5\sigma$ confidence level. 
\label{fig:data_info}}
\end{center}
\end{figure*}

\subsection{Data Reduction}
\label{sec:reduction}
Basically, the data is reduced in the same manner as \cite{fujimoto2016} 
In this process, we use the CASA versions from 4.2.0 to 4.7.2. 
Our CASA reduction has three major steps: bad data flagging, 
bandpass calibration, and 
gain calibration with flux scaling. 
These major steps are performed with the scripts provided by the ALMA observatory. 
We apply additional flaggings if we find problems in the final images that the noise level is significantly higher than the calibrated products provided by the ALMA observatory, or that there remain striped patterns. 
For the cycle 0 data, we update the flux model to 'Bulter-JPL-Horizons 2012', if the data is calibrated with the old version of the flux model of 'Bulter-JPL-Horizons 2010'.

Because the HUDF region is observed in two different ALMA projects (\#2013.1.00718.S and \#2013.1.00173.S), 
we combine these two data sets with the {\sc concat} task. 
Here, we recalculate the data weights with the {\sc statwt} task based on its visibility scatters which include the effects of integration time, 
channel widths, and systematic temperature. 

We perform Fourier transformations for the reduced $uv$-data to create "dirty" continuum maps. 
With the dirty maps, we estimate the standard deviation, $\sigma_{\rm d}$, of the pixel values. 

For the single-field data, 
we process the maps with the {\sc CLEAN} algorithm in the following three steps: 
1) identifying peak pixels with the $\geq$ 50$\sigma_{\rm d}$ level, 
2) making CLEAN boxes with the {\sc boxit} task for the peak pixels given in the step 1), and
3) performing the {\sc CLEAN} algorithm down to the depth of the 3$\sigma_{\rm d}$ level with the natural weighting in the {\sc CLEAN} boxes. 
We repeat 1)-3) for four times, changing from the 50$\sigma_{\rm d}$ level to the 20$\sigma_{\rm d}$, 10$\sigma_{\rm d}$, and 5$\sigma_{\rm d}$ levels in the step 1). 
We then obtain the final maps. 
The process from the dirty maps to the final maps is referred to as "CLEAN process". 
We define the 1$\sigma$ noise level of each ALMA map, $\sigma_{\rm f}$, with the standard deviation of the pixel values in the residual map.  

For the mosaic data, 
we find that the CLEAN process cannot be performed due to the large data size. 
We thus reduce the data size of the mosaic data, having the following three steps.
I) 
We extract the sources whose peak flux values exceed the $5\sigma_{\rm d}$ levels in the dirty maps. 
Here, we identify the source positions. 
II) 
We select pointings whose central positions are located within $20''$ from the source positions.  
III) 
We create a new dirty map only with the $uv$-visibility from these pointings. 
Completing above three steps, 
we carry out the CLEAN process to make final maps in the same manner as the single-field data. 

Since our scope is the dust continuum, it is desired to remove the contamination of atomic and molecular transition lines in the FIR wavelength. 
There are two cases that the FIR lines are included in the channels of ALMA Band 6/7. 
First case is that the initial ALMA data is taken for the FIR lines. 
In our ALMA data, the ALMA projects of \#2012.1.00076.S, \#2013.1.00668.S, and \#2012.1.00523.S are taken for the CO or [C {\sc ii}] 158 $\mu$m lines from high-$z$ star-forming galaxies. 
For those of three ALMA data, we remove the line channels 
if robust line profiles are identified in the data cubes. 
Second case is that the FIR lines are contaminated by chance. 
Among the FIR lines, the [C {\sc ii}] 158 $\mu$m line is one of the brightest FIR lines in the star-forming galaxies (Stacey et al. 1991), 
which is potentially contaminated in ALMA Band 6/7 with the source redshifts at $z\sim$ 4$-$7. 
In fact, \cite{swinbank2012} have identified two [C {\sc ii}] 158 $\mu$m lines from two SMGs at $z=4.42$ and $z=4.44$ in an ALMA survey of submillimeter galaxies in the Extended Chandra Deep Field-South (ALESS) with 126 ALMA data cubes. 
This indicates that the [C {\sc ii}] 158-$\mu$m line contamination is not significant with the chance of $\sim$ 1$-$2\%. 
Moreover, the flux densities of the [C {\sc ii}] 158-$\mu$m lines identified in \cite{swinbank2012} are calculated to contribute only $\lesssim10$\% of those of the dust continuum. 
Therefore, we assume that the second case is negligible in our statistical results, 
and do not take it into account through this paper. 

The 1627 final maps of the pointing and mosaic data achieve angular resolutions of $0\farcs18\times 0\farcs18$$-$$4\farcs08\times1\farcs07$ 
and the sensitivities of 0.01$-$1.4 mJy beam$^{-1}$ before the primary beam corrections.  
We summarize the properties of the final maps in Table \ref{tab:our_alma}. 
In Figure \ref{fig:data_info}, we also show the histograms of the circularized beam size, the sensitivity, and the observed frequency for the final maps. 

\section{Data Analysis}
\label{sec:data_analysis}

\subsection{Source Detection}
\label{sec:source_detection}
\tcb{
We conduct source extractions for our ALMA maps before primary beam corrections 
with {\small\sc sextractor} version 2.5.0 \citep{bertin1996}. 
The source extraction is carried out in the high sensitivity regions.
For the single-field data, 
we use the regions with the primary beam sensitivity greater than 50\%.
For the mosaic data, we perform the source extractions
where the relative sensitivity to the deepest part of the mosaic map is greater than 50\%.}

\tcb{
We identify sources with a positive peak count ($S^{\rm peak}_{\rm obs}$) 
above the 3$\sigma_{\rm f}$ level. 
We then select only the sources with 
$S^{\rm peak}_{\rm obs}\geq5\sigma_{\rm f}$ to make a 5$\sigma$-peak catalog. 
We obtain 1034 sources in the 5$\sigma$-peak catalog. 
The details of these 1034 sources are presented in Table \ref{tab:source_catalog}. 
To perform reliable size measurements, 
we select 780 sources with total flux densities ($S^{\rm total}_{\rm {\sc obs}}$)
above the 10$\sigma_{\rm f}$ level from the 5$\sigma_{\rm f}$-peak catalog, making a 10$\sigma_{\rm f}$-total catalog.
The total flux densities 
are measured with the {\sc imfit} task in CASA that carries out the fitting routine with the 2D Gaussian model on the image plane. 
We refer the $5\sigma_{\rm f}$-peak and $10\sigma_{\rm f}$-total catalogs as "ALL5S" and "ALL10S", respectively.  
The summary of the source catalogs is presented in Table \ref{tab:alma_catalog_sum}.
}

\begin{table}
{\scriptsize
\caption{Our Catalogs 
\label{tab:alma_catalog_sum}}
\begin{tabular}{ccc}
\hline
\hline
Catalog Name & Selection Criteria & Source Number \\
(1) &  (2) & (3) \\
\hline
ALL5S   & $S_{\rm obs}^{\rm peak}\geq 5\sigma_{\rm f}$  & 1034 \\
ALL10S &  $S_{\rm obs}^{\rm peak}\geq 5\sigma_{\rm f}\,\cap\,S_{\rm obs}^{\rm total}\geq 10\sigma_{\rm f}$ & 780 \\
OC5S   & ALL5S $\cap$ optical-NIR counterpart    & 577 \\
OC10S  & ALL10S $\cap$ optical-NIR counterpart & 444 \\ 
OC5S-mmT & OC5S $\cap$ FIR-mm target obs. & 131 \\ 
OC5S-optT & OC5S $\cap$ optical target obs. & 446 \\
\hline
\end{tabular}

\footnotesize{Notes: 
(1) Name of our ALMA source catalog. 
(2) Selection criteria of the ALMA sources for the catalog (see text).
(3) Source number in the catalog. 
}
}
\end{table} 

\begin{turnpage}
\begin{table*}
{\scriptsize
\caption{Examples of the 5$\sigma$-peak catalog (ALL5S)}
\begin{tabular}{cllllllllllllll}
\hline
\hline
 Source ID & R.A. & Dec. & $\lambda_{\rm obs}$ & SN$^{\rm peak}$ & $S_{\rm obs}^{\rm peak}$ & $S_{\rm pb}^{\rm imfit}$ & $S_{\rm pb}^{uv.{\rm fit}}$ & $S_{\rm corr}^{uv.{\rm fit}}$ & $R_{\rm e, obs}^{uv.{\rm fit}}$ & $R_{\rm e, crr}^{uv. {\rm fit}}$ & flag & offset & $z_{\rm phot}$ & $H$-mag  \\
          & (J2000) & (J2000)       &    (mm)                   &          &       (mJy/beam)      &      (mJy)      &      (mJy)   &    (mJy)    & ($''$)                                          & ($''$)                          &                   & ($''$) &                     & (mag)\\
          &    &        &      (1)         &   (2)  &                   (3)                    &              (4)                        &                  (5)                       &    			(6)                   &              		(7)                        &   (8)   &        (9) 	     &            (10)    & (11)  & (12)    \\
\hline
1 & 150.203888 & 2.50107 & 0.87 & 5.6 & 1.13 $\pm$ 0.2 & 2.22 $\pm$ 0.39 & $-$ & $-$ & $-$ & $-$ & $-$ & $-$ & $-$ & $-$ \\
2 & 150.192612 & 2.219835 & 0.87 & 17.5 & 3.61 $\pm$ 0.21 & 5.51 $\pm$ 0.21 & 5.23 $\pm$ 0.66 & 4.32 $\pm$ 0.55 & 0.23 $\pm$ 0.05 & 0.21 $\pm$ 0.04 & 2 & $-$ & $-$ & $-$ \\
3 & 149.882294 & 2.507151 & 0.87 & 9.3 & 1.92 $\pm$ 0.21 & 3.85 $\pm$ 0.35 & 3.66 $\pm$ 0.84 & 3.16 $\pm$ 0.73 & 0.13 $\pm$ 0.07 & 0.14 $\pm$ 0.08 & 2 & 0.03 & 2.38 & 23.1 \\
4 & 150.028015 & 2.43566 & 0.87 & 5.2 & 0.68 $\pm$ 0.13 & 0.8 $\pm$ 0.15 & $-$ & $-$ & $-$ & $-$ & $-$ & $-$ & $-$ & $-$ \\
5 & 150.064514 & 2.32913 & 0.87 & 9.6 & 1.24 $\pm$ 0.13 & 1.33 $\pm$ 0.13 & 1.33 $\pm$ 0.28 & 1.15 $\pm$ 0.24 & 0.09$\pm$ 0.07 & 0.10 $\pm$ 0.08 & 1 & $-$ & $-$ & $-$ \\
6 & 150.092575 & 2.398174 & 0.87 & 9.1 & 1.22 $\pm$ 0.13 & 1.68 $\pm$ 0.15 & 1.61 $\pm$ 0.37 & 1.4 $\pm$ 0.32 & 0.11 $\pm$ 0.07 & 0.12 $\pm$ 0.08 & 0 & $-$ & $-$ & $-$ \\
7 & 150.104691 & 2.243663 & 0.87 & 6.5 & 0.6 $\pm$ 0.09 & 0.58 $\pm$ 0.09 & $-$ & $-$ & $-$ & $-$ & $-$ & $-$ & $-$ & $-$ \\
8 & 150.018112 & 2.34992 & 0.87 & 25.6 & 2.38 $\pm$ 0.09 & 3.94 $\pm$ 0.14 & 4.02 $\pm$ 0.36 & 3.74 $\pm$ 0.33 & 0.14 $\pm$ 0.03 & 0.12 $\pm$ 0.02 & 0 & 0.09 & 2.27 & 28.3 \\
9 & 150.373688 & 2.112025 & 0.87 & 5.5 & 0.5 $\pm$ 0.09 & 0.46 $\pm$ 0.09 & $-$ & $-$ & $-$ & $-$ & $-$ & 0.13 & -98.99 & 20.7 \\
10 & 149.957428 & 2.028263 & 0.87 & 9.1 & 0.87 $\pm$ 0.1 & 1.37 $\pm$ 0.13 & 1.39 $\pm$ 0.34 & 1.2 $\pm$ 0.29 & 0.16 $\pm$ 0.08 & 0.17 $\pm$ 0.08 & 0 & $-$ & $-$ & $-$ \\
\hline
\end{tabular}
\tablecomments{
\footnotesize{
(1) Wavelength in the observed frame.
(2) Peak SNR in the ALMA maps before primary beam correction. 
(3) Peak value in the ALMA maps before primary beam correction. 
(4) Integrated flux density measured with the {\sc IMFIT} task after primary beam correction. 
(5) Integrated flux density measured with the {\sc UVMULTIFIT} task after primary beam correction.
(6) Integrated flux density measured with the {\sc UVMULTIFIT} task after the corrections of the primary beam and the Monte-Carlo simulation.
(7) FWHM of the source size obtained with the {\sc UVMULTIFIT} task. 
(8) FWHM of the source size derived with the {\sc UVMULTIFIT} task after the correction of the Monte-Carlo simulation. 
(9) Flag for the reliability of the {\sc UVMULTIFIT} fitting. 
(10) Offset between the centers of the ALMA source and the optical-NIR counterpart in unit of arcsecond. 
(11) Photometric redshift of the optical-NIR counterpart. 
(12) $H$-band AB magnitude of the optical-NIR counterpart.   
 \label{tab:source_catalog}}}
 }
\footnotesize{(The complete table is available in a machine-readable form in the online journal.)}
\end{table*}
\end{turnpage}

\subsection{Flux and Size Measurement}
\label{sec:flux_size_measure}

\begin{figure*}[h]
\begin{center}
\center{\includegraphics[trim=0.0cm 0cm 0cm 0cm, clip,angle=0,width=1.0 \textwidth]{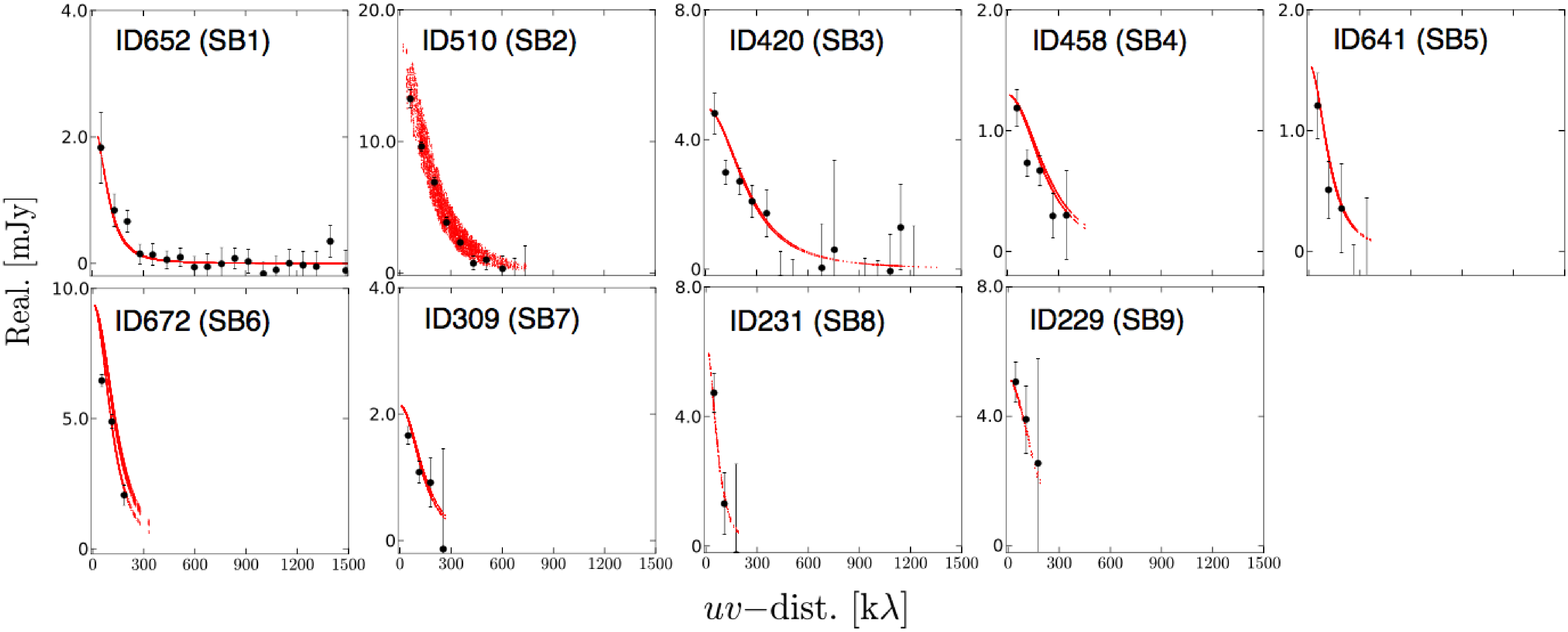}}
\caption{\tcb{
Examples of the best-fit result of the {\sc uvmultifit} task for ALL10S sources in the sub-datasets of SB1$-$SB9 on the $uv$-visibility. 
The black points denote the visibility amplitudes obtained from our ALMA data. 
The red dots indicate our best-fit results with the exponential disk model. 
Because there are no ALL10S sources with flag = 0 (reliable) in SB10, 
we do not show the example of the best-fit result of SB10. 
\label{fig:uv-dist_amp}}}
\end{center}
\end{figure*}

We estimate the flux densities and the sizes for our 780 objects in ALL10S. 
We perform {\sc uvmultifit} (\citealt{marti2014}) that is a simultaneous fitting tool for multiple objects on $uv$-visibility.
For the visibility fitting, 
we use the flux density and the size measurements obtained by the {\sc imfit} task as the initial values, 
while the source positions are fixed. 
Here we adopt a symmetric exponential disk model. 
\cite{hodge2016} estimate the median S$\acute{\rm e}$rsic index $n$ for 15 SMGs as $n=0.9\pm0.2$, 
which is close to our assumption of the exponential disk with $n=1$. 
The fitting result of the FWHM values are converted to the $R_{\rm e}$ values via the relation of $R_{\rm e}=0.826\,{\rm FWHM}$ in the case of $n=1$ \citep{macarthur2003}. 
Hereafter, the $R_{\rm e}$ values are referred to as our size measurements. 
Figure \ref{fig:uv-dist_amp} shows the several examples of our best-fit results with the {\sc uvmultifit} task. 
The flux densities for the rest of the 254 ($=$1034$-$780) sources in ALL5S are obtained by the {\sc imfit} task, 
and we do not estimate the sizes for those of the 254 sources.  

To remove the bad visibility data, 
we flag the ALL10S sources by visual inspection on the visibility amplitude plots such shown in Figure 2. 
We define flag = 0, 1, and 2 as a source whose fitting result is reliable, tentative, and bad, respectively.  
We use only the results with flag = 0 in the following analyses in this paper. 
In Table \ref{tab:source_catalog}, we summarize the flux and size measurements obtained by the {\sc uvmultifit} task together with the flag values. 
For the sources with flag = 1 and 2, 
we use the flux densities that are obtained by the {\sc imfit} task, 
and do not include the size measurement in this paper. 

\subsection{Simulation and Correction}
\label{sec:simulation}
We investigate the systematics in the flux density and size measurements with the {\sc uvmultifit} task for the ALL10S sources, 
performing Monte-Carlo simulations for the flux density and the size measurements. 
Because the angular resolution of the ALMA maps may produce different systematics in these measurements, 
we conduct the Monte-Carlo simulations based on each sub-dataset (Table \ref{tab:sub_data}).

\begin{figure*}
\begin{center}
\center{\includegraphics[trim=0.0cm 1.0cm 0cm 1.0cm, clip,angle=0,width=1.0 \textwidth]{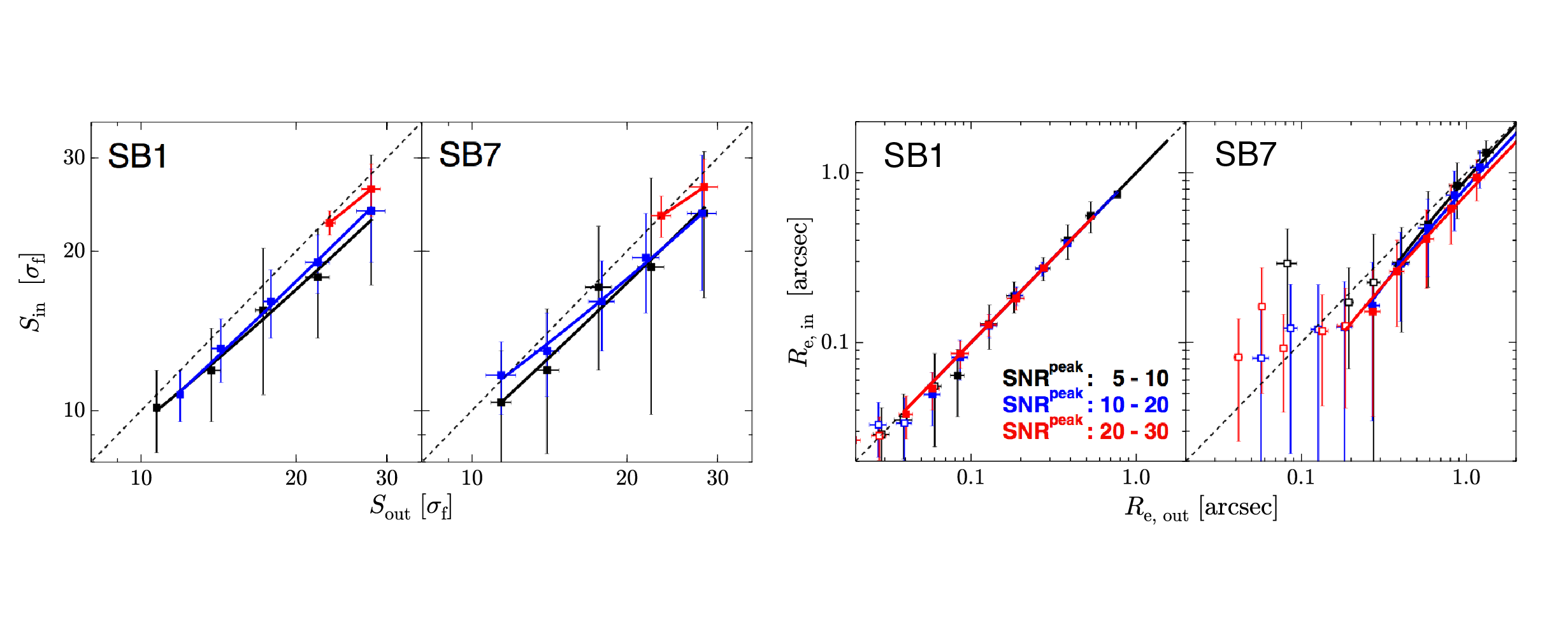}}
\vspace{-0.5cm}
\caption{
Relationship between input and output values of the flux density (left) and size (right) in two sub-datasets of SB1 and SB7, as for example. 
The black, blue, and red circles (curves) indicate the values (the best-fit functions) of the samples with the output peak SNR ranges of 5$-$10, 10$-$20, and 20$-$30, respectively. 
In the left panel, the flux density of the input and output values are normalized by $\sigma_{\rm f}$. 
In the right panel, the open squares indicate the output size measurements that are below the reliable size measurement limits (Section \ref{sec:comp_pre}). 
We do not include the open squares for the fitting. 
\label{fig:simulation}}
\end{center}
\end{figure*}

First, we select one ALMA map from each sub-dataset, making 10 representative ALMA maps. 
The representative ALMA map is randomly selected from the ALMA maps in which no sources are detected. 
Second, we create $26,600$ artificial sources 
with the uniform distribution of the total fluxes in the $5\sigma_{\rm f}-45\sigma_{\rm f}$ levels and the circularized source sizes of $0\farcs04-3\farcs0$. 
We then inject the artificial sources individually into the $uv$-visibilities of each representative ALMA map at random positions within a 10$''$ radius from the centers. 
Finally, we obtain the flux densities and sizes of the artificial sources in the same manner as the flux density and the size measurements for the real sources.  
Figure \ref{fig:simulation} shows two example results of the input and output values of the flux density and the size for the sub-datasets of SB1 and SB7. 

As shown in Figure \ref{fig:simulation}, 
we find that the output values of both flux density and size well recover the input values within the $1\sigma$ uncertainty. 
However, we also find that output values have a slight offset from the input values in some sub-datasets, 
which suggests that we need the corrections for the output values. 

To carry out the corrections for the flux density and size measurements, 
we model the input values of the flux density and the size in Figure \ref{fig:simulation} as a function of output value, 
\begin{eqnarray}
S_{\rm in} &=&  A_{0}\times S_{\rm out} + A_{1}, \\
R_{\rm e,\,in} &=& B_{0}\times R_{\rm e,\,out} + B_{1},
\end{eqnarray}
where $S_{\rm out}$ ($S_{\rm in}$) and ${\rm FWHM}_{\rm out}$ (${\rm FWHM}_{\rm in}$) are the output (input) values of the flux density and the size, respectively. 
$A_{0}, A_{1}, 
 B_{0},$ and $B_{1}$  are the free parameters. 
In the model fitting, we divide our simulation results into three bins with the output peak signal to noise ratio (SNR) of 5$-$10, 10$-$20, and 20$-$30.
In Figure \ref{fig:simulation}, we plot the best-fit models for these three peak SNR bins. 
Based on the peak SNR of each source and those of the best-fit models, 
we correct the values of the flux density and the size for the ALL10S sources. 
For the bright  ALL10S sources with the peak SNR $>$ 30, we use the best-fit model result of the peak SNR $=$ 20 $-$ 30. 
Table \ref{tab:source_catalog} lists the values of the flux density and the size after the corrections.

\subsection{Comparison with Previous Measurements}
\label{sec:comp_pre}
To investigate the potential systematics in our method, 
we compare our flux and size measurements with the previous ALMA results in the literature \citep{ikarashi2015, simpson2015a, tadaki2016, barro2016, hodge2016,rujopakarn2016,gonzalez2017}. 

\begin{figure}
\begin{center}
\includegraphics[trim=0cm 0cm 0cm 0cm, clip, angle=0,width=0.5\textwidth]{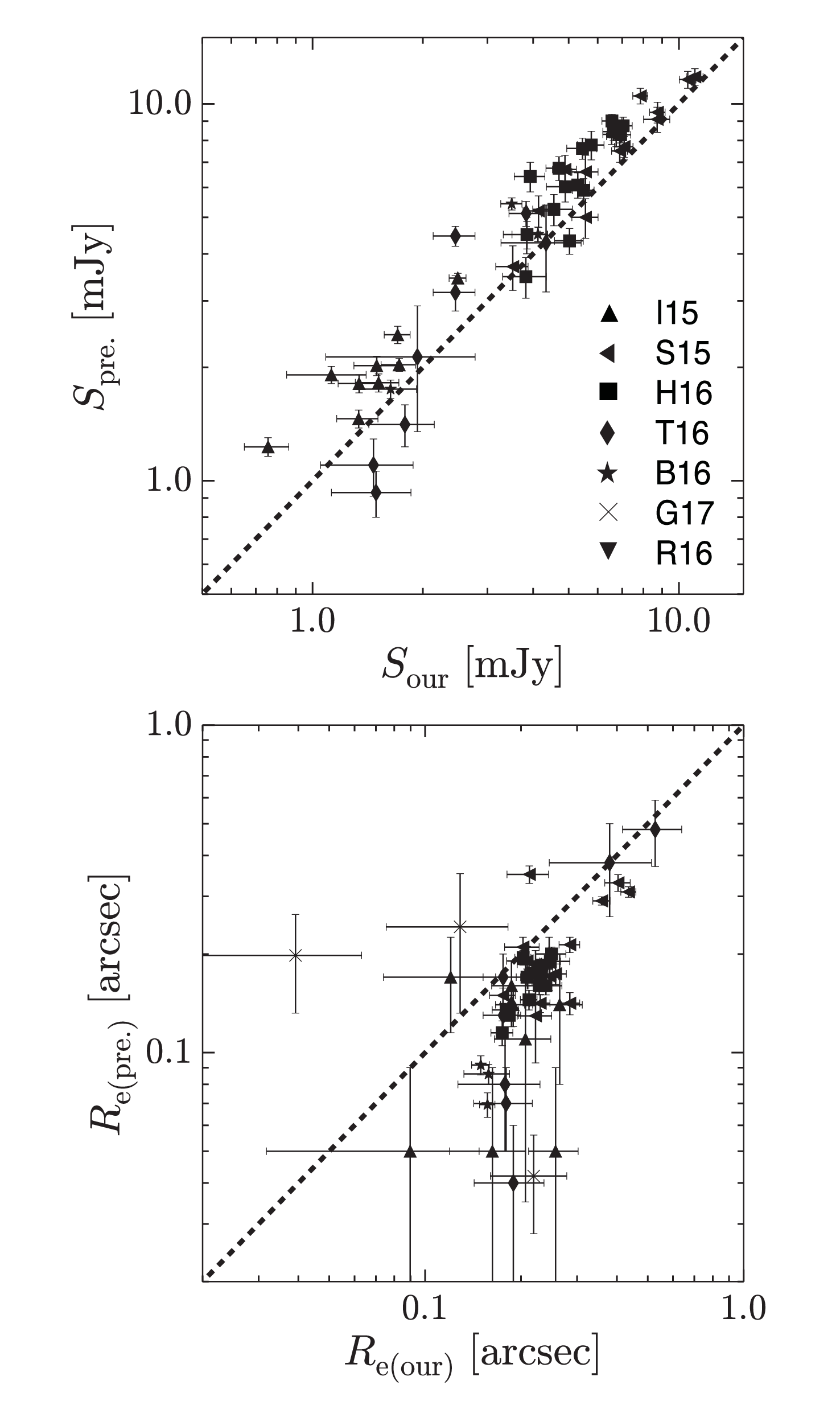}
\vspace{-0.5cm}
 \caption[]{
Flux (top) and size (bottom) comparison between our measurements and the previous ALMA results in the literature. 
The abscissa and ordinate axes provide our measurements ($S_{\rm our}$, $R_{\rm e(our)}$) and the previous ALMA results ($S_{\rm pre.}$, $R_{\rm e(pre.)}$), respectively, 
with  
black squares (H16; \citealt{hodge2016}), 
triangles (I15; \citealt{ikarashi2015}), 
sideways triangles (S15; \citealt{simpson2015a}), 
diamonds, (T16; \citealt{tadaki2016}), 
stars (B16; \citealt{barro2016}), 
crosses (G17; \citealt{gonzalez2017}), 
and inverse triangles (R16; \citealt{rujopakarn2016}). 
For the gravitationally lensed sources in HFF, we use the flux and size measurements before the lensing correction in the comparison. 
\label{fig:comp_fluxsize}}
\end{center}
\end{figure}

Figure \ref{fig:comp_fluxsize} presents our flux and size measurements and the previous ALMA results. 
For a fair comparison, the size values obtained in the previous ALMA results are converted into the $R_{\rm e}$ values. 
In Figure \ref{fig:comp_fluxsize}, 
our flux measurements show a good agreement with the previous ALMA results in the wide flux range. 
Our size measurements are also consistent with the majority of the previous ALMA results within the $\sim$ 1$-$2 $\sigma$ errors, 
and no systematic offsets are identified in the size comparison. 
Although the scatter in the size comparison seems to be larger than that in the flux comparison, 
the large scatter indicates that the size measurement is sensitive to the way of measurements, 
the assumptions of the model profile, the parameter range, and the initial values in the fitting process. 
Therefore, we conclude that both of our flux and size measurements are consistent with the previous ALMA results, and not biased by any systematics.  

\subsection{Selection and Measurement Completeness}
\label{sec:selection_comp}
\begin{figure*}
\begin{center}
\includegraphics[trim=0cm -0.2cm 0cm 0cm, clip, angle=0,width=1.0\textwidth]{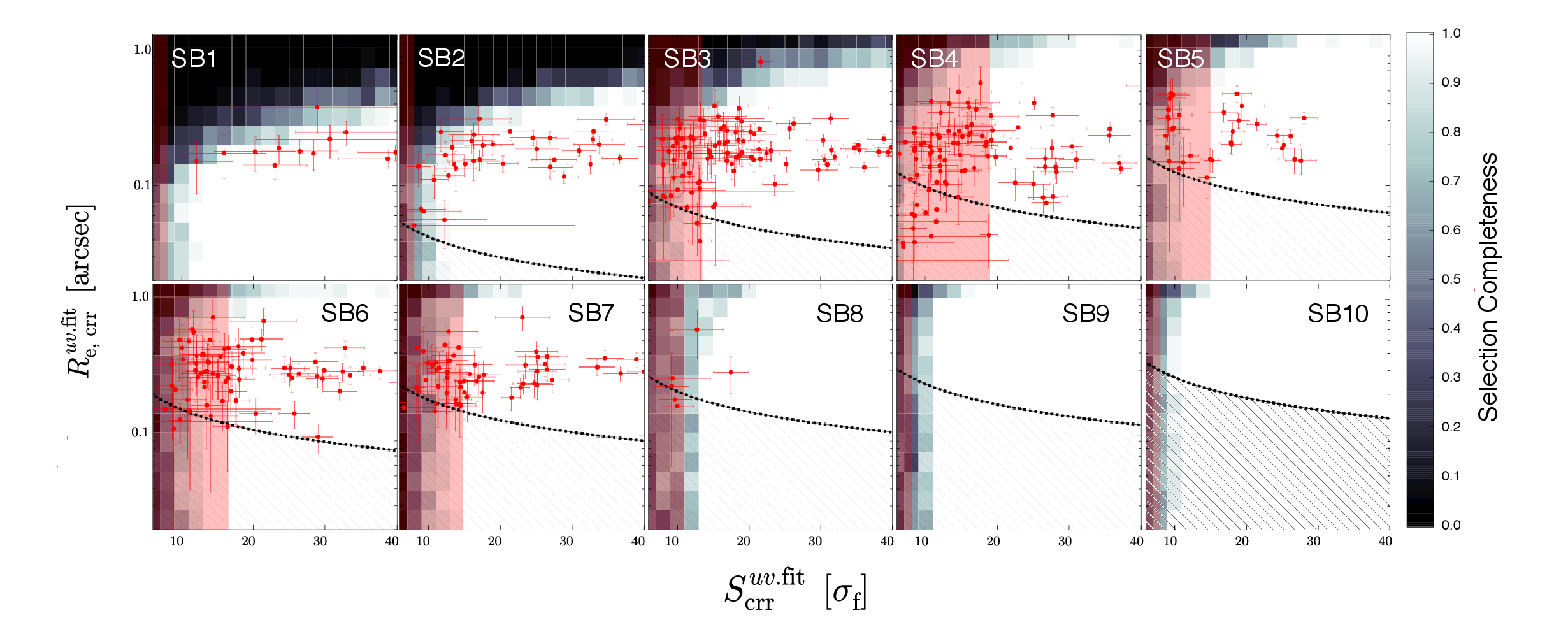}
\vspace{-0.5cm}
 \caption[]{
Selection and measurement completeness for the ALL10S sources in the 10 sub-datasets of SB1$-$SB10. 
The background gray scale denotes the selection completeness, which is defined with the color bar shown on the right-hand side. 
The dashed lines represent the limits of the reliable size measurements $\theta_{\rm min}$.  
For the $\theta_{\rm min}$ estimates from Equation \ref{eq:theta_min}, we adopt $\beta=0.75$ and $\lambda_{\rm c}=3.84$ for a $2\sigma$ cutoff, 
and assume that $\mathfrak{S}$ corresponds to the ratio of the total flux density to the pixel noise. 
The hatched areas show the regions where the measurements are not reliable. 
The red circles show the ALL10S sources with the flux density and the size measurements after the corrections. 
The red shades indicate the areas below the completeness thresholds (see text). 
\label{fig:selection_comp}}
\end{center}
\end{figure*}

We examine the selection and measurement completenesses for the ALL10S sources. 
Figure \ref{fig:selection_comp} shows the ALL10S sources in the 10 sub-datasets of SB1$-$SB10 on the flux density and size plane. 

We evaluate the selection completeness, 
performing the selection procedures same as those of ALL10S based on the Monte-Carlo simulation results (Section \ref{sec:simulation}). 
The background gray scale in Figure \ref{fig:selection_comp} denotes the selection completenesses of the 10 representative ALMA maps. 
In Figure \ref{fig:selection_comp}, 
there are two kinds of the selection incompletenesses. 
The first incompleteness is caused by one of ALL10S criteria, 
$S^{\rm total}_{\rm obs}\,\geq\,10\sigma_{\rm f}$ (Table \ref{tab:alma_catalog_sum}). 
Figure \ref{fig:selection_comp} shows the first incompleteness in the vertical region at the faint end. 
The second incompleteness is caused by another criterion of ALL10S, 
$S^{\rm peak}_{\rm obs}\,\geq\,5\sigma_{\rm f}$ (Table \ref{tab:alma_catalog_sum}).  
If a source is extended, the peak flux of the source does not reach the criterion especially in the high-resolution ALMA maps. 
Thus, the second incompleteness affects the identification of the extended sources in the high-resolution observations. 
In Figure \ref{fig:selection_comp}, the second incompleteness is shown in the diagonal region in each panel. 
Even with these two types of incompletenesses, 
our ALMA sources are placed in the parameter space with the selection completeness of $\gtrsim$ 60\%.

We next investigate the measurement completeness. 
A small size of the synthesized beam, $\theta_{\rm beam}$, enables us to measure a compact source size.
Moreover, if a source is bright, we can measure the small size accurately. 
In fact, the reliable size measurement limit with an interferometer, $\theta_{\rm min}$, is calculated as 
\begin{eqnarray}
\label{eq:theta_min}
\theta_{\rm min} = \beta\Bigl(\frac{\lambda_{c}}{2\mathfrak{S}^{2}}\Bigr)^{1/4} \times \theta_{\rm beam}, 
\end{eqnarray}
where $\mathfrak{S}$ is the SNR of the averaged visibilities, 
$\beta$ is a coefficient that typically takes the value in the range of $0.5-1.0$ weakly depending on the spatial distribution of the telescopes, 
and $\lambda_{\rm c}$ is related to the probability cutoff for a false size-detection \citep{marti2012}. 
We cannot obtain the reliable size measurements for the compact and/or faint sources in the low-resolution observations. 
In Figure \ref{fig:selection_comp}, the hatched areas show the regions in which the size values are below the reliable size measurement limit. 
The SB2$-$SB8 panels show that the source distributions overlap the hatched area, 
where the measurement completeness is not satisfied. 

To select the sources that are not biased by the selection nor measurement completenesses, 
we define the completeness threshold in each sub-dataset. 
In Figure \ref{fig:selection_comp}, 
the red shades represent the areas below the completeness thresholds. 
We use the completeness thresholds to verify the FIR size$-$luminosity relation in Section \ref{sec:fir_size_lumi}.

\subsection{Optical and NIR Counterpart}
\label{sec:counterpart}
We search the optical-NIR counterparts for the ALL5S and the ALL10S sources. 
Table \ref{tab:opt_catalog_summary} summarizes the $H$-band limiting magnitudes of the optical-NIR source catalogs 
of HUDF, HFF, COSMOS, SXDS, and GOODS-S that are used in this study. 
Recent studies with ALMA and HST have reported two types of the offsets between the ALMA and optical-NIR source centers. 
One is the intrinsic offset of $\sim0\farcs4$ between the rest-frame FIR and UV-optical emission \citep{chen2015}. 
The other is the astrometric uncertainty of HST in GOODS-S \citep[][see also Section 5.3]{dunlop2017,rujopakarn2016}, 
which we apply the corrections of $\Delta\alpha$ = $-80$ mas and $\Delta\delta$ = $+260$ mas \citep{rujopakarn2016} for the optical-NIR source centers. 
Since no one knows whether other fields also have the astrometric uncertainty same as GOODS-S, 
we do not apply this astrometry correction to the optical-NIR sources in other fields. 
Taking these potential offsets into account,  
we define the optical-NIR counterparts as the optical-NIR sources that locate within a search radius of $1\farcs$0 from our ALMA source centers. 
If there exist several optical-NIR objects within the search radius of $1\farcs0$, 
we regard the nearest one as the optical-NIR counterpart. 
We identify a total of 577 and 444 optical-NIR counterparts in ALL5S and the ALL10S that are referred to as "OC5S" and "OC10S", respectively. 
The OC5S and the OC10S catalogs are presented in Table \ref{tab:alma_catalog_sum}. 
We take the values of $z_{\rm phot}$ and $H$-band magnitude from the optical source catalogs in Table \ref{tab:source_catalog}, 
and estimate the spatial offset between the source centers in our ALMA maps and the optical catalogs. 
The $L_{\rm FIR}$ values is calculated with a modified black body following the analytical expression, 
\begin{eqnarray}
\label{eq:mbb}
\scalebox{0.85}{$\displaystyle
L_{\rm FIR} = 4\pi D_{\rm L}^{2} \frac{S(\nu_{\rm obs})}{\nu_{0}^{\beta}B(\nu_{0},T_{\rm d})} \frac{2h}{c^{2}}(\frac{kT_{d}}{h})^{4+\beta}\Gamma(4+\beta)\zeta(4+\beta), 
$}
\end{eqnarray}
where $h$ and $k$ are Planck and Boltzmann constant, respectively, 
$c$ is the light velocity, 
$\nu_{0}$ and $\nu_{\rm rest}$ are the rest- and observed-frame frequency, respectively,  
$S$ is the observed-flux density, 
$D_{\rm L}$ is the luminosity distance, 
$B$, $\Gamma$, $\zeta$ are the black body, the Gamma, and the Riemann zeta function, respectively, 
$\beta$ is the spectral index, 
and $T_{\rm d}$ is the dust temperature. 
Here, we assume $\beta=1.8$ \citep[e.g.,][Planck Collaboration \citeyear{planck2011}]{chapin2009} and $T_{\rm d}=35$ K \citep[e.g.,][]{kovacs2006,coppin2008}, 
and take the cosmic microwave background (CMB) effect \citep{cunha2013} into account. 
Table \ref{tab:source_catalog} summarize $z_{\rm phot}$, spatial offset, $H$-band, and $L_{\rm FIR}$. 

Note that our ALMA sources identified behind the HFF clusters are gravitationally lensed. 
For these lensed ALMA sources, 
we use the intrinsic  $R_{\rm e(FIR)}$, $L_{\rm FIR}$, and $H$-band magnitudes estimated by dividing the observed values with the magnification factors that are obtained from \cite{castellano2016}. 
For the other ALMA sources identified in the fields, 
we assume that the contamination of the lensed sources is negligible, 
because \cite{simpson2016} show that the contamination of the potential lensed sources is $\sim$ 8\%. 
We examine the effect of the potential lensed sources to our results in Section \ref{sec:fir_size_lumi}.

\begin{table}
{\scriptsize
\caption{Optical Catalog Summary
\label{tab:opt_catalog_summary}}
\begin{tabular}{ccc}
\hline
\hline
Field & Catalog Reference & Detection Limit (ap.) \\
(1) &  (2) & (3) \\
\hline
SXDS   	 & \cite{santini2015}  & $H<27.5$ $(0\farcs2)$            \\ 
GOODS-S Wide & \cite{santini2015}  & $H<27.4$  $(0\farcs17)$  \\ 
GOODS-S Deep & \cite{santini2015}  & $H<28.2$ $(0\farcs17)$  \\ 
HUDF	 & \cite{santini2015}  & $H<29.7$ $(0\farcs17)$           \\ 
COSMOS Wide \footnotemark[1] & \cite{ilbert2013}     & $H<23.9$ $(2\farcs0)$\tablenotemark{$\dagger$}      \\ 
COSMOS Deep \footnotemark[2] & \cite{momcheva2016} & $H<26.4$ $(1\farcs0)$\tablenotemark{$\dagger\dagger$} \\
HFF 		 & \cite{castellano2016} & $H<$28.5$-$29.0 $(0\farcs35)$      \\ 
\hline
\end{tabular}
\footnotesize{Notes: 
(1) Field name.
(2) Optical-NIR catalog that is used in our optical-NIR counterpart identification. 
(3) Optical-NIR source $5\sigma$ detection limit. The aperture size is presented in the parentheses. 
}
$\dagger$ The catalog is constructed with a detection image from the sum of the UltraVISTA DR1 $YJHK_{\rm s}$ images, where the 5$\sigma$ detection limit of the $H$-band image is $H<23.9$ in a $2\farcs0$ aperture.
$\dagger\dagger$The catalog is constructed with the HST $JH$-band that has the 5$\sigma$ detection limit of $JH<26.0$ in a $1\farcs0$ aperture.
\footnotetext[1]{COSMOS field overlapped with 3D-HST. }
\footnotetext[2]{COSMOS field not overlapped with 3D-HST. }
}
\end{table}

\subsection{Our Sample on the SFR$-M_{\rm star}$ plane}
\label{sec:sf_mode}
\begin{figure}
\begin{center}
\includegraphics[trim=0cm 0cm 0cm 0cm, clip, angle=0,width=0.5\textwidth]{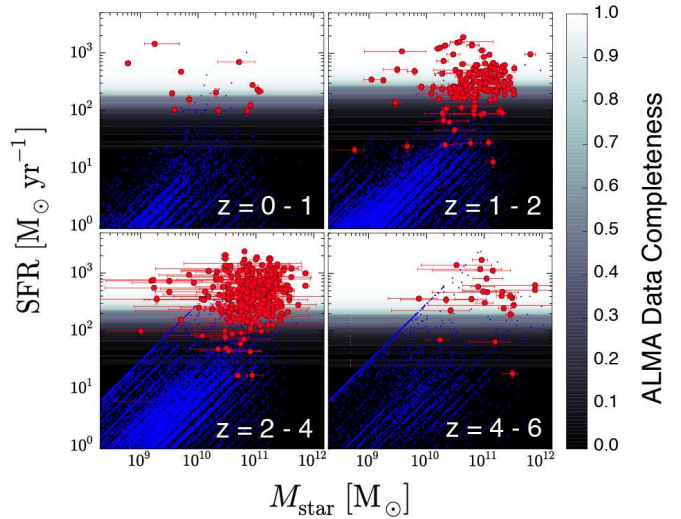}
\caption[]
{
$M_{\rm star}-$SFR relation at $z=$ 0$-$6. 
The red and blue circles denote the OC5S and the 3D-HST sources, respectively. 
The background gray-scale represents the completeness of 
the ALMA data depending on the flux density limits that correspond to SFRs by Equation \ref{eq:sfr_ir}. 
\label{fig:ms_z0-6}}
\end{center}
\end{figure}

We examine the star-formation properties of our ALMA sources. 
Figure \ref{fig:ms_z0-6} presents the stellar mass ($M_{\rm star}$) $-$ star formation rate (SFR) relation for the OC5S sources.  
For comparison, Figure \ref{fig:ms_z0-6} also displays the 3D-HST sources whose SFR and $M_{\rm star}$ values are estimated in  \cite{momcheva2016}. 

For our ALMA sources, the $M_{\rm star}$ values are obtained from the optical source catalogs in Table \ref{tab:opt_catalog_summary}.  
In this paper, we assume the \cite{chabrier2003} initial mass function (IMF). 
To obtain the Chabrier IMF values of $M_{\rm star}$, 
we divide the \cite{salpeter1955} IMF values that are used in \cite{castellano2016} by a factor of 1.8. 
The SFR values are estimated from the sum of the dust obscured (${\rm SFR}_{\rm IR}$) 
and un-obscured (${\rm SFR}_{\rm UV}$) star-formation rates, and given by 
\begin{eqnarray}
\label{eq:sfr_tot}
{\rm SFR} = {\rm SFR}_{\rm UV} + {\rm SFR}_{\rm IR}.
\end{eqnarray}
The SFR$_{\rm UV}$ and SFR$_{\rm IR}$ values are calculated with the fomulae of \cite{murphy2011}, 
\begin{eqnarray}
\label{eq:sfr_uv}
{\rm SFR}_{\rm UV}&=& 4.42\times10^{-44}L_{\rm UV} ({\rm erg\,s}^{-1}), \\
\label{eq:sfr_ir}
{\rm SFR}_{\rm IR}&=& 3.88\times10^{-44}L_{\rm FIR} ({\rm erg\,s}^{-1}).   
\end{eqnarray}
Note that the different detection limits in our ALMA maps ($S_{\rm obs}^{\rm peak}\geq5\sigma_{\rm f}$) produces the different limits of the SFR$_{\rm IR}$ values.  
We calculate the SFR$_{\rm IR}$ limit of each ALMA map by Equation \ref{eq:sfr_ir}, 
where the $L_{\rm FIR}$ value is estimated in the same manner as Section \ref{sec:counterpart} from the ALMA detection limit. 
In Figure \ref{fig:ms_z0-6}, the background gray-scale corresponds to the fraction of our ALMA maps whose SFR$_{\rm IR}$ limits are below the SFR values on the ordinate axis, 
referred to as "ALMA data completeness". 

In Figure \ref{fig:ms_z0-6}, 
most of our ALMA sources fall in the ranges of $M_{\rm star}\sim10^{10}-10^{11.5}\,M_{\odot}$ and SFR $\sim100-1000\,M_{\odot}$yr$^{-1}$. 
Due to the ALMA data completeness, 
our ALMA sources are limited by the SFR values, 
which indicates that our ALMA sources are not the complete sample at a fixed $M_{\rm star}$. 

To investigate the star-formation mode of our ALMA sources, 
we compare the specific SFR ($\equiv$ SFR/$M_{\rm star}$; sSFR) of our ALMA sources with the main sequence in four redshift bins of $z=$ 0$-$1, 1$-$2, 2$-$4, and 4$-$6. 
We define the starbursts as the sources with sSFR over a factor of 4 larger than that of the main sequence \citep{rodighiero2011}.
The sSFR value of the main sequence is calculated by the median sSFR value of the 3D-HST sources in each redshift bin. 
We find that $\sim$100\%, $\sim$ 58\%, $\sim$49\%, and $\sim$26\% of our ALMA sources are classified as the starbursts at $z=$ 0$-$1, 1$-$2, 2$-$4, and 4$-$6, respectively.  
With the redshift bins all together, the starburst fraction is $\sim$52\%. 
The decreasing trend of the starburst fraction toward high redshifts indicates that the massive sources on the main-sequence are also detected above the SFR$_{\rm IR}$ limits at the high redshifts. 
This is because the SFR values of the main sequence are increased at high redshifts.
The starburst fraction and the trend are generally consistent with the ALESS result 
that shows the starburst fraction of $\sim$ 49\% and 27\% at $z=$ 1.5$-$2.5 and 2.5$-$4.5, respectively \citep{da-cunha2015}.  
Therefore, we conclude that our ALMA sources are the general population of the dusty starbursts similar to the ALESS sample, 
having the half of them are the starbursts, while the rest of half are the high-mass end of the main sequence.

\section{Results}
\label{sec:result}

\begin{figure*}
\begin{center}
\includegraphics[trim=0cm 0cm 0cm 0cm, clip, angle=0,width=1.0\textwidth]{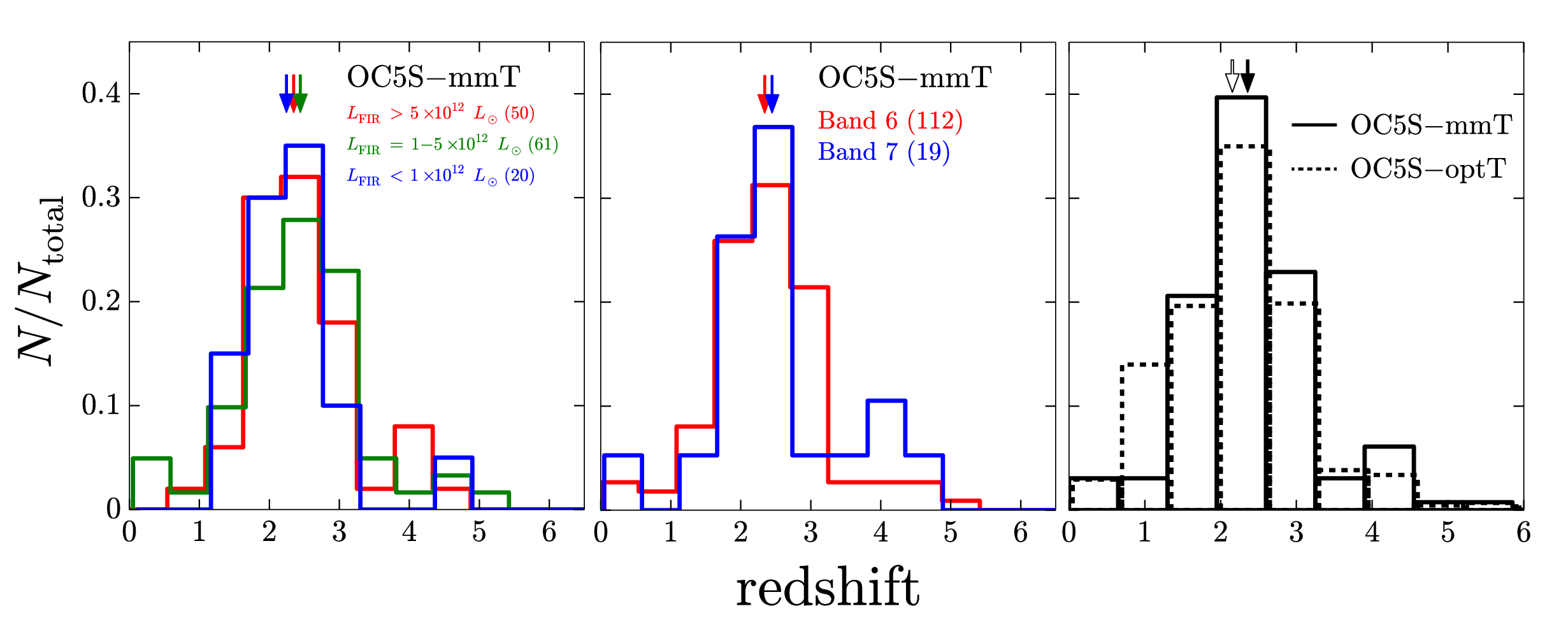}
\vspace{-0.2cm}
 \caption[]
{
Normalized redshift distribution of the OC5S sources. The source number in each sub-sample is presented in the parentheses.
{\it Left}: The red, green, and blue histograms denote the samples with the $L_{\rm FIR}$ ranges of $L_{\rm FIR}>5\times10^{12}\,L_{\rm \odot}$, 
$L_{\rm FIR}=1-5\times10^{12}\,L_{\rm \odot}$, and $L_{\rm FIR} < 1\times10^{12}\,L_{\rm \odot}$, respectively. 
The color arrows indicate the median redshifts of the three samples. 
{\it Middle}: The red and blue histograms present the samples of ALMA Band 6 and 7, respectively. 
The color arrows show the median redshifts of the two samples. 
{\it Right}: Redshift distribution of OC5S-mmT (solid histogram) and OC5S-optT (dashed histogram). 
The filled (open) arrow denotes the median redshift of OC5S-mmT (OC5S-optT). 
\label{fig:z_distribution}}
\end{center}
\end{figure*}

\subsection{Redshift Distribution}
\label{sec:nz}
We estimate the redshift distribution for our ALMA sources, 
and verify the potential bias of the target selection in the initial ALMA observations by comparing the redshift distributions. 
In this analysis, 
we divide the OC5S sources into two types of the sources.  
The sources whose central targets in the initial ALMA observations are selected in the FIR-mm and optical-NIR bands are referred to as "OC5S-mmT" and "OC5S-optT", respectively. 
We also regard the sources as OC5S-mmT, if the initial ALMA observation is the blind survey. 
On the other hand, if the central target selection include the redshift properties even based on the FIR-mm bands, 
we classify the sources as OC5S-optT. 
We assume that OC5S-mmT is not biased by the target selection in the initial ALMA observations. 
We summarize the source number in the new catalogs of OC5S-mmT and OC5S-optT in Table \ref{tab:alma_catalog_sum}. 

First, we derive the redshift distribution with the OC5S-mmT sources. 
In the left and middle panels of Figure \ref{fig:z_distribution}, the redshift distributions are plotted in three bins of $L_{\rm FIR}$ and two bins of ALMA Band 6/7, respectively.  
We perform the Kolmogorov-Smirnov test (KS-test) to investigate whether the redshift distributions of our ALMA sources are changed by $L_{\rm FIR}$ or ALMA Band 6/7. 
The result of the KS-test suggests that we cannot rule out the possibility that the redshift distributions are produced from the same parent sample in the both cases. 
We thus conclude that neither $L_{\rm FIR}$ nor ALMA Band 6/7 significantly affect the redshift distribution for the ALMA sources. 
The median redshift is estimated to be $z_{\rm med}=2.36$, which is consistent with the previous results of the blind submm/mm surveys \citep{chapman2005,yun2012,simpson2014,dunlop2017}. 

Second, we also derive the redshift distribution of the OC5S-optT sources to test whether the OC5S-optT sources are biased by the target selection in the initial ALMA observations. 
The right panel of Figure \ref{fig:z_distribution} shows the redshift distributions of the OC5S-mmT and OC5S-optT sources. 
We perform the KS-test for these two redshift distributions. 
The result of the KS-test suggests that we cannot reject the possibility that these two redshift distributions are made from the same parent sample. 
The median redshift of the OC5S-optT sources is estimated to be $z=2.15$ that is almost same as $z_{\rm med}=2.36$ obtained from the OC5S-mmT sources. 
The little difference of the redshift properties between the OC5S-mmT and OC5S-optT sources implies that the OC5S-optT sources are not biased by target selection in the initial ALMA observations. 
Therefore, we use both OC5S-mmT and OC5S-optT in the following analyses with $z_{\rm med}=2.36$ except for Section \ref{sec:morophology}. 

\subsection{FIR Size and Luminosity Relation}
\label{sec:fir_size_lumi}
Before we investigate the $R_{\rm e(FIR)}$ and $L_{\rm FIR}$ relation, 
we perform two tests to verify whether there exist potential biases in the $R_{\rm e(FIR)}$ and $L_{\rm FIR}$ measurements our ALMA sources.  

\begin{figure}
\begin{center}
\includegraphics[trim=0cm 0.6cm 0cm 0cm, clip, angle=0,width=0.5\textwidth]{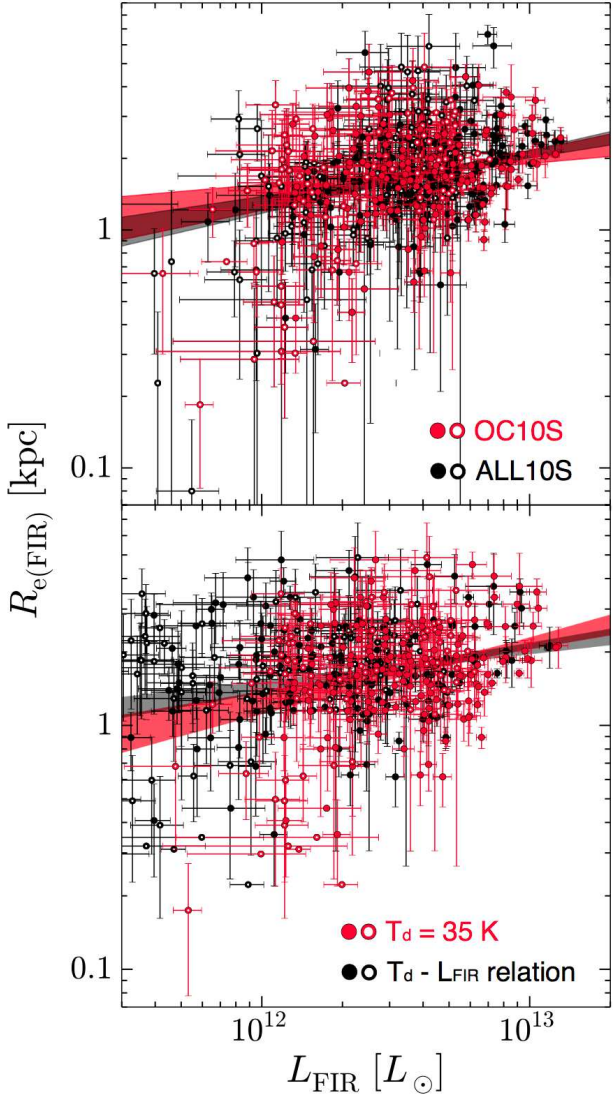}
\vspace{0.5cm}
 \caption[]
{Comparison of the FIR size$-$luminosity relations. 
The filled and open circles indicate our ALMA sources above and below the completeness thresholds (Section \ref{sec:selection_comp}), respectively. 
{\it Top}: 
The red and black circles represent the OC10S and ALL10S sources, respectively. 
For the $R_{\rm FIR}$ and $L_{\rm FIR}$ estimates, we assume that all of the OC10S and ALL10S sources reside at $z_{\rm med}=2.36$. 
{\it Bottom}: 
The red and black circles denote the OC10S sources whose $L_{\rm FIR}$ values are 
derived with the assumptions of $T_{\rm d}=$ 35 K and the $T_{\rm d}$$-$$L_{\rm FIR}$ relation (see text), respectively. 
\label{fig:size_test}}
\end{center}
\end{figure}

First, we examine whether there is a difference between OC10S and ALL10S on the $R_{\rm e(FIR)}$$-$$L_{\rm FIR}$ plane. 
In the top panel of Figure \ref{fig:size_test}, 
we show the $R_{\rm e (FIR)}$ estimates as a function of $L_{\rm FIR}$ for the ALL10S and OC10S sources with the black and red circles, respectively. 
To calculate the $R_{\rm e(FIR)}$ and $L_{\rm FIR}$ values for the ALL10S and OC10S sources, 
we assume that all of the sources reside at $z_{\rm med}=2.36$ (Section \ref{sec:nz}). 
To evaluate the distributions of the ALL10S and OC10S sources, 
we fit a power-law function to the $R_{\rm e(FIR)}$ and $L_{\rm FIR}$ values. 
The power-law function is defined as 
\begin{eqnarray}
\label{eq:pow-law}
R_{\rm e(FIR)} = R_{0}\Bigl(\frac{L_{\rm FIR}}{L_{0}}\Bigr)^{\alpha^{\rm IR}},
\end{eqnarray}
where $R_{0}$ and $\alpha^{\rm FIR}$ are free parameters. 
The $R_{0}$ value presents the effective radius at a luminosity of $L_{0}$. 
We select the $L_{0}$ value to the best-fit Schechter parameter $L_{*}$ at $z\sim3$.
For the fitting, 
we adopt the completeness thresholds (Section \ref{sec:selection_comp}) 
to select the ALMA sources that are not biased neither by the selection nor measurement completeness. 
In Figure \ref{fig:size_test}, 
open and filled red circles represent our ALMA sources below and above the completeness thresholds, respectively. 
We use our ALMA sources above the completeness thresholds alone in the following analyses in this paper. 
We estimate the 1$\sigma$-error range of the power-law function by the perturbation method \citep{curran2014}. 
We generate 1000 data sets of the $R_{\rm e (FIR)}$ and $L_{\rm FIR}$ for the ALL10S and OC10S sources 
based on the random perturbations following the Gaussian distribution
whose sigma is defined by the observational errors. 
We obtain 1000 best-fit power-law functions for the 1000 data sets. 
We define the error of the power-law function as the range of 68\% distribution for the $R_{\rm e (FIR)}$ values in the 1000 best-fit power-law functions. 
We refer this process to evaluate the 1$\sigma$-error range as the perturbation method. 
In the top panel of Figure \ref{fig:size_test}, 
the black and red shades denote the estimated 1$\sigma$-error ranges of the power-law functions for the ALL10S and OC10S sources, respectively. 
We find that the red shade shows a agreement with the black shade, 
suggesting that there is no significant difference in the distributions between the ALL10S and OC10S sources. 
We thus conclude that the OC10S sources generally represent the ALL10S sources. 
Because the ALL10S sources with no optical-NIR counterparts do not allow us to examine the physical properties, 
we investigate the OC10S sources alone with the individual photometric redshifts in the following analyses in this subsection. 

Second, 
we test whether our assumption of the single dust temperature, $T_{\rm d}=35$ K, makes a bias on the $R_{\rm e (FIR)}$$-$$L_{\rm FIR}$ plane. 
\cite{symeonidis2013} report that $T_{\rm d}$ has a positive correlation with $L_{\rm FIR}$ for the $Herschel$-selected galaxies at $0.1<z<2$. 
We estimate the $L_{\rm FIR}$ values with the $T_{\rm d}$$-$$L_{\rm FIR}$ relation for the OC10S sources, 
and compare the source distribution on the $R_{\rm e (FIR)}$$-$$L_{\rm FIR}$ plane to that obtained with $T_{\rm d}=35$ K. 
For evaluating the $T_{\rm d}$$-$$L_{\rm FIR}$ relation, 
we model $T_{\rm d}$ as a function of $L_{\rm FIR}$ with a linear function, 
\begin{eqnarray}
\label{eq:Td-Lir}
T_{\rm d} = C_{0} \log(L_{\rm FIR}) + C_{1}, 
\end{eqnarray}
where $C_{0}$ and $C_{1}$ are the free parameters. 
Note that it is not clear that  $Herschel$- and ALMA-selected galaxies have the same $T_{\rm d}$$-$$L_{\rm FIR}$ relation, 
because the wavelength coverage of ALMA Band 6/7 is sensitive to the galaxies with the dust temperatures colder than that of $Herschel$ does at a fixed $L_{\rm FIR}$ \citep[e.g.,][]{simpson2016}. 
We thus evaluate the $T_{\rm d}$$-$$L_{\rm FIR}$ relation with following 3 steps: 
a) fitting the linear function to the $Herschel$-selected galaxies in \cite{symeonidis2013}, 
b) fixing the $C_{0}$ value of the best-fit linear function obtained in a), and 
c) fitting the linear function with the fixed $C_{0}$ to the ALMA-selected galaxies whose 
$T_{\rm d}$ and $L_{\rm FIR}$ values are well determined with $Herschel$$+$ALMA bands in \cite{swinbank2014} and \cite{simpson2016}. 
We then obtain the $T_{\rm d}$ and $L_{\rm FIR}$ values from our best estimate of the $T_{\rm d}$$-$$L_{\rm FIR}$ relation and Equation \ref{eq:mbb}. 
In the bottom panel of Figure \ref{fig:size_test}, 
we show the OC10S sources whose $L_{\rm FIR}$ values are estimated from the $T_{\rm d}$$-$$L_{\rm FIR}$ relation and $T_{\rm d}=35$ K 
with the black and red circles, respectively. 
The color shades denote the 1$\sigma$-error ranges of the power-law functions for the $R_{\rm e (FIR)}$ and $L_{\rm FIR}$ values obtained by the perturbation method, 
which shows that the power-law functions are consistent within the 1$\sigma$ errors. 
We thus conclude that our assumption of $T_{\rm d}=35$ K does not produce the significant systematics on the $R_{\rm e (FIR)}$$-$$L_{\rm FIR}$ plane. 
Because the evaluation and the application processes of the $T_{\rm d}$$-$$L_{\rm FIR}$ relation may contain the systematics in the $L_{\rm FIR}$ estimates rather than the assumption of the single dust temperature, 
we adopt the $L_{\rm FIR}$ values estimated by $T_{\rm d}=35$ K in the following analyses. 
\\

\begin{figure*}
\begin{center}
\includegraphics[trim=0.2cm 0cm 0cm 0cm, clip, angle=0,width=0.95\textwidth]{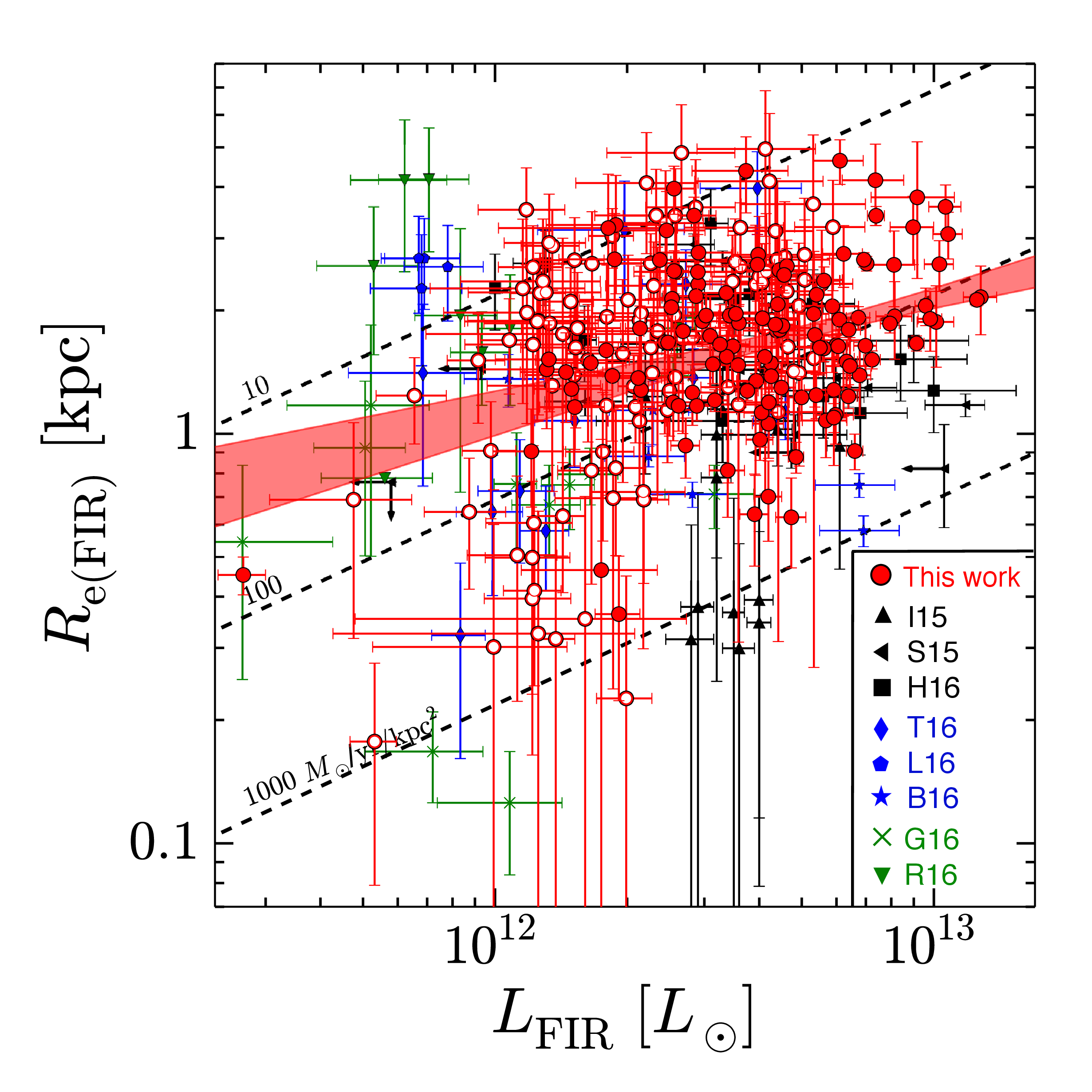}
 \caption
{
FIR size$-$luminosity relation of our best estimates with OC10S representing ALL10S (see text). 
The red filled and open circles are the OC10S sources above and below the completeness thresholds (Section \ref{sec:selection_comp}), respectively.  
The shade region indicates 1$\sigma$ uncertainty range of the best-fit power-law function that is calculated by the perturbation method. 
In the power-law fitting, we do not include the OC10S sources shown with the open circles. 
The black dashed lines denote the constant surface SFR density $\Sigma_{\rm SFR}$ =10, 100, and 1000 $M_{\odot}/{\rm yr}/{\rm kpc^{2}}$ 
estimated from Equation \ref{eq:sfr_ir} and the assumption of the uniform surface density with $R_{\rm e(FIR)}$. 
Other symbols denote the previous ALMA results in the same assignment as Figure \ref{fig:comp_fluxsize}, 
except for the blue pentagons (L16; \citealt{lindroos2016}). 
The symbol colors indicate that the types in the initial ALMA observations of 
the submm/mm bands selected target (black), 
the optical/NIR bands selected target (blue), 
and the blind survey (green). 
\label{fig:size_previous}}
\end{center}
\end{figure*}

Figure \ref{fig:size_previous} shows our best estimates of the $R_{\rm e(FIR)}-L_{\rm FIR}$ relation with the red circles. 
In Figure \ref{fig:size_previous}, our ALMA sources fall in the range of $R_{\rm e(FIR)}\sim0.3-5$ kpc. 
From Equation \ref{eq:sfr_ir} and the assumption of the uniform surface density with $R_{\rm e(FIR)}$, 
the SFR surface densities $\Sigma_{\rm SFR}$ of our ALMA sources are estimated to be $\sim$ 8$-$800 $M_{\odot}/{\rm yr}/{\rm kpc^{2}}$.

We compare our $R_{\rm e(FIR)}$ estimates to previous ALMA results. 
In Figure \ref{fig:size_previous}, we show the previous ALMA results of the FIR size studies with various symbols. 
The black symbols denote the high-resolution (0\farcs16$-$0\farcs3) observations for the SMGs at $z\gtrsim2$ \citep{ikarashi2015,simpson2015a,hodge2016}. 
The FIR size for the SMGs are estimated to be $R_{\rm e(FIR)}\sim0.3-3.0$ kpc. 
The green symbols represent the results of the deep observations for one blank field of HUDF and three gravitational lensing clusters of HFF \citep{rujopakarn2016,gonzalez2017}. 
These deep observations explore the FIR size measurements for the fainter sources than SMGs. 
The typical FIR sizes of these faint sources are estimated to be $R_{\rm e (FIR)}\sim 1.5-2.0$ kpc. 
The optically-selected galaxies with blue symbols are also examined for the FIR size properties. 
\cite{lindroos2016} estimate the FIR sizes by the $uv$-stacking method for the star-forming galaxies at $z\sim2$. 
The stacking result shows the typical FIR sizes of $R_{\rm e(FIR)}\sim2.5$ kpc. 
\cite{tadaki2016} and \cite{barro2016} conduct the ALMA observations for massive ($M_{\rm star}\sim10^{10.5-11.5}\,M_{\odot}$) star-forming galaxies at $z\sim2.5$, 
and report the rest-frame FIR size range of $R_{\rm e(FIR)}=0.3-4.0$ kpc. 

These results of the previous studies indicate that the $R_{\rm e(FIR)}-L_{\rm FIR}$ relation has a large scatter. 
In Figure \ref{fig:size_previous}, we find that our $R_{\rm e(FIR)}$ and $\Sigma_{\rm SFR}$ estimates are generally consistent with the previous studies within the scatter. 

We next evaluate the $R_{\rm e(FIR)}-L_{\rm FIR}$ relation. 
We divide our sample into four redshift samples of $z=$ 0$-$1, 1$-$2, 2$-$4, and 4$-$6. 
Spearman's rank test is used to examine the correlation between $R_{\rm e(FIR)}$ and $L_{\rm FIR}$. 
We find that there exist positive correlations between $R_{\rm e(FIR)}$ and $L_{\rm FIR}$ with $\sim$90$-$98\% significance levels in the four redshift samples. 
With the redshift samples all together, the significance level becomes $>$99\%. 
To estimate the slope of the positive correlation, 
we fit the power-law function of Equation \ref{eq:pow-law} to the $R_{\rm e(FIR)}-L_{\rm FIR}$ relation. 
We estimate the $\alpha^{\rm FIR}$ value from the redshift sample of $z=$ 2$-$4 that has the largest source number among the redshift samples. 
We obtain $\alpha^{\rm FIR} = 0.23\,\pm\,0.07$.  
With the redshift samples all together, 
the $\alpha^{\rm FIR}$ value is estimated to be 0.28 $\pm$ 0.07, 
suggesting that $\alpha^{\rm FIR}$ is $\sim0.2-0.3$ in any cases. 
Hereafter, we adopt a fiducial value of $\alpha^{\rm FIR}= 0.28\,\pm\,0.07$. 
Note that 
we obtain the $\alpha^{\rm FIR}$ value that is consistent with the fiducial value within the error,
if we perform the power-law fitting for the redshift samples all together with no completeness thresholds. 
We also confirm that the $\alpha^{\rm FIR}$ value is unchanged from the fiducial value within the error, 
if we adopt another fitting model with the S$\acute{\rm e}$rsic index different from $n=1$ in the size measurements in Section \ref{sec:flux_size_measure}. 

\begin{figure}
\begin{center}
\includegraphics[trim=0cm 0cm 0cm 0cm, clip, angle=0,width=0.5\textwidth]{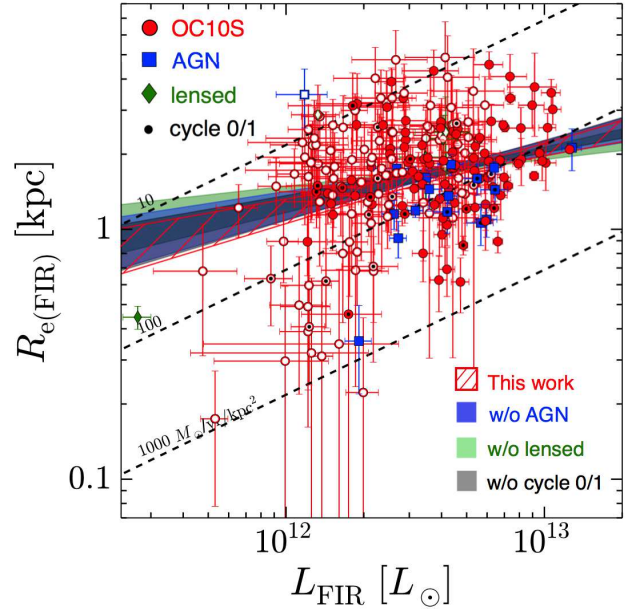}
 \caption[]
{Comparison of the FIR size$-$luminosity relations.  
The red circles present the OC10S sources in the same assignment as Figure \ref{fig:size_previous}. 
The blue, green, and black shade regions denote the 1$\sigma$ uncertainty range of the best-fit $R_{\rm e(FIR)}$$-$$L_{\rm FIR}$ relation for 
the OC10S sources without the AGNs (blue squares), potential lensed sources (green diamonds), and the data in ALMA cycles 0/1 (black inner circles), respectively, that are estimated by the perturbation method. 
The red hatched region represents the 1$\sigma$ uncertainty range of the best-fit $R_{\rm e(FIR)}$$-$$L_{\rm FIR}$ relation that is shown in Figure \ref{fig:size_previous}. 
The dashed lines are the constant surface SFR density in the same assignment as Figure \ref{fig:size_previous}. 
\label{fig:cont_test}}
\end{center}
\end{figure}

To test the influence by the AGNs, 
the gravitationally lensed sources, 
and the potentially low quality data in early ALMA cycles,  
we also derive the $R_{\rm e(FIR)}-L_{\rm FIR}$ relation without these sources. 
For the AGN identifications, 
we cross-match the OC10S sources with the X-ray AGNs in the literature \citep{ueda2008,xue2011,civano2012}. 
If the OC10S sources correspond to the central target AGNs in the initial ALMA observations, 
we also classify these OC10S sources as AGNs. 
For the lensed sources, 
we regard the OC10S sources as the potential lensed sources, 
if the optical-NIR counterparts of the OC10S sources have the photometric redshifts at $z\leq1$. 
This is because the lensed sources typically have the lensing pairs at $z\leq$1 \citep[e.g.,][]{simpson2016,sonnenfeld2017}. 
For the data in early ALMA cycles, 
we assume the all of the ALMA data in cycle 0/1 as the potentially low-quality data. 
Figure \ref{fig:cont_test} presents the $R_{\rm e(FIR)}-L_{\rm FIR}$ relations for the OC10S sources 
without the AGNs, the potential lensed sources, the potentially low-quality data of ALMA cycle 0/1 with the 
blue, green, and black shades, respectively, that are obtained from the perturbation method. 
For comparison, we also show our best estimate of the $R_{\rm e(FIR)}-L_{\rm FIR}$ relation with the red hatched region. 
In Figure \ref{fig:cont_test}, 
we find that our best estimate of the $R_{\rm e(FIR)}-L_{\rm FIR}$ relation is consistent with any of these three cases within the errors. 
We thus conclude that the $R_{\rm e(FIR)}-L_{\rm FIR}$ relation is unchanged by the contaminations of the AGNs, the lensed sources, and the early ALMA cycles. 
Interestingly, in Figure \ref{fig:cont_test}, we also find that the X-ray AGNs are likely to be placed at the region where the $\Sigma_{\rm SFR}$ values are higher than the general values of the other ALMA sources. 
Although there remain the uncertainties such as the $L_{\rm FIR}$ measurements and the completeness of the X-ray source catalogs, 
it may indicate that the high $\Sigma_{\rm SFR}$ values are related to the fueling mechanism of the AGNs.
\\

We compare the $\alpha^{\rm FIR}$ value with that of UV wavelength, $\alpha^{\rm UV}$. 
One of the most extensive UV size study is performed by \cite{shibuya2015} with $\sim 190,000$ star-forming galaxies at $z=$ 0$-$8 from deep HST images. 
This comprehensive study obtain the $\alpha^{\rm UV}$ value to be 0.27 $\pm$ 0.01. 
Our estimate of $\alpha^{\rm FIR}=0.28\,\pm\,0.07$ shows a good agreement with the $\alpha^{\rm UV}$ measurement. 
We discuss the physical origins of this $\alpha^{\rm FIR}$ value in Section \ref{sec:dis5-1}. 

\begin{figure*}
\begin{center}
\includegraphics[trim=0cm 0cm 0cm 0cm, clip, angle=0,width=1.0\textwidth]{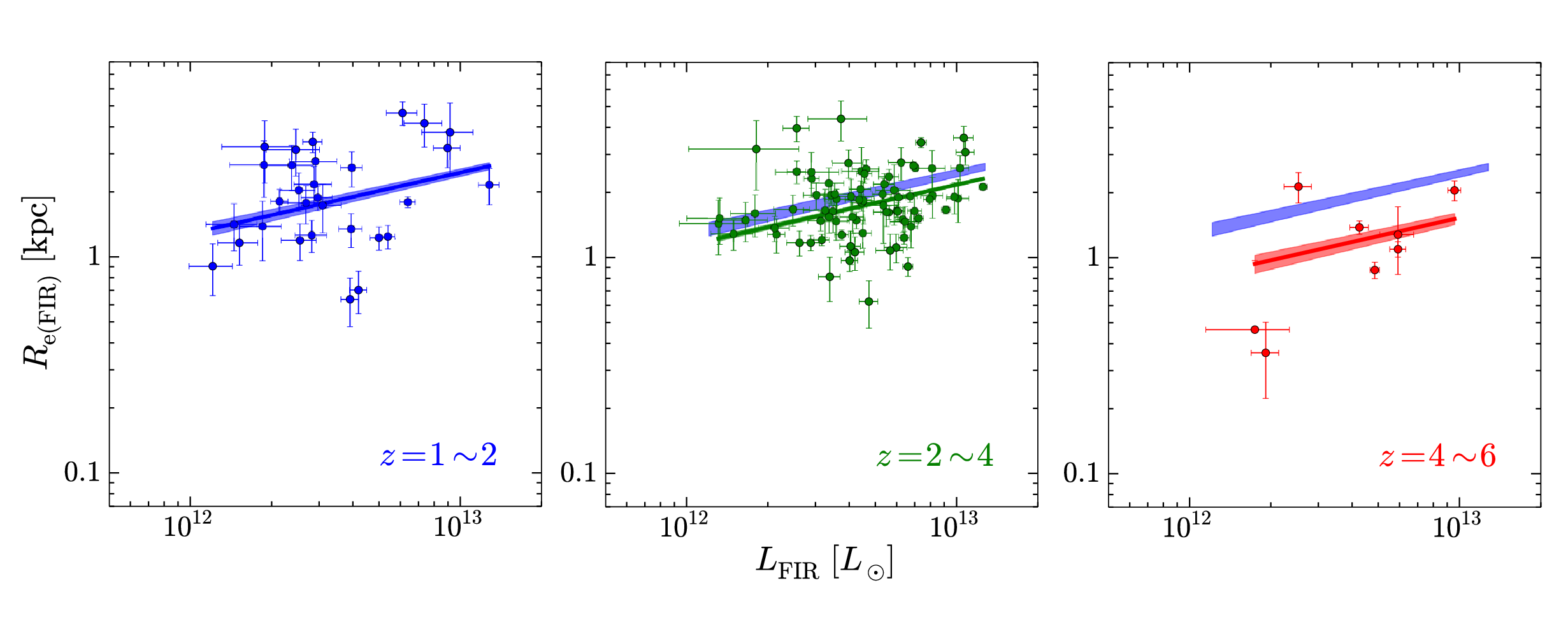}
 \caption[]{
Redshift evolution of the rest-frame FIR size$-$luminosity relation for the OC10S sources at $z=1-6$. 
The colors indicate the redshift ranges described at the bottom right of each panel. 
The solid lines and the shade regions present the best-fit power law functions and the associated 1$\sigma$ errors, respectively. 
In the power law function fitting, 
the $\alpha^{\rm FIR}$ vlaue of each redshift is fixed to $\alpha^{\rm FIR}=0.28$ that is the best-fit value form the redshift samples all together. 
\label{fig:FIR_size_lumi}}
\end{center}
\end{figure*}

We also investigate the redshift evolution of the $R_{\rm e(FIR)}-L_{\rm FIR}$ relation. 
Because the source number is not enough in the redshift sample of $z=$ 0$-$1, 
here we use the redshift samples of $z=$ 1$-$2, 2$-$4, and 4$-$6. 
To overcome the small statistics in the individual redshift samples, 
we fix the $\alpha^{\rm FIR}$ value to estimate the $R_{0}$ values for all of the redshift samples. 
Figure \ref{fig:FIR_size_lumi} shows the best-fit functions and the associated $1\sigma$ errors with solid lines and the shade regions, respectively. 
For comparison, the 1$\sigma$-error region of $z=$ 1$-$2 sample is presented in all of the redshift panels. 
We find that the best-fit $R_{0}$ values are tend to decrease toward high redshift. 
This trend is also consistent with that of the rest-frame UV and optical studies \citep[e.g.,][]{shibuya2015}.  

Note that the CMB effect may produce a bias in the $R_{\rm e(FIR)}$ measurements towards high redshifts \citep{zhang2016}.  \
To test whether the decreasing trend is contributed by the CMB effect, 
we compare the $R_{\rm e(FIR)}$ values between the sources identified in ALMA Band 6 and Band 7 in the redshift sample of $z=$ 4$-$6. 
This is because ALMA Band 6 is sensitive to the CMB effect more than Band 7, especially at $z>4$.  
We find that there is no significant difference in the $R_{\rm e(FIR)}$ values between ALMA Band 6 and Band 7, 
indicating that the CMB effect is negligible in our results. 

\section{Discussion}
\label{sec:discussion}
In this section, we discuss the physical origins of the dusty starbursts 
based on the properties in the rest-frame FIR, optical, and the UV wavelengths.  
\subsection{Slope of FIR Size and Luminosity Relation}
\label{sec:dis5-1}

Here we discuss the physical origins of the slope of $\alpha^{\rm FIR}= 0.28\,\pm\,0.07$ that is obtained in our rest-frame FIR study of the $R_{\rm e(FIR)}-L_{\rm FIR}$ relation. 
In the rest-frame UV and optical studies, the size$-$mass relation shows the different power-law slopes in the star-forming and the compact quiescent galaxies. \citep[e.g.,][]{shen2003,huang2013,vanderwel2014,van-dokkum2015}. 
For the star-forming galaxies, the power-law slope is estimated to be $\sim0.2-0.3$ that is consistent with the predictions from the disk formation models  \citep{shen2003,huang2013,vanderwel2014}. 
For the compact quiescent galaxies, the relatively steep power-law slope of $\sim0.5-0.7$ is obtained, which can be explained by the repeated mergers \citep{shen2003,van-dokkum2015}. 
Our best estimate of $\alpha^{\rm FIR}$ shows the good agreement with the rest-frame UV size$-$luminosity relation for the star-forming galaxies with $\alpha^{\rm UV}=0.27\,\pm\,0.01$ \citep[e.g.,][]{huang2013,shibuya2015}, suggesting that the origin of the slope $\alpha^{\rm FIR}$ is also explained by the disk formation. 
In fact, \cite{hodge2016} and \cite{barro2016} report that the dusty star-forming galaxies has the disk-like morphologies in the rest-frame FIR wavelength. 
The CO observations support the existence of the gas disks with the rotating kinematics in the dusty star-forming galaxies \citep[e.g.,][]{hodge2012}. 
Therefore, our result may indicate that our ALMA sources occur in the disk formation process that makes the slope of $\alpha^{\rm FIR}= 0.28\,\pm\,0.07$. 

It is worth mentioning that we find no ALMA sources with $\Sigma_{\rm SFR}>$ 800 $M_{\odot}$/yr/kpc$^{2}$. 
Based on the balance between the radiation pressure from the star-formation and the self-gravitation, 
\cite{simpson2015a} estimate the Eddington limit of the SMGs to be $\sim$ 1000 $M_{\odot}$/yr/kpc$^{2}$, 
which is generally consistent with the maximum $\Sigma_{\rm SFR}$ values among our ALMA sources. 
Although not all of our ALMA sources are likely to reach the Eddington limit, 
the slope of the $R_{\rm e(FIR)}-L_{\rm FIR}$ relation may also be contributed by the Eddington limit. 

Note that \cite{lutz2016} show the negative $\alpha^{\rm FIR}$ slope in the $R_{\rm e(FIR)}-L_{\rm FIR}$ relation for local (U)LIRGs. 
In the local (U)LIRGs, the strong gas compression such as major mergers is needed to occur the dusty starbursts due to the low gas fractions \citep[e.g.,][]{casey2014}. 
The different mechanisms of the dusty starbursts between local and high redshifts probably provides the different $\alpha^{\rm FIR}$ slopes.

\subsection{Sizes in UV, Optical, and FIR}
\label{sec:uv-opt-fir}

\begin{figure*}
\begin{center}
\includegraphics[trim=0cm 0cm 0cm 0cm, clip, angle=0,width=1.0\textwidth]{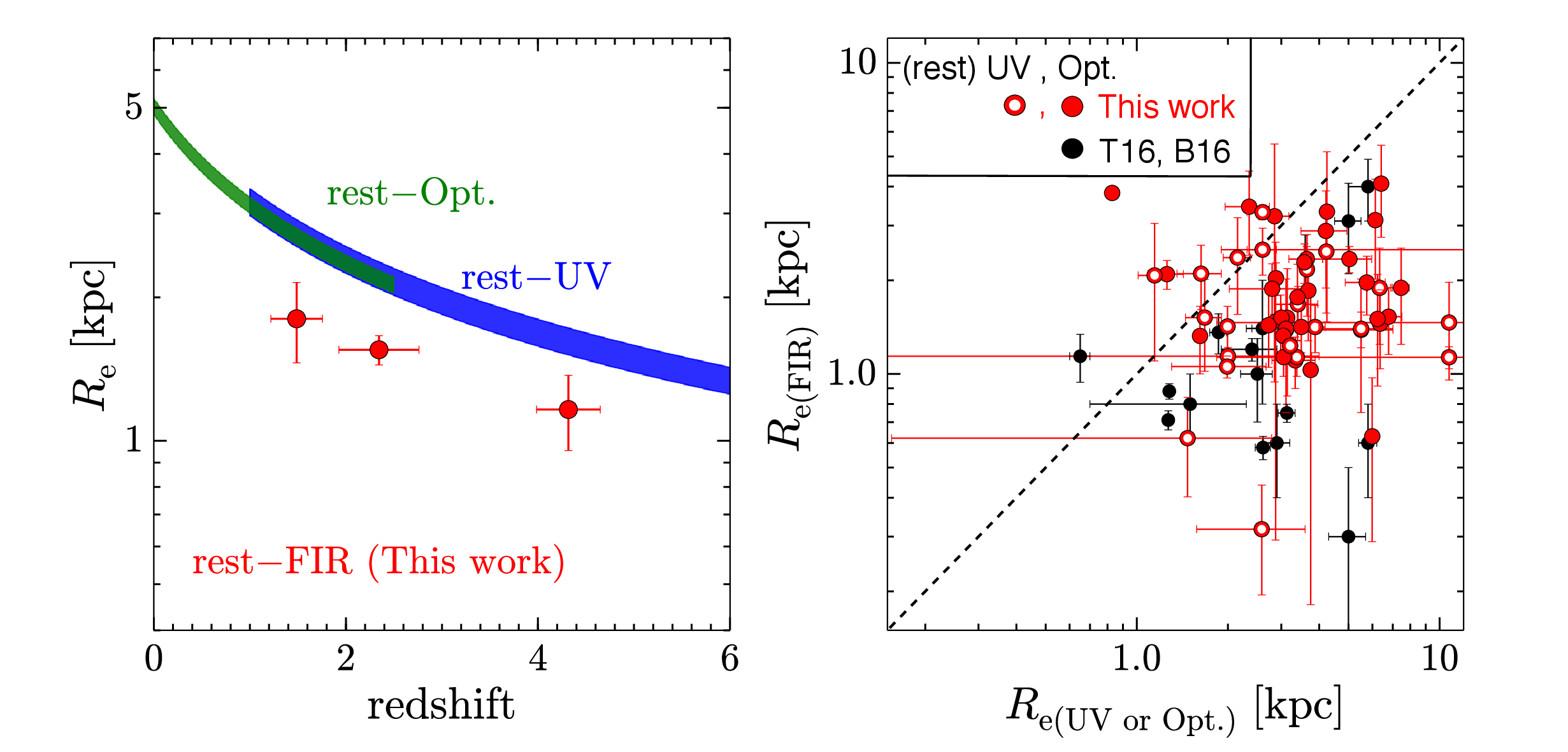}
 \caption[]{
{\it Left:} Redshift evolution of the average values of $R_{\rm e(FIR)}$ (red cycles), $R_{\rm e(UV)}$ (blue shade) and $R_{\rm e(opt.)}$ (green shade) obtain in this work and the literature (see Section \ref{sec:uv-opt-fir}). 
The average $R_{\rm e(UV)}$ and $R_{\rm e(opt.)}$ values are estimated from the results of \cite{shibuya2015}.  
{\it Right:} Rest-frame FIR and UV-optical sizes in the individual galaxies. 
The filled and open circles present the $R_{\rm e(Opt.)}$ and $R_{\rm e(UV)}$ values, respectively, in the abscissa axis. 
The red symbols show the OC10S sources whose $R_{\rm e(Opt.)}$ and/or $R_{\rm e(UV)}$ values are estimated with flag = 0 (reliable) in \cite{vanderwel2012}. 
The black symbols are given in the results of T16 and B16. 
\label{fig:uv-opt_size}}
\end{center}
\end{figure*}

We compare the $R_{\rm e(FIR)}$ values with the $R_{\rm e(UV)}$ and $R_{\rm e(Opt.)}$ values 
to investigate the properties of the dusty starbursts. 
We perform two approaches for this comparison: the statistical and individual approaches. 

First in the statistical approach, we compare the average $R_{\rm e}$ values in the rest-frame FIR and UV-optical wavelengths as a function of redshift.  
For the average $R_{\rm e(FIR)}$ measurements, we use our best estimates of the $R_{\rm e(FIR)}-L_{\rm FIR}$ relation at $z=$ 1$-$2, 2$-$4, and 4$-$6, 
, where the $L_{\rm FIR}$ range is defined as $L_{\rm FIR}=$ $10^{12}$$-$$10^{13}\,L_{\odot}$. 
The left panel of Figure \ref{fig:uv-opt_size} shows the average $R_{\rm e(FIR)}$ values as a function of redshift with the 1$\sigma$ errors. 
For comparison, 
the left panel of Figure \ref{fig:uv-opt_size} also presents the average values of $R_{\rm e (UV)}$ and $R_{\rm e (Opt.)}$ that are estimated from the result of \cite{shibuya2015}. 
The average values of $R_{\rm e (UV)}$ and $R_{\rm e (Opt.)}$ are obtained by fitting a function of $B_{z}(1+z)^{\beta_{z}}$, where $B_{z}$ and $\beta_{z}$ are free parameters. 
In the fitting, we use the results for star-forming galaxies with $M_{\rm star}=10^{10.5}-10^{11}\,M_{\odot}$ that is similar to the $M_{\rm star}$ range of our ALMA sources (Section \ref{sec:sf_mode}). 
In the left panel of Figure \ref{fig:uv-opt_size}, we find that the average values of $R_{\rm e (UV)}$ and $R_{\rm e(Opt.)}$ are constantly higher than those of $R_{\rm e (FIR)}$
in the wide redshift range at $z=1-6$. 

Second in the individual approach, 
we carry out direct comparison of $R_{\rm e}$ between the rest-frame FIR and UV-optical bands for the individual sources. 
We cross-match the OC10S sources with the 3D-HST sources whose
$R_{\rm e}$ values are measured with the HST/$J$- and $H$-bands in \cite{vanderwel2012}. 
In the cross-matching, we only use the 3D-HST sources with the measurement flag = 0 (reliable) in \cite{vanderwel2012}.
We identify 53 sources by this cross-matching that are presented in the right panel of Figure \ref{fig:uv-opt_size}. 
Here, we regard the $R_{\rm e}$ values estimated in the $J$-band at $z=2.5-6$ as $R_{\rm e(UV)}$, 
while the $R_{\rm e}$ values in the $H$-band at $z=1-3$ as $R_{\rm e(Opt.)}$. 

In the right panel of Figure \ref{fig:uv-opt_size}, 
most of our ALMA sources fall in the region of $R_{\rm e(FIR)}\,\lesssim$ $R_{\rm e(UV)}$ ($R_{\rm e(Opt.)}$) within the errors. 
The trend of the small $R_{\rm e(FIR)}$ value is consistent with the results of T16 and B16 that 
are also presented in the right panel of Figure \ref{fig:uv-opt_size}. 
Although a few ALMA sources exceptionally have $R_{\rm e(UV)}$ or $R_{\rm e(Opt.)}$ smaller than $R_{\rm e(FIR)}$ over the errors, 
the small $R_{\rm e(UV)}$ or $R_{\rm e(Opt.)}$ values are probably caused by the strong dust obscuration or the misidentification for the optical-NIR counterparts. 
Therefore, we conclude that the $R_{\rm e(FIR)}$ values are generally smaller than the $R_{\rm e(UV)}$ and $R_{\rm e(Opt.)}$ values, 
which is consistent with the result of the first statistical approach. 

Both results of the statistical and individual approaches suggest
that the $R_{\rm e(FIR)}$ values are smaller than the $R_{\rm e(UV)}$ and $R_{\rm e(Opt.)}$ values. 
This trend would indicate that the dusty starbursts take place in the compact regions.

\subsection{Size and Stellar Mass Relation}
\label{sec:re_Ms}
We examine the $R_{\rm e}$ and $M_{\rm star}$ relation among the different galaxy populations. 
Note that here we do not derive the best-fit function of the FIR size$-$stellar mass relation for our ALMA sources. 
This is because our ALMA sample is not the complete sample at a fixed stellar mass (Section \ref{sec:sf_mode}). 
Moreover, the positive FIR size$-$luminosity relation implies that the sources with small FIR sizes are difficult to be detected at a fixed $M_{\rm star}$ due to their faintness, 
which causes a bias on the FIR size$-$stellar mass plane. 
 
\begin{figure}
\begin{center}
\includegraphics[trim=0cm 0cm 0cm 0cm, clip, angle=0,width=0.5\textwidth]{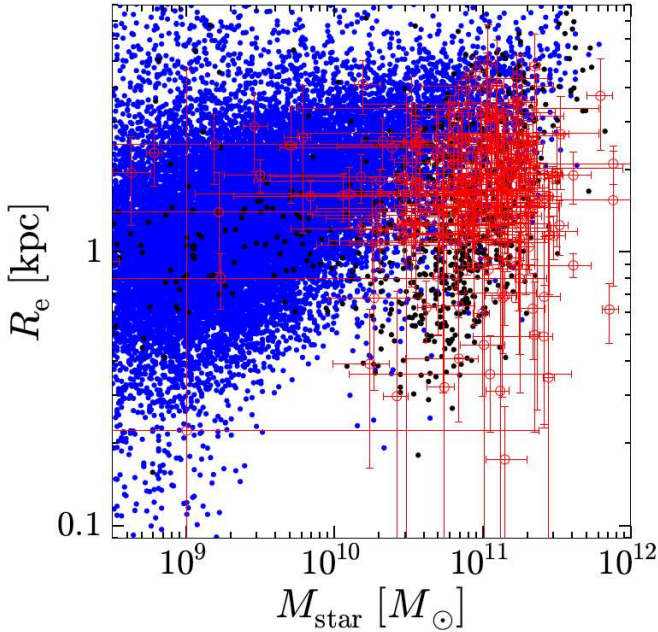}
 \caption[]{
Size$-$stellar mass relation for our ALMA sources and other galaxy populations. 
The red open circles present the $R_{\rm e}$ values in the rest-frame FIR for the OC10S sources. 
The blue and black circles denote the $R_{\rm e}$ values in the rest-frame optical wavelength for the star-forming and quiescent galaxies at $z=$ 1$-$3, respectively, in the 3D-HST regions obtained by \cite{shibuya2015}. 
\label{fig:size_mstar}}
\end{center}
\end{figure}

Figure \ref{fig:size_mstar} presents the $R_{\rm e}$ values in the rest-frame FIR as a function of $M_{\rm star}$ for the OC10S sources with red circles. 
For comparison, Figure \ref{fig:size_mstar} also shows the $R_{\rm e}$ values in the rest-frame optical wavelength for the star-forming and the quiescent galaxies at $z=$ 1$-$3 with blue and black circles, respectively. 
These blue and black circles are obtained from the HST results \citep{shibuya2015} 
which show the sequence of the quiescent galaxies is placed below that of the star-forming galaxies on the $R_{\rm e}$$-$$M_{\rm star}$ plane. 
In Figure \ref{fig:size_mstar}, we find that the majority of our ALMA sources fall on the sequence of the quiescent galaxies. 
Since the intense star formation in the dusty starbursts are potentially transformed into the major part of the stellar mass distribution in the host galaxies, 
it may be natural that our ALMA sources and the quiescent galaxies have the similar $R_{\rm e}$ values in the rest-frame FIR and optical wavelength, respectively. 
This connection is consistent with the evolutionary scenario of the local elliptical galaxies from the high-$z$ dusty starbursts through the compact quiescent galaxies at $z\sim2$ \citep[e.g.,][]{lilly1999,genzel2003,tacconi2008,hickox2012,toft2014,chen2015,simpson2015a,barro2016}. 
For some of our ALMA sources that are located between the sequences of the sta-forming and the quiescent galaxies, 
we may witness the transition phase from the star-forming to the quiescent galaxies under the rapid mass assembly through the dusty starbursts. 

\subsection{Spatial Offset between FIR and UV-Optical Emission}
\label{sec:offset}
\begin{figure}
\begin{center}
\includegraphics[trim=0cm 0.5cm 0.3cm 0.5cm, clip, angle=0,width=0.5\textwidth]{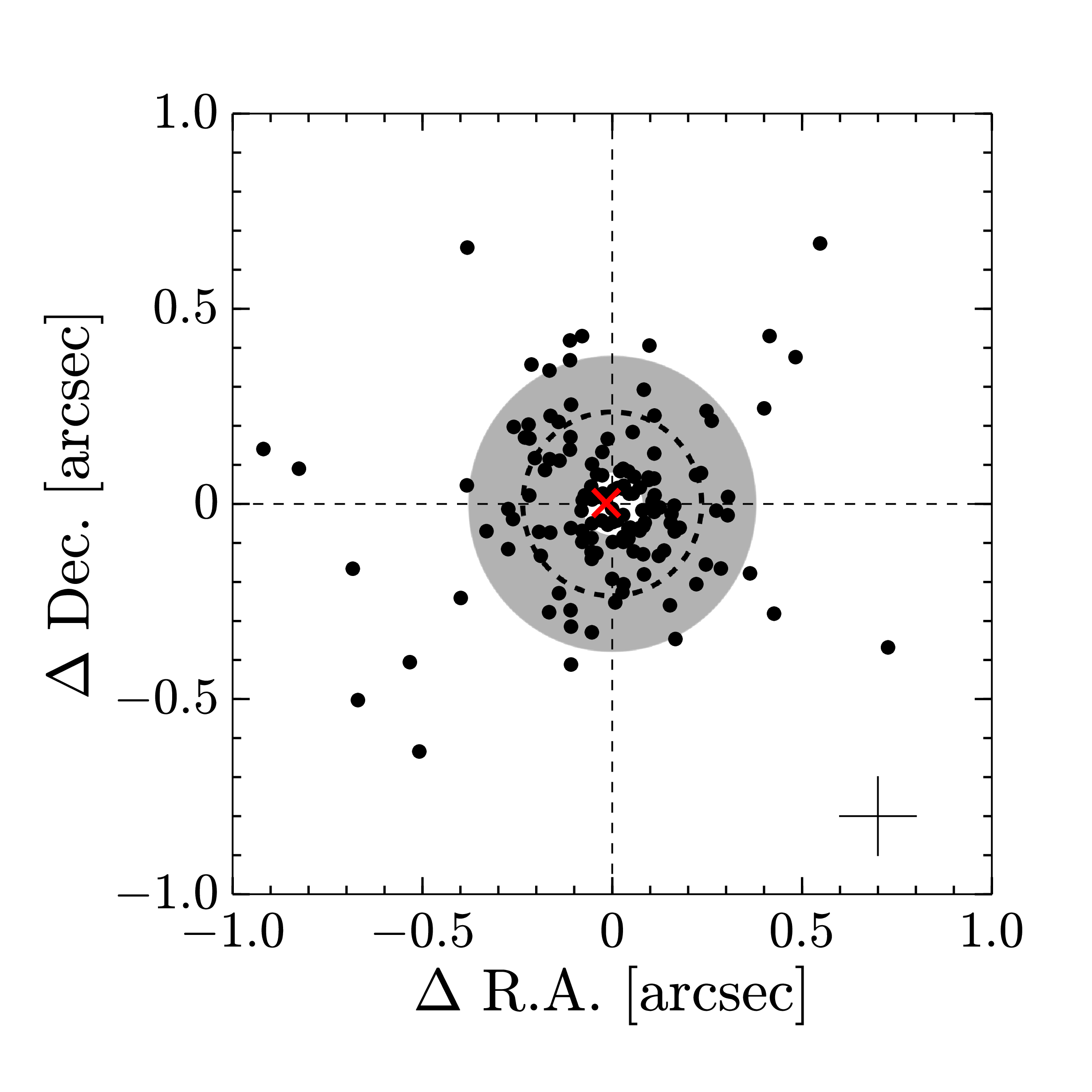}
 \caption{
Spatial offsets between source centers of ALMA and HST/$H$-band counterparts. 
The black circles show the 136 OC5S sources that are identified in the 3D-HST regions. 
The typical error scale is presented at the bottom right. 
The red cross indicates the average offset, ($\Delta$R.A,$\Delta$Dec) = ($-0\farcs02$,$0\farcs00$). 
The dashed circle and the shade region denote the median and 16$-$84th percentiles of the data distribution, respectively. 
\label{fig:offset}}
\end{center}
\end{figure}

We estimate the offsets between the rest-frame FIR and UV-optical emission from our ALMA sources. 
Because the offset measurements require a high resolution in the rest-frame UV-optical emission data, 
we use OC5S sources that have an optical-NIR counterpart in the 3D-HST images. 
We cross-match the OC5S sources with the 3D-HST sources \citep{momcheva2016}, 
and identify 136 OC5S sources  that are detected in the deep HST images. 
Figure \ref{fig:offset} presents the offsets between the centers of the OC5S sources in the ALMA images and their optical-NIR counterparts in the HST/$H$-band images.

In Figure \ref{fig:offset}, we find that the source distribution on average shows almost no offset from the center. 
The average offset is estimated to be ($\Delta$R.A, $\Delta$Dec) = ($-0\farcs02$,$0\farcs00$) that resides within the typical error scale of the individual offsets ($\sim0\farcs1$). 
In Figure \ref{fig:offset}, we also find that the some individual plots show the large scatters from the center with the maximum offset of $0\farcs92$. 
There are two possibilities for the large scatters. 

First possibility is the astrometry uncertainty. 
\cite{dunlop2017} report the astrometric uncertainty of HST that causes a systematic offset of $\sim0\farcs25$ between the ALMA and HST images in GOODS-S. 
In fact, the median value of the individual scatters is estimated to be $0\farcs24$. 
Although the good agreement between the average offset and the center in Figure \ref{fig:offset} indicates that there are no similar astrometric uncertainties of HST in other fields, we cannot rule out the possibility that the part of the individual scatters is caused by the astrometric uncertainty. 

Second possibility is the intrinsic offset between the rest-frame FIR and UV-optical emission. 
About the half of the sources has the scatters larger than the potential astrometric uncertainty of $0\farcs25$. 
\cite{chen2015} estimate the typical offset of $0\farcs4\pm0\farcs05$ with the ALMA and HST observations for the SMGs, 
which also indicates the existence of the intrinsic offsets beyond the astrometric uncertainty. 

In the second case, one interpretation is that the heavy dust obscuration causes the offsets between the rest-frame FIR and UV-optical emission. 
The rest-frame FIR emission is produced in the dusty star-bursting area, 
while the rest-frame UV-optical emission comes from the moderate star-forming area with the small amount of dust. 
Another interpretation is that we identify the optical-NIR counterparts that are not physically related to the ALMA sources.  
We calculate the probability of the misidentification due to the chance projection with the offset range of $0\farcs25$$-$$0\farcs92$ 
whose lower and upper limits are defined by the astrometric uncertainty and the maximum offset. 
Following the calculation method of \cite{downes1986}, we estimate the P-value of the chance projection to be $\sim1\%-13\%$
with the number density of the $H$-band sources given in \cite{momcheva2016}. 
We then obtain the expected number of the misidentification $\sim$ 3 by integrating the P-values. 
This suggests that the contribution of the misidentification is negligible in our statistical result. 
We thus conclude that majority of the large scatter over the potential astrometric uncertainty is caused by the intrinsic offset between the rest-frame FIR and  UV-optical emission. 

\subsection{Do Mergers trigger Dusty Starbursts ?}
\label{sec:morophology}
We test whether our ALMA sources are major mergers based on the rest-frame UV and optical morphology. 
To remove the pre-selection bias in the rest-frame UV and optical bands, 
here we use the OC5S-mmT sources whose central targets in the initial ALMA observations are selected in the submm/mm bands (Section \ref{sec:nz}). 

\begin{figure*}
\begin{center}
\includegraphics[trim=0cm 0cm 0cm 0cm, clip, angle=0,width=1.0\textwidth]{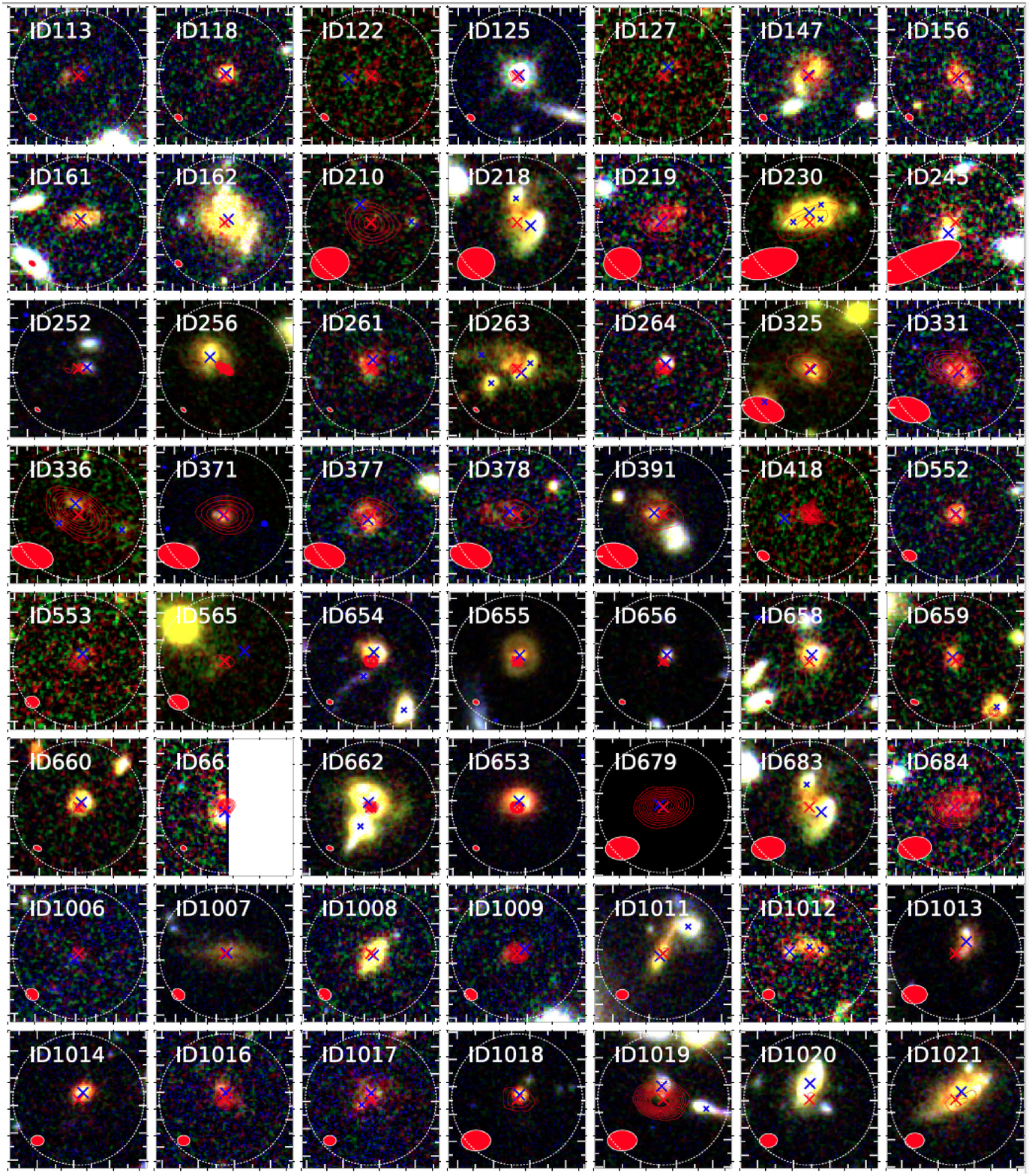}
 \caption
{$5''\times5''$ fake-color HST images of the 56 OC5S-mmT sources that are identified in the 3D-HST regions (red: $H_{160}$, green: $J_{125}$, blue: $I_{814}$). 
The red contours indicate ALMA mm band intensity from the 5 to 30$\sigma$ levels with a 2$\sigma$-level step. 
The red cross represents the ALMA source center. 
The red ellipse shows the synthesized beam size. 
The white circle denotes the $2\farcs5$ search radius for the optical-NIR objects. 
The large blue cross presents the optical-NIR counterparts of the ALMA sources, 
while the small blue cross indicates the major merger pair.
Note that ID661 is placed at the edge of the coverage of the HST observations. 
\label{fig:postage}}
\end{center}
\end{figure*}

Firstly, we cross-match the OC5S-mmT sources with the 3D-HST sources  \citep{momcheva2016}. 
The 3D-HST sources in the high spatial resolution HST images are necessary to perform the homogeneous morphological analysis. 
We identify 56 OC5S-mmT sources in this process, 
and Figure \ref{fig:postage} shows the false-color HST images for these 56 OC5S-mmT sources. 

Secondly, we classify whether the 56 OC5S-mmT sources have the major mergers.  
We regard the OC5S-mmT sources that have major merger pairs as the major mergers. 
We identify the major merger pairs with the following three steps: 
i) selecting the optical-NIR objects that are located within an offset of $2\farcs5$ from the OC5S-mmT source centers, 
ii) removing the optical-NIR objects with the $\Delta z_{\rm phot}\geq1$ from the optical-NIR counterpart of the OC5S-mmT source, and
iii) removing the optical-NIR objects with $M_{\rm star}$ less than 10\% of that of the optical-NIR counterpart.
We find that 
27\% ($=$15/56) of the OC5S-mmT sources have the major merger pairs.  
Because our i)$-$iii) steps cannot investigate close major merger pairs within the spatial resolution of HST/$H$-band ($0\farcs$18), 
we refer to these 27\% of the OC5S-mmT sources as "early/mid-stage major merger". 
The rest of 73\% ($=$41/56) of the OC5S-mmT sources are referred to as "isolated galaxy". 
Note that the radius of $2\farcs5$ in the step of i) corresponds to $\sim20$ kpc at $z=2.5$ that is used in \cite{lefevre2000} to identify the major merger system. 
We identify no additional OS5S-mmT sources that have the major merger pairs with an extended radius of 6$''$ ($\sim$ 50 kpc at $z=2.5$), 
ensuring that we have not missed any potential major mergers. 

\begin{figure*}
\begin{center}
\includegraphics[trim=0cm 0cm 0cm 0cm, clip, angle=0,width=1.0\textwidth]{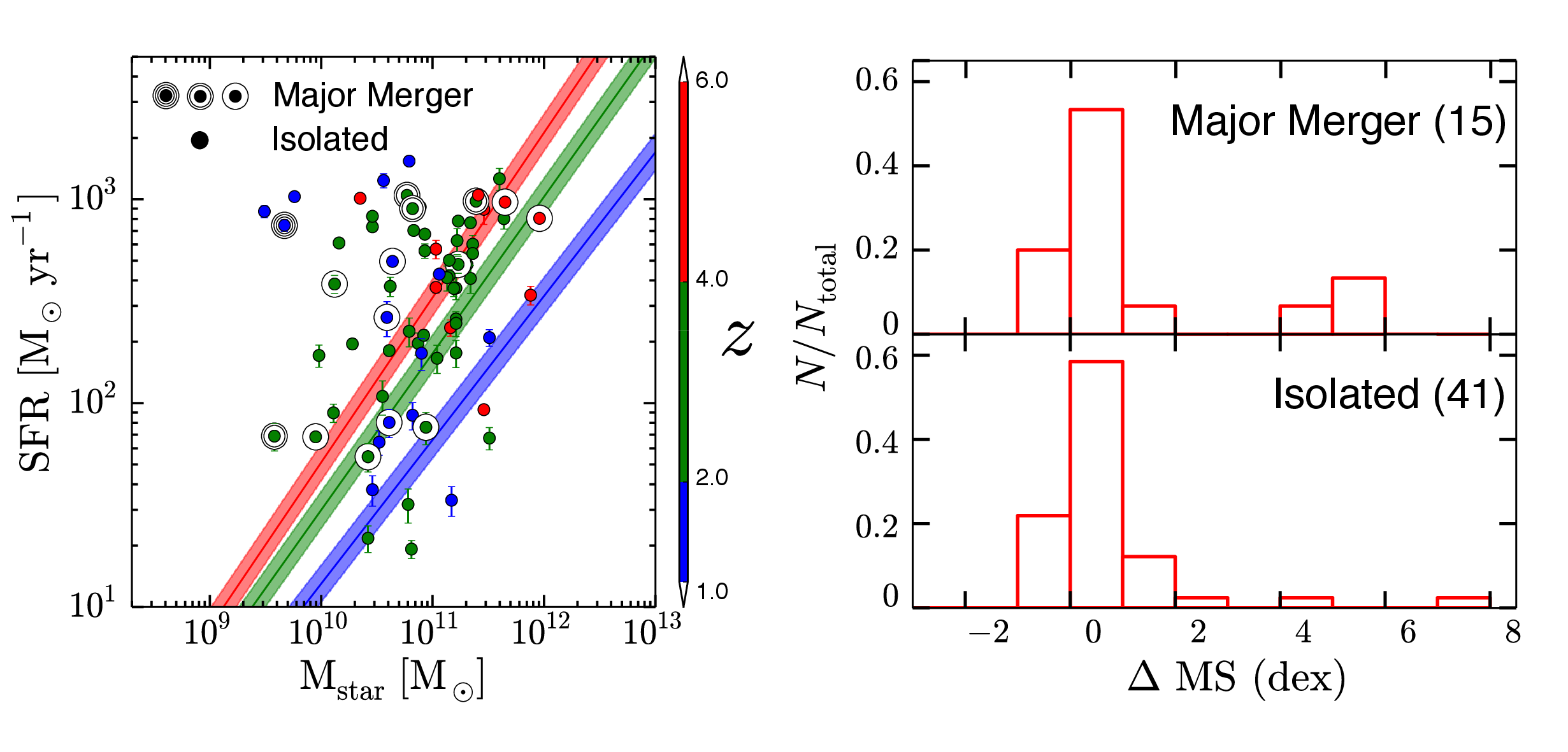}
 \caption[]
{{\it Left:} 
Our ALMA sources in the $M_{\rm star}$$-$SFR plane for investigating the star-formation mode. 
The circles represent the 56 OC5S-mmT sources whose optical-NIR counterparts are detected in the 3D-HST regions. 
The number of the circles indicates the number of the multiple components in the major merger system that is indicated by the rest-frame UV-optical morphologies. 
The solid lines and the shade regions are the main sequences of the $M_{\rm star}$$-$SFR relation and the associated $1\sigma$ errors, respectively, at $z=$ 1$-$2, 2$-$4, and 4$-$6 that are estimated by \cite{speagle2014}. 
The colors correspond to the redshift ranges in the color-bar scale. 
{\it Right:} Histograms of the isolated galaxies (bottom) and the major mergers (top) as a function of $\Delta$MS, where $\Delta$MS is the difference in sSFR from the main sequence. 
\label{fig:ms_merger}}
\end{center}
\end{figure*}

Thirdly, we investigate the star-formation modes in the early/mid-stage major mergers and the isolated galaxies. 
The left panel of Figure \ref{fig:ms_merger} presents the $M_{\rm star}-{\rm SFR}$ plots of the 56 OC5S-mmT sources.  
In the right panel of Figure \ref{fig:ms_merger}, 
we show the histograms for the early/mid-stage major mergers and the isolated galaxies as a function of $\Delta$MS in units of sSFR, 
where $\Delta$MS is the offset from the main sequence of the $M_{\rm star}-{\rm SFR}$ relation. 
The right panel of Figure \ref{fig:ms_merger} indicates that the histogram of the early/mid-stage major mergers is similar to that of the isolated galaxies. 
We perform the KS-test for these two histograms. 
The result of the KS-test shows that one cannot rule out the possibility that the samples of the early/mid-stage major mergers and the isolated galaxies are made from the same parent sample. 

One explanation for the similar histograms between the early/mid-stage major mergers and the isolated galaxies 
is that the isolated galaxies are dominated by the major mergers such as late-stage one. 
To examine whether the isolated galaxies have the merger-like morphology, 
we cross-match the isolated galaxies with the morphology catalog of \cite{huertas-company2015} that 
estimate the probabilities of the spheroid, disk, irregular, compact, and unclassifiable morphologies with the Convolutional Neural Network technique. 
Here, we adopt the threshold of the irregular probability over 50\% to select the isolated galaxies with the merger-like morphology. 
We identify twenty-five isolate galaxies in the morphology catalog, 
and find that only four out of the twenty-five have the merger-like morphology. 
Changing the threshold of the irregular probability, we confirm that the fraction of 
the isolated galaxies with the merger-like morphology is not significantly changed. 
This suggests that the isolated galaxies are not dominated by the major mergers. 
If we combine our selections of the early/mid-stage major mergers and the isolated galaxies with the merger-like morphology, 
only $\sim$30$-$40\% of our ALMA sources are classified as the major mergers. 
We thus conclude that the dusty starbursts are triggered not only by the major mergers but also the other mechanism(s).

Note that the visual-like morphology classification depends on the spatial resolution. 
In \cite{huertas-company2015}, 
the morphology classification is performed with the HST/$H$-band images that have the spatial resolution of $0\farcs18$. 
In recent ALMA observations, \cite{iono2016} and \cite{oteo2016} achieve ultra-high resolutions of $0\farcs015-0\farcs05$. 
Although the wavelengths are different in our analysis and in the recent ALMA observations, 
it may be possible to obtain the hints for understanding the dusty starbursts. 
The ALMA images with the ultra-high resolutions reveal that four out of five high-$z$ SMGs have several $\sim200-300$ pc scale clumps with the SFR 
density of $\sim300-3000\,M_{\odot}$yr$^{\rm -1}$kpc$^{\rm -2}$ in the central kiloparcsec. 
Since the SFR density values correspond to those of the local major merger of Arp220, 
the physical origins can be explained by the major mergers \citep{iono2016}. 
Although the origins of the compact clumps at the central regions are still under debated \citep[cf.][]{hodge2016}, 
there remains the possibility that some of the isolated galaxies are also resolved into the merger-like morphologies by future observations with the ultra-high resolution. 

\section{Summary}
\label{sec:summary}
In this paper, we study the statistics of the effective radius in the rest-frame FIR, $R_{\rm e(FIR)}$, 
based on the 1627 deep ALMA maps in Band 6/7 that are open for public by 2017 July. 
The sample consists of the 1034 submm/mm continuum sources, which is the largest ALMA source sample ever made. 
The 577 optical-NIR counterparts are identified with the photometric redshifts at $z=$ 0$-$6. 
The redshift distribution shows that there exists no pre-selection bias by the initial ALMA observations. 
Our homogeneous $uv$-visibility size analyses with the exponential disk model ($n=1$) allow us to measure the $R_{\rm e(FIR)}$ values for the large sample by the same technique. 
Evaluating the selection and the measurement completeness carefully with the Monte-Carlo simulations, 
we identify the $R_{\rm e(FIR)}-L_{\rm FIR}$ relation and the $R_{\rm e(FIR)}$ evolution.  
Comparing our results with the morphological properties in the rest-frame UV-optical bands, 
we discuss the physical origins of the dusty starbursts. 
The major findings of this paper are summarized below.
\begin{enumerate}
\item 
Our ALMA sources typically have the SFR values of $\sim100$$-$$1000\,M_{\odot}{\rm yr^{-1}}$ and the $M_{\rm star}$ values of $\sim10^{10}$$-$$10^{11.5}\,M_{\odot}$.  
We find that the half of our ALMA sources are the starbursts, while the rest of the half are the high-mass end of the main sequence. 
The starburst fraction decreases towards high redshifts, which is consistent with the ALESS result. 
\item The redshift distribution is not changed by $L_{\rm FIR }$ nor ALMA Band 6/7.  
The median redshift is estimated to be $z_{\rm med}=2.36$ that is consistent with the previous studies. 
\item We derive the $R_{\rm e(FIR)}$$-$$L_{\rm FIR}$ relation over the wide redshift range of $z=$ 0$-$6, 
and find that there exists a positive correlation at the $>$99\% significance level. 
\item The power-law fitting of $R_{\rm e(FIR)}\propto L_{\rm FIR}^{\alpha^{\rm IR}}$ allows us to obtain the best-fit slope of $\alpha^{\rm FIR}=0.28\pm0.07$.  
The best-fit slope value is not changed within the error 
even without the AGNs, potential lensed sources, and the potentially low-quality data of early ALMA cycles. 
Moreover, the $R_{\rm e(FIR)}$ values at a fixed $L_{\rm FIR}$ decreases toward high redshifts.  
Both results of the best-fit $\alpha^{\rm FIR}$ and the redshift evolution trend are consistent with those of the galaxy effective radius in the rest-frame UV (optical) wavelength $R_{\rm e(UV)}$ ($R_{\rm e(Opt.)}$). 
\item On the statistical basis, the average values of $R_{\rm e(FIR)}$ at $z=$ 1$-$6 are generally smaller than those of $R_{\rm e(UV)}$ and $R_{\rm e(Opt.)}$ in the similar $M_{\rm star}$ range. 
On the individual source basis, the direct comparison between the $R_{\rm e(FIR)}$ and $R_{\rm e(UV)}$ ($R_{\rm e(Opt.)}$) for individual galaxies 
also supports the relation of $R_{\rm e(FIR)} \lesssim$ $R_{\rm e(UV)}$ ($R_{\rm e(Opt.)}$). 
Both statistical and individual results suggest that the dusty starbursts take place in the compact regions. 
\item On the size$-$stellar mass plane, $R_{\rm e(FIR)}$ of our ALMA sources and $R_{\rm e(opt.)}$ of the quiescent galaxies at $z=$ 1$-$3 shows the similar distribution. 
This implies that the intense star-formation of the dusty starbursts is transformed into the major part of the stellar distribution of the host galaxies that are evolved to the local elliptical galaxies through the compact quiescent galaxies. 
\item The median spatial offset between the rest-frame FIR and UV-optical emission is estimated to be $0\farcs24$ that may be explained by the astrometric uncertainty of $\sim0\farcs25$. 
The offset values larger than the astrometric uncertainty is not caused by the misidentification for the optical-NIR counterparts due to the chance projection, 
but the intrinsic spatial offsets between the rest-frame FIR and UV star-forming regions. 
\item We test whether our ALMA sources are major mergers based on the rest-frame UV and optical morphology with the deep HST images. 
We identify 27\% of our ALMA sources as the early/mid-stage major mergers, while the rest of the 73\% as the isolated galaxies. 
The total fraction of the major mergers including the isolated galaxies with the merger-like morphology is estimated to be $\sim$30$-$40\% in our ALMA sources.
This indicates that the dusty starbursts are triggered not only by the major mergers but also the other mechanism(s). 
\end{enumerate}

We are grateful to Soh Ikarashi and Ken-ichi Tadaki for giving us helpful advice on analyzing the data, 
and to James Simpson for providing us the data with the useful information. 
We appreciate Rob Ivison and Zhi-Yu Zhang for giving us helpful suggestions and comments. 
We also thank 
Yoshiaki Ono, Ivan Marti-Vidal, Sebastien Muller, Yoichi Tamura, Nicholas Z. Scoville, Jacqueline A. Hodge, Kristen K. Knudsen, Chian-Chou Chen, Andrea Ferrara, Hiroyuki Hirashita, Ryota Kawamata, Inoue Shigeki, Daisuke Iono, Takuma Izumi, Junko Ueda, Bunyo Hatsukade, Graziano Ucci, Sune Toft, and Carlos G$\acute{\rm o}$mez-Guijarro for useful comments and discussions. 
We are indebted to Alex Hagen, Yuichi Harikane, Akira Konno, Yoshiaki Ono, and Ryo Higuchi for helping to name the project title. 
We appreciate the support of the staffs at the ALMA Regional Center. 
This paper makes us use of following ALMA data: ADS/JAO. ALMA 
\#2011.0.00097.S, 
\#2011.0.00294.S, 
\#2012.1.00076.S, 
\#2012.1.00090.S, 
\#2012.1.00245.S, 
\#2012.1.00307.S, 
\#2012.1.00326.S, 
\#2012.1.00523.S, 
\#2012.1.00756.S, 
\#2012.1.00775.S, 
\#2012.1.00869.S, 
\#2012.1.00979.S, 
\#2012.1.00983.S, 
\#2013.1.00034.S, 
\#2013.1.00118.S, 
\#2013.1.00151.S, 
\#2013.1.00173.S, 
\#2013.1.00718.S, 
\#2013.1.00884.S, 
\#2013.1.00205.S, 
\#2013.1.00208.S, 
\#2013.1.00566.S, 
\#2013.1.00668.S, 
\#2013.1.00781.S, 
\#2013.1.00999.S,
\#2013.1.01292.S, 
\#2015.1.00137.S, 
\#2015.1.00540.S, 
\#2015.1.00664.S, and 
\#2015.1.01495.S.
ALMA is a partnership of the ESO (representing its member states), 
NSF (USA), and NINS (Japan), together with the NRC (Canada) and NSC and ASIAA (Taiwan), 
in cooperation with the Republic of Chile. 
The Joint ALMA Observatory is operated by the ESO, AUI/NRAO, and NAOJ. 
This work is supported by 
World Premier International Research Center Initiative (WPI Inititative), MEXT, Japan, and KAKENHI (15H02064), 
Grant-in-Aid for JSPS Research Fellow, a grant from the Hayakawa Satio Fund awarded by the Astronomical Society of Japan, 
and the ALMA Japan Research Grant of NAOJ Chile Observatory, NAOJ-ALMA-164. 
This work has made use of the Rainbow Cosmological Surveys Database, which is operated by the Universidad Complutense de Madrid (UCM), 
partnered with the University of California Observatories at Santa Cruz (UCO/Lick, UCSC).

\bibliographystyle{apj}
\bibliography{apj-jour,reference}

\begin{thebibliography}{}
\expandafter\ifx\csname natexlab\endcsname\relax\def\natexlab#1{#1}\fi

\bibitem[{{Barro} {et~al.}(2016){Barro}, {Kriek}, {P{\'e}rez-Gonz{\'a}lez},
  {Trump}, {Koo}, {Faber}, {Dekel}, {Primack}, {Guo}, {Kocevski},
  {Mu{\~n}oz-Mateos}, {Rujopakarn}, \& {Seth}}]{barro2016}
{Barro}, G., {Kriek}, M., {P{\'e}rez-Gonz{\'a}lez}, P.~G., {et~al.} 2016,
  \apjl, 827, L32

\bibitem[{{Bertin} \& {Arnouts}(1996)}]{bertin1996}
{Bertin}, E., \& {Arnouts}, S. 1996, \aap, 117, 393

\bibitem[{{Casey} {et~al.}(2014){Casey}, {Narayanan}, \& {Cooray}}]{casey2014}
{Casey}, C.~M., {Narayanan}, D., \& {Cooray}, A. 2014, \physrep, 541, 45

\bibitem[{{Castellano} {et~al.}(2016){Castellano}, {Amor{\'{\i}}n}, {Merlin},
  {Fontana}, {McLure}, {M{\'a}rmol-Queralt{\'o}}, {Mortlock}, {Parsa},
  {Dunlop}, {Elbaz}, {Balestra}, {Boucaud}, {Bourne}, {Boutsia}, {Brammer},
  {Bruce}, {Buitrago}, {Capak}, {Cappelluti}, {Ciesla}, {Comastri}, {Cullen},
  {Derriere}, {Faber}, {Giallongo}, {Grazian}, {Grillo}, {Mercurio},
  {Micha{\l}owski}, {Nonino}, {Paris}, {Pentericci}, {Pilo}, {Rosati},
  {Santini}, {Schreiber}, {Shu}, \& {Wang}}]{castellano2016}
{Castellano}, M., {Amor{\'{\i}}n}, R., {Merlin}, E., {et~al.} 2016, \aap, 590,
  A31

\bibitem[{{Chabrier}(2003)}]{chabrier2003}
{Chabrier}, G. 2003, \pasp, 115, 763

\bibitem[{{Chapin} {et~al.}(2009){Chapin}, {Pope}, {Scott}, {Aretxaga},
  {Austermann}, {Chary}, {Coppin}, {Halpern}, {Hughes}, {Lowenthal},
  {Morrison}, {Perera}, {Scott}, {Wilson}, \& {Yun}}]{chapin2009}
{Chapin}, E.~L., {Pope}, A., {Scott}, D., {et~al.} 2009, \mnras, 398, 1793

\bibitem[{{Chapman} {et~al.}(2005){Chapman}, {Blain}, {Smail}, \&
  {Ivison}}]{chapman2005}
{Chapman}, S.~C., {Blain}, A.~W., {Smail}, I., \& {Ivison}, R.~J. 2005, \apj,
  622, 772

\bibitem[{{Chen} {et~al.}(2015){Chen}, {Smail}, {Swinbank}, {Simpson}, {Ma},
  {Alexander}, {Biggs}, {Brandt}, {Chapman}, {Coppin}, {Danielson},
  {Dannerbauer}, {Edge}, {Greve}, {Ivison}, {Karim}, {Menten}, {Schinnerer},
  {Walter}, {Wardlow}, {Wei{\ss}}, \& {van der Werf}}]{chen2015}
{Chen}, C.-C., {Smail}, I., {Swinbank}, A.~M., {et~al.} 2015, \apj, 799, 194

\bibitem[{{Civano} {et~al.}(2012){Civano}, {Elvis}, {Brusa}, {Comastri},
  {Salvato}, {Zamorani}, {Aldcroft}, {Bongiorno}, {Capak}, {Cappelluti},
  {Cisternas}, {Fiore}, {Fruscione}, {Hao}, {Kartaltepe}, {Koekemoer}, {Gilli},
  {Impey}, {Lanzuisi}, {Lusso}, {Mainieri}, {Miyaji}, {Lilly}, {Masters},
  {Puccetti}, {Schawinski}, {Scoville}, {Silverman}, {Trump}, {Urry},
  {Vignali}, \& {Wright}}]{civano2012}
{Civano}, F., {Elvis}, M., {Brusa}, M., {et~al.} 2012, \apjs, 201, 30

\bibitem[{{Coppin} {et~al.}(2008){Coppin}, {Halpern}, {Scott}, {Borys},
  {Dunlop}, {Dunne}, {Ivison}, {Wagg}, {Aretxaga}, {Battistelli}, {Benson},
  {Blain}, {Chapman}, {Clements}, {Dye}, {Farrah}, {Hughes}, {Jenness}, {van
  Kampen}, {Lacey}, {Mortier}, {Pope}, {Priddey}, {Serjeant}, {Smail},
  {Stevens}, \& {Vaccari}}]{coppin2008}
{Coppin}, K., {Halpern}, M., {Scott}, D., {et~al.} 2008, \mnras, 384, 1597

\bibitem[{{Curran}(2014)}]{curran2014}
{Curran}, P.~A. 2014, ArXiv e-prints, arXiv:1411.3816

\bibitem[{{da Cunha} {et~al.}(2013){da Cunha}, {Groves}, {Walter}, {Decarli},
  {Weiss}, {Bertoldi}, {Carilli}, {Daddi}, {Elbaz}, {Ivison}, {Maiolino},
  {Riechers}, {Rix}, {Sargent}, \& {Smail}}]{cunha2013}
{da Cunha}, E., {Groves}, B., {Walter}, F., {et~al.} 2013, \apj, 766, 13

\bibitem[{{da Cunha} {et~al.}(2015){da Cunha}, {Walter}, {Smail}, {Swinbank},
  {Simpson}, {Decarli}, {Hodge}, {Weiss}, {van der Werf}, {Bertoldi},
  {Chapman}, {Cox}, {Danielson}, {Dannerbauer}, {Greve}, {Ivison}, {Karim}, \&
  {Thomson}}]{da-cunha2015}
{da Cunha}, E., {Walter}, F., {Smail}, I.~R., {et~al.} 2015, \apj, 806, 110

\bibitem[{{Downes} {et~al.}(1986){Downes}, {Peacock}, {Savage}, \&
  {Carrie}}]{downes1986}
{Downes}, A.~J.~B., {Peacock}, J.~A., {Savage}, A., \& {Carrie}, D.~R. 1986,
  \mnras, 218, 31

\bibitem[{{Dunlop} {et~al.}(2017){Dunlop}, {McLure}, {Biggs}, {Geach},
  {Micha{\l}owski}, {Ivison}, {Rujopakarn}, {van Kampen}, {Kirkpatrick},
  {Pope}, {Scott}, {Swinbank}, {Targett}, {Aretxaga}, {Austermann}, {Best},
  {Bruce}, {Chapin}, {Charlot}, {Cirasuolo}, {Coppin}, {Ellis}, {Finkelstein},
  {Hayward}, {Hughes}, {Ibar}, {Jagannathan}, {Khochfar}, {Koprowski},
  {Narayanan}, {Nyland}, {Papovich}, {Peacock}, {Rieke}, {Robertson},
  {Vernstrom}, {Werf}, {Wilson}, \& {Yun}}]{dunlop2017}
{Dunlop}, J.~S., {McLure}, R.~J., {Biggs}, A.~D., {et~al.} 2017, \mnras, 466,
  861

\bibitem[{{Fall}(1983)}]{fall1983}
{Fall}, S.~M. 1983, in Internal Kinematics and Dynamics of Galaxies, ed.
  E.~{Athanassoula}, Vol. 100, 391--398

\bibitem[{{Fall}(2002)}]{fall2002}
{Fall}, S.~M. 2002, in Astronomical Society of the Pacific Conference Series,
  Vol. 273, The Dynamics, Structure, and History of Galaxies: A Workshop in
  Honour of Professor Ken Freema, ed. G.~S. {Da Costa}, E.~M. {Sadler}, \&
  H.~{Jerjen}, 289

\bibitem[{{Ferguson} {et~al.}(2004){Ferguson}, {Dickinson}, {Giavalisco},
  {Kretchmer}, {Ravindranath}, {Idzi}, {Taylor}, {Conselice}, {Fall},
  {Gardner}, {Livio}, {Madau}, {Moustakas}, {Papovich}, {Somerville},
  {Spinrad}, \& {Stern}}]{ferguson2004}
{Ferguson}, H.~C., {Dickinson}, M., {Giavalisco}, M., {et~al.} 2004, \apjl,
  600, L107

\bibitem[{{Fujimoto} {et~al.}(2016){Fujimoto}, {Ouchi}, {Ono}, {Shibuya},
  {Ishigaki}, {Nagai}, \& {Momose}}]{fujimoto2016}
{Fujimoto}, S., {Ouchi}, M., {Ono}, Y., {et~al.} 2016, \apjs, 222, 1

\bibitem[{{Furusawa} {et~al.}(2008){Furusawa}, {Kosugi}, {Akiyama}, {Takata},
  {Sekiguchi}, {Tanaka}, {Iwata}, {Kajisawa}, {Yasuda}, {Doi}, {Ouchi},
  {Simpson}, {Shimasaku}, {Yamada}, {Furusawa}, {Morokuma}, {Ishida}, {Aoki},
  {Fuse}, {Imanishi}, {Iye}, {Karoji}, {Kobayashi}, {Kodama}, {Komiyama},
  {Maeda}, {Miyazaki}, {Mizumoto}, {Nakata}, {Noumaru}, {Ogasawara}, {Okamura},
  {Saito}, {Sasaki}, {Ueda}, \& {Yoshida}}]{furusawa2008}
{Furusawa}, H., {Kosugi}, G., {Akiyama}, M., {et~al.} 2008, \apjs, 176, 1

\bibitem[{{Genzel} {et~al.}(2003){Genzel}, {Baker}, {Tacconi}, {Lutz}, {Cox},
  {Guilloteau}, \& {Omont}}]{genzel2003}
{Genzel}, R., {Baker}, A.~J., {Tacconi}, L.~J., {et~al.} 2003, \apj, 584, 633

\bibitem[{{Gonz{\'a}lez-L{\'o}pez} {et~al.}(2017){Gonz{\'a}lez-L{\'o}pez},
  {Bauer}, {Romero-Ca{\~n}izales}, {Kneissl}, {Villard}, {Carvajal}, {Kim},
  {Laporte}, {Anguita}, {Aravena}, {Bouwens}, {Bradley}, {Carrasco}, {Demarco},
  {Ford}, {Ibar}, {Infante}, {Messias}, {Mu{\~n}oz Arancibia}, {Nagar},
  {Padilla}, {Treister}, {Troncoso}, \& {Zitrin}}]{gonzalez2017}
{Gonz{\'a}lez-L{\'o}pez}, J., {Bauer}, F.~E., {Romero-Ca{\~n}izales}, C.,
  {et~al.} 2017, \aap, 597, A41

\bibitem[{{Hathi} {et~al.}(2008){Hathi}, {Malhotra}, \& {Rhoads}}]{hathi2008}
{Hathi}, N.~P., {Malhotra}, S., \& {Rhoads}, J.~E. 2008, \apj, 673, 686

\bibitem[{{Hickox} {et~al.}(2012){Hickox}, {Wardlow}, {Smail}, {Myers},
  {Alexander}, {Swinbank}, {Danielson}, {Stott}, {Chapman}, {Coppin}, {Dunlop},
  {Gawiser}, {Lutz}, {van der Werf}, \& {Wei{\ss}}}]{hickox2012}
{Hickox}, R.~C., {Wardlow}, J.~L., {Smail}, I., {et~al.} 2012, \mnras, 421, 284

\bibitem[{{Hodge} {et~al.}(2012){Hodge}, {Carilli}, {Walter}, {de Blok},
  {Riechers}, {Daddi}, \& {Lentati}}]{hodge2012}
{Hodge}, J.~A., {Carilli}, C.~L., {Walter}, F., {et~al.} 2012, \apj, 760, 11

\bibitem[{{Hodge} {et~al.}(2016){Hodge}, {Swinbank}, {Simpson}, {Smail},
  {Walter}, {Alexander}, {Bertoldi}, {Biggs}, {Brandt}, {Chapman}, {Chen},
  {Coppin}, {Cox}, {Dannerbauer}, {Edge}, {Greve}, {Ivison}, {Karim},
  {Knudsen}, {Menten}, {Rix}, {Schinnerer}, {Wardlow}, {Weiss}, \& {van der
  Werf}}]{hodge2016}
{Hodge}, J.~A., {Swinbank}, A.~M., {Simpson}, J.~M., {et~al.} 2016, \apj, 833,
  103

\bibitem[{{Huang} {et~al.}(2013){Huang}, {Ferguson}, {Ravindranath}, \&
  {Su}}]{huang2013}
{Huang}, K.-H., {Ferguson}, H.~C., {Ravindranath}, S., \& {Su}, J. 2013, \apj,
  765, 68

\bibitem[{{Huertas-Company} {et~al.}(2015){Huertas-Company}, {Gravet},
  {Cabrera-Vives}, {P{\'e}rez-Gonz{\'a}lez}, {Kartaltepe}, {Barro}, {Bernardi},
  {Mei}, {Shankar}, {Dimauro}, {Bell}, {Kocevski}, {Koo}, {Faber}, \&
  {Mcintosh}}]{huertas-company2015}
{Huertas-Company}, M., {Gravet}, R., {Cabrera-Vives}, G., {et~al.} 2015, \apjs,
  221, 8

\bibitem[{{Ikarashi} {et~al.}(2015){Ikarashi}, {Ivison}, {Caputi}, {Aretxaga},
  {Dunlop}, {Hatsukade}, {Hughes}, {Iono}, {Izumi}, {Kawabe}, {Kohno}, {Lagos},
  {Motohara}, {Nakanishi}, {Ohta}, {Tamura}, {Umehata}, {Wilson}, {Yabe}, \&
  {Yun}}]{ikarashi2015}
{Ikarashi}, S., {Ivison}, R.~J., {Caputi}, K.~I., {et~al.} 2015, \apj, 810, 133

\bibitem[{{Ilbert} {et~al.}(2013){Ilbert}, {McCracken}, {Le F{\`e}vre},
  {Capak}, {Dunlop}, {Karim}, {Renzini}, {Caputi}, {Boissier}, {Arnouts},
  {Aussel}, {Comparat}, {Guo}, {Hudelot}, {Kartaltepe}, {Kneib}, {Krogager},
  {Le Floc'h}, {Lilly}, {Mellier}, {Milvang-Jensen}, {Moutard}, {Onodera},
  {Richard}, {Salvato}, {Sanders}, {Scoville}, {Silverman}, {Taniguchi},
  {Tasca}, {Thomas}, {Toft}, {Tresse}, {Vergani}, {Wolk}, \&
  {Zirm}}]{ilbert2013}
{Ilbert}, O., {McCracken}, H.~J., {Le F{\`e}vre}, O., {et~al.} 2013, \aap, 556,
  A55

\bibitem[{{Iono} {et~al.}(2016){Iono}, {Yun}, {Aretxaga}, {Hatsukade},
  {Hughes}, {Ikarashi}, {Izumi}, {Kawabe}, {Kohno}, {Lee}, {Matsuda},
  {Nakanishi}, {Saito}, {Tamura}, {Ueda}, {Umehata}, {Wilson}, {Michiyama}, \&
  {Ando}}]{iono2016}
{Iono}, D., {Yun}, M.~S., {Aretxaga}, I., {et~al.} 2016, \apjl, 829, L10

\bibitem[{{Kov{\'a}cs} {et~al.}(2006){Kov{\'a}cs}, {Chapman}, {Dowell},
  {Blain}, {Ivison}, {Smail}, \& {Phillips}}]{kovacs2006}
{Kov{\'a}cs}, A., {Chapman}, S.~C., {Dowell}, C.~D., {et~al.} 2006, \apj, 650,
  592

\bibitem[{{Le F{\`e}vre} {et~al.}(2000){Le F{\`e}vre}, {Abraham}, {Lilly},
  {Ellis}, {Brinchmann}, {Schade}, {Tresse}, {Colless}, {Crampton},
  {Glazebrook}, {Hammer}, \& {Broadhurst}}]{lefevre2000}
{Le F{\`e}vre}, O., {Abraham}, R., {Lilly}, S.~J., {et~al.} 2000, \mnras, 311,
  565

\bibitem[{{Lilly} {et~al.}(1999){Lilly}, {Eales}, {Gear}, {Hammer}, {Le
  F{\`e}vre}, {Crampton}, {Bond}, \& {Dunne}}]{lilly1999}
{Lilly}, S.~J., {Eales}, S.~A., {Gear}, W.~K.~P., {et~al.} 1999, \apj, 518, 641

\bibitem[{{Lindroos} {et~al.}(2016){Lindroos}, {Knudsen}, {Fan}, {Conway},
  {Coppin}, {Decarli}, {Drouart}, {Hodge}, {Karim}, {Simpson}, \&
  {Wardlow}}]{lindroos2016}
{Lindroos}, L., {Knudsen}, K.~K., {Fan}, L., {et~al.} 2016, \mnras, 462, 1192

\bibitem[{{Lutz} {et~al.}(2016){Lutz}, {Berta}, {Contursi}, {F{\"o}rster
  Schreiber}, {Genzel}, {Graci{\'a}-Carpio}, {Herrera-Camus}, {Netzer},
  {Sturm}, {Tacconi}, {Tadaki}, \& {Veilleux}}]{lutz2016}
{Lutz}, D., {Berta}, S., {Contursi}, A., {et~al.} 2016, \aap, 591, A136

\bibitem[{{MacArthur} {et~al.}(2003){MacArthur}, {Courteau}, \&
  {Holtzman}}]{macarthur2003}
{MacArthur}, L.~A., {Courteau}, S., \& {Holtzman}, J.~A. 2003, \apj, 582, 689

\bibitem[{{Mart{\'{\i}}-Vidal} {et~al.}(2012){Mart{\'{\i}}-Vidal},
  {P{\'e}rez-Torres}, \& {Lobanov}}]{marti2012}
{Mart{\'{\i}}-Vidal}, I., {P{\'e}rez-Torres}, M.~A., \& {Lobanov}, A.~P. 2012,
  \aap, 541, A135

\bibitem[{{Mart{\'{\i}}-Vidal} {et~al.}(2014){Mart{\'{\i}}-Vidal}, {Vlemmings},
  {Muller}, \& {Casey}}]{marti2014}
{Mart{\'{\i}}-Vidal}, I., {Vlemmings}, W.~H.~T., {Muller}, S., \& {Casey}, S.
  2014, \aap, 563, A136

\bibitem[{{Mo} {et~al.}(1998){Mo}, {Mao}, \& {White}}]{mo1998}
{Mo}, H.~J., {Mao}, S., \& {White}, S.~D.~M. 1998, \mnras, 295, 319

\bibitem[{{Momcheva} {et~al.}(2016){Momcheva}, {Brammer}, {van Dokkum},
  {Skelton}, {Whitaker}, {Nelson}, {Fumagalli}, {Maseda}, {Leja}, {Franx},
  {Rix}, {Bezanson}, {Da Cunha}, {Dickey}, {F{\"o}rster Schreiber},
  {Illingworth}, {Kriek}, {Labb{\'e}}, {Ulf Lange}, {Lundgren}, {Magee},
  {Marchesini}, {Oesch}, {Pacifici}, {Patel}, {Price}, {Tal}, {Wake}, {van der
  Wel}, \& {Wuyts}}]{momcheva2016}
{Momcheva}, I.~G., {Brammer}, G.~B., {van Dokkum}, P.~G., {et~al.} 2016, \apjs,
  225, 27

\bibitem[{{Murphy} {et~al.}(2011){Murphy}, {Condon}, {Schinnerer}, {Kennicutt},
  {Calzetti}, {Armus}, {Helou}, {Turner}, {Aniano}, {Beir{\~a}o}, {Bolatto},
  {Brandl}, {Croxall}, {Dale}, {Donovan Meyer}, {Draine}, {Engelbracht},
  {Hunt}, {Hao}, {Koda}, {Roussel}, {Skibba}, \& {Smith}}]{murphy2011}
{Murphy}, E.~J., {Condon}, J.~J., {Schinnerer}, E., {et~al.} 2011, \apj, 737,
  67

\bibitem[{{Oesch} {et~al.}(2010){Oesch}, {Bouwens}, {Carollo}, {Illingworth},
  {Magee}, {Trenti}, {Stiavelli}, {Franx}, {Labb{\'e}}, \& {van
  Dokkum}}]{oesch2010}
{Oesch}, P.~A., {Bouwens}, R.~J., {Carollo}, C.~M., {et~al.} 2010, \apjl, 725,
  L150

\bibitem[{{Ono} {et~al.}(2013){Ono}, {Ouchi}, {Curtis-Lake}, {Schenker},
  {Ellis}, {McLure}, {Dunlop}, {Robertson}, {Koekemoer}, {Bowler}, {Rogers},
  {Schneider}, {Charlot}, {Stark}, {Shimasaku}, {Furlanetto}, \&
  {Cirasuolo}}]{ono2013}
{Ono}, Y., {Ouchi}, M., {Curtis-Lake}, E., {et~al.} 2013, \apj, 777, 155

\bibitem[{{Oteo} {et~al.}(2016){Oteo}, {Zwaan}, {Ivison}, {Smail}, \&
  {Biggs}}]{oteo2016}
{Oteo}, I., {Zwaan}, M.~A., {Ivison}, R.~J., {Smail}, I., \& {Biggs}, A.~D.
  2016, ArXiv e-prints, arXiv:1607.06464

\bibitem[{{Planck Collaboration} {et~al.}(2011){Planck Collaboration},
  {Abergel}, {Ade}, {Aghanim}, {Arnaud}, {Ashdown}, {Aumont}, {Baccigalupi},
  {Balbi}, {Banday}, {Barreiro}, {Bartlett}, {Battaner}, {Benabed},
  {Beno{\^\i}t}, {Bernard}, {Bersanelli}, {Bhatia}, {Bock}, {Bonaldi}, {Bond},
  {Borrill}, {Bouchet}, {Boulanger}, {Bucher}, {Burigana}, {Cabella},
  {Cardoso}, {Catalano}, {Cay{\'o}n}, {Challinor}, {Chamballu}, {Chiang},
  {Chiang}, {Christensen}, {Colombi}, {Couchot}, {Coulais}, {Crill}, {Cuttaia},
  {Dame}, {Danese}, {Davies}, {Davis}, {de Bernardis}, {de Gasperis}, {de
  Rosa}, {de Zotti}, {Delabrouille}, {Delouis}, {D{\'e}sert}, {Dickinson},
  {Donzelli}, {Dor{\'e}}, {D{\"o}rl}, {Douspis}, {Dupac}, {Efstathiou},
  {En{\ss}lin}, {Finelli}, {Forni}, {Frailis}, {Franceschi}, {Galeotta},
  {Ganga}, {Giard}, {Giardino}, {Giraud-H{\'e}raud}, {Gonz{\'a}lez-Nuevo},
  {G{\'o}rski}, {Gratton}, {Gregorio}, {Grenier}, {Gruppuso}, {Hansen},
  {Harrison}, {Henrot-Versill{\'e}}, {Herranz}, {Hildebrandt}, {Hivon},
  {Hobson}, {Holmes}, {Hovest}, {Hoyland}, {Huffenberger}, {Jaffe}, {Jaffe},
  {Jones}, {Juvela}, {Keih{\"a}nen}, {Keskitalo}, {Kisner}, {Kneissl}, {Knox},
  {Kurki-Suonio}, {Lagache}, {L{\"a}hteenm{\"a}ki}, {Lamarre}, {Lasenby},
  {Laureijs}, {Lawrence}, {Leach}, {Leonardi}, {Leroy}, {Lilje},
  {Linden-V{\o}rnle}, {L{\'o}pez-Caniego}, {Lubin}, {Mac{\'{\i}}as-P{\'e}rez},
  {MacTavish}, {Maffei}, {Mandolesi}, {Mann}, {Maris}, {Marshall},
  {Mart{\'{\i}}nez-Gonz{\'a}lez}, {Masi}, {Matarrese}, {Matthai}, {Mazzotta},
  {McGehee}, {Meinhold}, {Melchiorri}, {Mendes}, {Mennella},
  {Miville-Desch{\^e}nes}, {Moneti}, {Montier}, {Morgante}, {Mortlock},
  {Munshi}, {Murphy}, {Naselsky}, {Natoli}, {Netterfield},
  {N{\o}rgaard-Nielsen}, {Noviello}, {Novikov}, {Novikov}, {Osborne}, {Pajot},
  {Paladini}, {Pasian}, {Patanchon}, {Perdereau}, {Perotto}, {Perrotta},
  {Piacentini}, {Piat}, {Plaszczynski}, {Pointecouteau}, {Polenta}, {Ponthieu},
  {Poutanen}, {Pr{\'e}zeau}, {Prunet}, {Puget}, {Rachen}, {Reach}, {Rebolo},
  {Reich}, {Renault}, {Ricciardi}, {Riller}, {Ristorcelli}, {Rocha}, {Rosset},
  {Rubi{\~n}o-Mart{\'{\i}}n}, {Rusholme}, {Sandri}, {Santos}, {Savini},
  {Scott}, {Seiffert}, {Shellard}, {Smoot}, {Starck}, {Stivoli}, {Stolyarov},
  {Stompor}, {Sudiwala}, {Sygnet}, {Tauber}, {Terenzi}, {Toffolatti}, {Tomasi},
  {Torre}, {Tristram}, {Tuovinen}, {Umana}, {Valenziano}, {Varis}, {Vielva},
  {Villa}, {Vittorio}, {Wade}, {Wandelt}, {Wilkinson}, {Ysard}, {Yvon},
  {Zacchei}, \& {Zonca}}]{planck2011}
{Planck Collaboration}, {Abergel}, A., {Ade}, P.~A.~R., {et~al.} 2011, \aap,
  536, A21

\bibitem[{{Rodighiero} {et~al.}(2011){Rodighiero}, {Daddi}, {Baronchelli},
  {Cimatti}, {Renzini}, {Aussel}, {Popesso}, {Lutz}, {Andreani}, {Berta},
  {Cava}, {Elbaz}, {Feltre}, {Fontana}, {F{\"o}rster Schreiber},
  {Franceschini}, {Genzel}, {Grazian}, {Gruppioni}, {Ilbert}, {Le Floch},
  {Magdis}, {Magliocchetti}, {Magnelli}, {Maiolino}, {McCracken}, {Nordon},
  {Poglitsch}, {Santini}, {Pozzi}, {Riguccini}, {Tacconi}, {Wuyts}, \&
  {Zamorani}}]{rodighiero2011}
{Rodighiero}, G., {Daddi}, E., {Baronchelli}, I., {et~al.} 2011, \apjl, 739,
  L40

\bibitem[{{Rujopakarn} {et~al.}(2016){Rujopakarn}, {Dunlop}, {Rieke}, {Ivison},
  {Cibinel}, {Nyland}, {Jagannathan}, {Silverman}, {Alexander}, {Biggs},
  {Bhatnagar}, {Ballantyne}, {Dickinson}, {Elbaz}, {Geach}, {Hayward},
  {Kirkpatrick}, {McLure}, {Micha{\l}owski}, {Miller}, {Narayanan}, {Owen},
  {Pannella}, {Papovich}, {Pope}, {Rau}, {Robertson}, {Scott}, {Swinbank}, {van
  der Werf}, {van Kampen}, {Weiner}, \& {Windhorst}}]{rujopakarn2016}
{Rujopakarn}, W., {Dunlop}, J.~S., {Rieke}, G.~H., {et~al.} 2016, \apj, 833, 12

\bibitem[{{Salpeter}(1955)}]{salpeter1955}
{Salpeter}, E.~E. 1955, \apj, 121, 161

\bibitem[{{Santini} {et~al.}(2015){Santini}, {Ferguson}, {Fontana}, {Mobasher},
  {Barro}, {Castellano}, {Finkelstein}, {Grazian}, {Hsu}, {Lee}, {Lee},
  {Pforr}, {Salvato}, {Wiklind}, {Wuyts}, {Almaini}, {Cooper}, {Galametz},
  {Weiner}, {Amorin}, {Boutsia}, {Conselice}, {Dahlen}, {Dickinson},
  {Giavalisco}, {Grogin}, {Guo}, {Hathi}, {Kocevski}, {Koekemoer},
  {Kurczynski}, {Merlin}, {Mortlock}, {Newman}, {Paris}, {Pentericci},
  {Simons}, \& {Willner}}]{santini2015}
{Santini}, P., {Ferguson}, H.~C., {Fontana}, A., {et~al.} 2015, \apj, 801, 97

\bibitem[{{Scoville} {et~al.}(2007){Scoville}, {Aussel}, {Brusa}, {Capak},
  {Carollo}, {Elvis}, {Giavalisco}, {Guzzo}, {Hasinger}, {Impey}, {Kneib},
  {LeFevre}, {Lilly}, {Mobasher}, {Renzini}, {Rich}, {Sanders}, {Schinnerer},
  {Schminovich}, {Shopbell}, {Taniguchi}, \& {Tyson}}]{scoville2007}
{Scoville}, N., {Aussel}, H., {Brusa}, M., {et~al.} 2007, \apjs, 172, 1

\bibitem[{{Shen} {et~al.}(2003){Shen}, {Mo}, {White}, {Blanton}, {Kauffmann},
  {Voges}, {Brinkmann}, \& {Csabai}}]{shen2003}
{Shen}, S., {Mo}, H.~J., {White}, S.~D.~M., {et~al.} 2003, \mnras, 343, 978

\bibitem[{{Shibuya} {et~al.}(2015){Shibuya}, {Ouchi}, \&
  {Harikane}}]{shibuya2015}
{Shibuya}, T., {Ouchi}, M., \& {Harikane}, Y. 2015, \apjs, 219, 15

\bibitem[{{Simpson} {et~al.}(2014){Simpson}, {Swinbank}, {Smail}, {Alexander},
  {Brandt}, {Bertoldi}, {de Breuck}, {Chapman}, {Coppin}, {da Cunha},
  {Danielson}, {Dannerbauer}, {Greve}, {Hodge}, {Ivison}, {Karim}, {Knudsen},
  {Poggianti}, {Schinnerer}, {Thomson}, {Walter}, {Wardlow}, {Wei{\ss}}, \&
  {van der Werf}}]{simpson2014}
{Simpson}, J.~M., {Swinbank}, A.~M., {Smail}, I., {et~al.} 2014, \apj, 788, 125

\bibitem[{{Simpson} {et~al.}(2015){Simpson}, {Smail}, {Swinbank}, {Almaini},
  {Blain}, {Bremer}, {Chapman}, {Chen}, {Conselice}, {Coppin}, {Danielson},
  {Dunlop}, {Edge}, {Farrah}, {Geach}, {Hartley}, {Ivison}, {Karim}, {Lani},
  {Ma}, {Meijerink}, {Micha{\l}owski}, {Mortlock}, {Scott}, {Simpson},
  {Spaans}, {Thomson}, {van Kampen}, \& {van der Werf}}]{simpson2015a}
{Simpson}, J.~M., {Smail}, I., {Swinbank}, A.~M., {et~al.} 2015, \apj, 799, 81

\bibitem[{{Simpson} {et~al.}(2016){Simpson}, {Smail}, {Swinbank}, {Ivison},
  {Dunlop}, {Geach}, {Almaini}, {Arumugam}, {Bremer}, {Chen}, {Conselice},
  {Coppin}, {Farrah}, {Ibar}, {Hartley}, {Ma}, {Michalowski}, {Spaans},
  {Thomson}, \& {van der Werf}}]{simpson2016}
---. 2016, ArXiv e-prints, arXiv:1611.03084

\bibitem[{{Sonnenfeld} {et~al.}(2017){Sonnenfeld}, {Chan}, {Shu}, {More},
  {Oguri}, {Suyu}, {Wong}, {Lee}, {Coupon}, {Yonehara}, {Bolton}, {Jaelani},
  {Tanaka}, {Miyazaki}, \& {Komiyama}}]{sonnenfeld2017}
{Sonnenfeld}, A., {Chan}, J.~H.~H., {Shu}, Y., {et~al.} 2017, ArXiv e-prints,
  arXiv:1704.01585

\bibitem[{{Speagle} {et~al.}(2014){Speagle}, {Steinhardt}, {Capak}, \&
  {Silverman}}]{speagle2014}
{Speagle}, J.~S., {Steinhardt}, C.~L., {Capak}, P.~L., \& {Silverman}, J.~D.
  2014, \apjs, 214, 15

\bibitem[{{Swinbank} {et~al.}(2012){Swinbank}, {Karim}, {Smail}, {Hodge},
  {Walter}, {Bertoldi}, {Biggs}, {de Breuck}, {Chapman}, {Coppin}, {Cox},
  {Danielson}, {Dannerbauer}, {Ivison}, {Greve}, {Knudsen}, {Menten},
  {Simpson}, {Schinnerer}, {Wardlow}, {Wei{\ss}}, \& {van der
  Werf}}]{swinbank2012}
{Swinbank}, A.~M., {Karim}, A., {Smail}, I., {et~al.} 2012, \mnras, 427, 1066

\bibitem[{{Swinbank} {et~al.}(2014){Swinbank}, {Simpson}, {Smail}, {Harrison},
  {Hodge}, {Karim}, {Walter}, {Alexander}, {Brandt}, {de Breuck}, {da Cunha},
  {Chapman}, {Coppin}, {Danielson}, {Dannerbauer}, {Decarli}, {Greve},
  {Ivison}, {Knudsen}, {Lagos}, {Schinnerer}, {Thomson}, {Wardlow}, {Wei{\ss}},
  \& {van der Werf}}]{swinbank2014}
{Swinbank}, A.~M., {Simpson}, J.~M., {Smail}, I., {et~al.} 2014, \mnras, 438,
  1267

\bibitem[{{Symeonidis} {et~al.}(2013){Symeonidis}, {Vaccari}, {Berta}, {Page},
  {Lutz}, {Arumugam}, {Aussel}, {Bock}, {Boselli}, {Buat}, {Capak}, {Clements},
  {Conley}, {Conversi}, {Cooray}, {Dowell}, {Farrah}, {Franceschini},
  {Giovannoli}, {Glenn}, {Griffin}, {Hatziminaoglou}, {Hwang}, {Ibar},
  {Ilbert}, {Ivison}, {Floc'h}, {Lilly}, {Kartaltepe}, {Magnelli}, {Magdis},
  {Marchetti}, {Nguyen}, {Nordon}, {O'Halloran}, {Oliver}, {Omont},
  {Papageorgiou}, {Patel}, {Pearson}, {P{\'e}rez-Fournon}, {Pohlen}, {Popesso},
  {Pozzi}, {Rigopoulou}, {Riguccini}, {Rosario}, {Roseboom}, {Rowan-Robinson},
  {Salvato}, {Schulz}, {Scott}, {Seymour}, {Shupe}, {Smith}, {Valtchanov},
  {Wang}, {Xu}, {Zemcov}, \& {Wuyts}}]{symeonidis2013}
{Symeonidis}, M., {Vaccari}, M., {Berta}, S., {et~al.} 2013, \mnras, 431, 2317

\bibitem[{{Tacconi} {et~al.}(2008){Tacconi}, {Genzel}, {Smail}, {Neri},
  {Chapman}, {Ivison}, {Blain}, {Cox}, {Omont}, {Bertoldi}, {Greve},
  {F{\"o}rster Schreiber}, {Genel}, {Lutz}, {Swinbank}, {Shapley}, {Erb},
  {Cimatti}, {Daddi}, \& {Baker}}]{tacconi2008}
{Tacconi}, L.~J., {Genzel}, R., {Smail}, I., {et~al.} 2008, \apj, 680, 246

\bibitem[{{Tadaki} {et~al.}(2016){Tadaki}, {Genzel}, {Kodama}, {Wuyts},
  {Wisnioski}, {F{\"o}rster Schreiber}, {Burkert}, {Lang}, {Tacconi}, {Lutz},
  {Belli}, {Davies}, {Hatsukade}, {Hayashi}, {Herrera-Camus}, {Ikarashi},
  {Inoue}, {Kohno}, {Koyama}, {Mendel}, {Nakanishi}, {Shimakawa}, {Suzuki},
  {Tamura}, {Tanaka}, {{\"U}bler}, \& {Wilman}}]{tadaki2016}
{Tadaki}, K.-i., {Genzel}, R., {Kodama}, T., {et~al.} 2016, ArXiv e-prints,
  arXiv:1608.05412

\bibitem[{{Toft} {et~al.}(2014){Toft}, {Smol{\v c}i{\'c}}, {Magnelli}, {Karim},
  {Zirm}, {Michalowski}, {Capak}, {Sheth}, {Schawinski}, {Krogager}, {Wuyts},
  {Sanders}, {Man}, {Lutz}, {Staguhn}, {Berta}, {Mccracken}, {Krpan}, \&
  {Riechers}}]{toft2014}
{Toft}, S., {Smol{\v c}i{\'c}}, V., {Magnelli}, B., {et~al.} 2014, \apj, 782,
  68

\bibitem[{{Ueda} {et~al.}(2008){Ueda}, {Watson}, {Stewart}, {Akiyama},
  {Schwope}, {Lamer}, {Ebrero}, {Carrera}, {Sekiguchi}, {Yamada}, {Simpson},
  {Hasinger}, \& {Mateos}}]{ueda2008}
{Ueda}, Y., {Watson}, M.~G., {Stewart}, I.~M., {et~al.} 2008, \apjs, 179, 124

\bibitem[{{van der Wel} {et~al.}(2012){van der Wel}, {Bell}, {H{\"a}ussler},
  {McGrath}, {Chang}, {Guo}, {McIntosh}, {Rix}, {Barden}, {Cheung}, {Faber},
  {Ferguson}, {Galametz}, {Grogin}, {Hartley}, {Kartaltepe}, {Kocevski},
  {Koekemoer}, {Lotz}, {Mozena}, {Peth}, \& {Peng}}]{vanderwel2012}
{van der Wel}, A., {Bell}, E.~F., {H{\"a}ussler}, B., {et~al.} 2012, \apjs,
  203, 24

\bibitem[{{van der Wel} {et~al.}(2014){van der Wel}, {Franx}, {van Dokkum},
  {Skelton}, {Momcheva}, {Whitaker}, {Brammer}, {Bell}, {Rix}, {Wuyts},
  {Ferguson}, {Holden}, {Barro}, {Koekemoer}, {Chang}, {McGrath},
  {H{\"a}ussler}, {Dekel}, {Behroozi}, {Fumagalli}, {Leja}, {Lundgren},
  {Maseda}, {Nelson}, {Wake}, {Patel}, {Labb{\'e}}, {Faber}, {Grogin}, \&
  {Kocevski}}]{vanderwel2014}
{van der Wel}, A., {Franx}, M., {van Dokkum}, P.~G., {et~al.} 2014, \apj, 788,
  28

\bibitem[{{van Dokkum} {et~al.}(2015){van Dokkum}, {Nelson}, {Franx}, {Oesch},
  {Momcheva}, {Brammer}, {F{\"o}rster Schreiber}, {Skelton}, {Whitaker}, {van
  der Wel}, {Bezanson}, {Fumagalli}, {Illingworth}, {Kriek}, {Leja}, \&
  {Wuyts}}]{van-dokkum2015}
{van Dokkum}, P.~G., {Nelson}, E.~J., {Franx}, M., {et~al.} 2015, \apj, 813, 23

\bibitem[{{Vanzella} {et~al.}(2005){Vanzella}, {Cristiani}, {Dickinson},
  {Kuntschner}, {Moustakas}, {Nonino}, {Rosati}, {Stern}, {Cesarsky}, {Ettori},
  {Ferguson}, {Fosbury}, {Giavalisco}, {Haase}, {Renzini}, {Rettura}, {Serra},
  \& {GOODS Team}}]{vanzella2005}
{Vanzella}, E., {Cristiani}, S., {Dickinson}, M., {et~al.} 2005, \aap, 434, 53

\bibitem[{{Xue} {et~al.}(2011){Xue}, {Luo}, {Brandt}, {Bauer}, {Lehmer},
  {Broos}, {Schneider}, {Alexander}, {Brusa}, {Comastri}, {Fabian}, {Gilli},
  {Hasinger}, {Hornschemeier}, {Koekemoer}, {Liu}, {Mainieri}, {Paolillo},
  {Rafferty}, {Rosati}, {Shemmer}, {Silverman}, {Smail}, {Tozzi}, \&
  {Vignali}}]{xue2011}
{Xue}, Y.~Q., {Luo}, B., {Brandt}, W.~N., {et~al.} 2011, \apjs, 195, 10

\bibitem[{{Yun} {et~al.}(2012){Yun}, {Scott}, {Guo}, {Aretxaga}, {Giavalisco},
  {Austermann}, {Capak}, {Chen}, {Ezawa}, {Hatsukade}, {Hughes}, {Iono},
  {Johnson}, {Kawabe}, {Kohno}, {Lowenthal}, {Miller}, {Morrison}, {Oshima},
  {Perera}, {Salvato}, {Silverman}, {Tamura}, {Williams}, \&
  {Wilson}}]{yun2012}
{Yun}, M.~S., {Scott}, K.~S., {Guo}, Y., {et~al.} 2012, \mnras, 420, 957

\bibitem[{{Zhang} {et~al.}(2016){Zhang}, {Papadopoulos}, {Ivison}, {Galametz},
  {Smith}, \& {Xilouris}}]{zhang2016}
{Zhang}, Z.-Y., {Papadopoulos}, P.~P., {Ivison}, R.~J., {et~al.} 2016, Royal
  Society Open Science, 3, 160025

\end{thebibliography}

\end{document}